\documentclass[aps,prb,reprint,superscriptaddress,longbibliography]{revtex4-2}

\usepackage[T1]{fontenc}
\usepackage{amsmath,amssymb,amsthm,bm,graphicx,physics,mathtools}
\usepackage{xcolor}
\usepackage{hyperref}
\usepackage[capitalize]{cleveref}
\usepackage{multirow}
\usepackage{booktabs}
\hypersetup{colorlinks=true,linkcolor=blue,citecolor=blue,urlcolor=blue}
\usepackage{comment}

\begin{document}

\title{\textcolor{black}{Strain-controlled crossover between Majorana and Andreev bound states in disordered superconductor-semiconductor heterostructures}}

\author{Shubhanshu Karoliya}
\thanks{These two authors contributed equally to this work.}
\affiliation{Department of Physics, Indian Institute of Technology Mandi, Kamand, Mandi, H.P. 175005, India.}
\author{Ekta}
\thanks{These two authors contributed equally to this work.}
\affiliation{Department of Physics, Indian Institute of Technology Delhi, Hauz Khas, New Delhi 110016, India.}
\author{Gargee Sharma}
\affiliation{Department of Physics, Indian Institute of Technology Delhi, Hauz Khas, New Delhi 110016, India.}

\begin{abstract}

The unambiguous identification of topological Majorana-bound states (MBSs) in superconducting hybrid systems is hindered by the presence of trivial low-energy excitations, particularly partially separated Andreev bound states (psABSs), which mimic the Majorana physics. In this work, we demonstrate that spatially nonuniform strain provides a route to systematically control and interconvert these low-energy states. Using tight-binding Bogoliubov--de Gennes simulations, we investigate (i) one-dimensional semiconductor nanowires, and (ii) graphene nanoribbons, incorporating superconductivity, Rashba spin--orbit coupling, Zeeman fields, and disorder. We show that even a small amount of strain can qualitatively reshape the low-energy spectrum by modifying the effective band parameters and redistributing the wavefunction weight. In nanowires, strain tunes the spatial overlap of Majorana components and shifts the topological phase boundary, enabling controlled transitions between trivial states, psABSs, and topological MBSs. In graphene nanoribbons, where multiband effects and edge states generate a dense and hybridized low-energy spectrum, strain suppresses subband mixing, lifts degeneracies, and stabilizes boundary-localized modes. In both systems, we identify regimes in which disorder-induced psABSs are converted into well-separated and robust Majorana-bound states through strain-induced enhancement of nonlocality. Furthermore, we develop an analytical framework based on a position-dependent topological mass and strain-driven domain-wall dynamics, which captures the mechanism underlying these strain-induced crossovers and provides a real-space criterion for the emergence and stability of Majorana modes. Our results suggest that strain can serve as a useful tuning parameter for distinguishing and stabilizing topological MBSs in realistic disordered systems, and offer a compelling pathway toward better control, identification, and eventual utilization of Majorana modes in complex, disordered environments for topological quantum computation.
\end{abstract}

\maketitle

\section{Introduction}

Majorana fermions (MFs), originally proposed as real solutions of the Dirac equation~\cite{Majorana1937}, occupy a central position in modern condensed matter physics due to their emergence as zero-energy quasiparticles in superconducting systems and their potential application in fault-tolerant topological quantum computation (TQC)~\cite{moore1991nonabelions,read2000paired,nayak19962n,Kitaev2003,Nayak2008}. In engineered solid-state platforms, a minimal route to realizing Majorana-bound states (MBSs) involves combining spin-orbit coupling, proximity-induced superconductivity, and a Zeeman field~\cite{sau2010generic,sau2010non,oreg2010helical,lutchyn2010majorana}. This paradigm has been extensively explored in one-dimensional (1D) semiconductor nanowires, where MBSs are expected to localize at the system boundaries and produce quantized zero-bias conductance signatures~\cite{sengupta2001midgap,law2009majorana,flensberg2010tunneling}.

Despite significant experimental progress~\cite{mourik2012signatures,deng2012anomalous,das2012zero,rokhinson2012fractional,churchill2013superconductor,finck2013anomalous,deng2016majorana,zhang2017ballistic,chen2017experimental,nichele2017scaling,albrecht2017transport,o2018hybridization,shen2018parity,sherman2017normal,vaitiekenas2018selective,albrecht2016exponential,Yu_2021,microsoft2025interferometric,mondal2025distinguishing,glodzik2020measure,rossi2020majorana,tanaka2024theory,tanaka2011symmetry,sharma2016tunneling,tanaka2009manipulation}, the unambiguous identification of topological MBSs remains an outstanding challenge. A central difficulty arises from the presence of trivial low-energy states that mimic Majorana signatures~\cite{Brouwer2012,Mi2014,Bagrets2012,pikulin2012zero,ramon2012transport,pan2020physical,moore2018two,moore2018quantized,vuik2018reproducing,Stanescu_Robust,ramon_Jorge2106exceptional,ramon2019nonhermitian,Jorge2019supercurrent,sharma2020hybridization,zeng2020feasibility,zeng2022partially,ramon2020from,Jorge2021distinguishing,zhang2021,DasSarma2021,Frolov2021}. In particular, partially separated Andreev bound states (psABSs), also known as quasi-Majorana modes, can exhibit near-zero-energy behavior and strong Majorana-like characteristics while lacking a clear topological protection~\cite{moore2018two,moore2018quantized}. These states arise naturally in the presence of disorder, smooth confinement, or spatial inhomogeneity, and pose a fundamental obstacle to distinguishing true MBSs from trivial excitations.

While 1D nanowires provide a conceptually clean single-channel platform, graphene nanoribbons offer a complementary and fundamentally richer setting for exploring Majorana physics. Owing to their bipartite lattice structure, edge-dependent states, and multiband nature~\cite{nakada1996edge,castro2009electronic}, graphene systems host a dense and highly hybridized low-energy spectrum when proximitized by an $s$-wave superconductor and subjected to Rashba spin--orbit coupling and a Zeeman field~\cite{ma2025graphene,kaladzhyan2017formation,kaladzhyan2017majorana,PhysRevX.5.041042,laubscher2020majorana,wang2018strain,manesco2019effective}. This multiband character enhances the likelihood of generating psABSs and other trivial near-zero modes, making graphene both a challenging and promising platform for studying the interplay between topology, disorder, and spatial structure.

A key question that emerges in both systems is whether one can \textit{systematically control} the nature of low-energy bound states--namely, whether the trivial states, psABSs, and the true MBSs can be smoothly tuned into one another through experimentally accessible parameters. In this context, we find that spatially nonuniform strain provides control over this crossover. 
There is substantial experimental evidence that strain modifies the electronic and superconducting properties of both one-dimensional nanowires and two-dimensional materials~\cite{signorello2017manipulating,zeng2018correlation,chen2020strain}. 
In graphene, for example, strain engineering is known to reconstruct the low-energy electronic spectrum through emergent gauge fields and pseudomagnetic Landau levels, resulting in strongly inhomogeneous spatial distributions of electronic states observed experimentally~\cite{guinea2010energy,vozmediano2010gauge,pereira2009strain,levy2010strain,kang2021pseudo}. While a few works have explored strain manipulation of Majorana modes in low-dimensional systems~\cite{wang2018strain}, its role in controlling and distinguishing Majorana and Andreev bound states remains largely unexplored.

In this work, we demonstrate that even relatively small spatially varying strain can have a profound impact on the low-energy physics of proximitized systems. Using tight-binding Bogoliubov-de Gennes Hamiltonians, we investigate both one-dimensional nanowires and graphene nanoribbons on equal footing, incorporating proximitized superconductivity, Rashba spin-orbit coupling, Zeeman fields, and spatial disorder. 
We show that strain acts as a unifying control parameter that enables continuous tuning between different classes of low-energy states. In particular, we find that strain can (i) drive a crossover from topological Majorana-bound states to partially separated Andreev bound states by enhancing wave-function overlap and shifting the topological boundary, (ii) drive a converse crossover, i.e., disorder-induced psABSs into well-separated and topologically robust MBSs by restoring nonlocality, and (iii) induce re-entrant behavior between trivial states, psABSs, and MBSs. To uncover the underlying mechanism, we develop an analytical theory of strain-driven topological crossovers based on a local topological mass and its associated domain-wall dynamics. This framework reveals how strain reshapes the effective chemical potential landscape, thereby controlling the formation, motion, and hybridization of Majorana components in real space.

Importantly, the underlying mechanism depends on the dimensionality and band structure of the system. In effectively single-channel nanowires, strain primarily controls the spatial overlap and topological phase boundary, enabling direct manipulation of Majorana components. In contrast, in multiband graphene nanoribbons, strain suppresses hybridization between competing low-energy channels, lifts accidental degeneracies, and reorganizes the spectrum to favor boundary-localized modes. Despite these differences, the central outcome remains universal: \textit{strain provides a controlled pathway for manipulating and interconverting psABSs and Majorana-bound states}. Strain-driven manipulation of low-energy bound states provides a compelling route to enhanced control, identification, and ultimately practical exploitation of Majorana modes in complex, disordered settings relevant for topological quantum computation.

The remainder of the paper is organized as follows. In Sec.~\ref{sec:models}, we introduce the theoretical model for both nanowire and graphene systems, including the implementation of disorder and strain. In Sec.~\ref{sec:polarization}, we review the concept of Majorana polarization in different symmetry classes. In Sec.~\ref{sec:results}, we present our results on strain-driven manipulation of low-energy bound states, highlighting the conversion between psABSs and Majorana modes. In Sec.~\ref{sec:analytical}, we develop a comprehensive analytical theory of strain-driven crossovers, formulating a real-space description in terms of local topological masses, domain-wall motion, and subband projection. 
Finally, Sec.~\ref{conclude} summarizes our conclusions.

\section{Theoretical Model}
\label{sec:models}

We investigate the interplay between proximitized $s-$wave superconductivity, spin-orbit coupling, disorder, and spatially nonuniform strain in low-dimensional systems by constructing tight-binding Bogoliubov-de Gennes (BdG) Hamiltonians for both semiconductor nanowires and graphene nanoribbons. This combination provides a minimal and well-established platform for realizing topological superconductivity and Majorana-bound states. Throughout this work, we employ a lattice discretization and work in the Nambu basis. 
\begin{figure*}
    \centering
    \includegraphics[width=1.99\columnwidth]{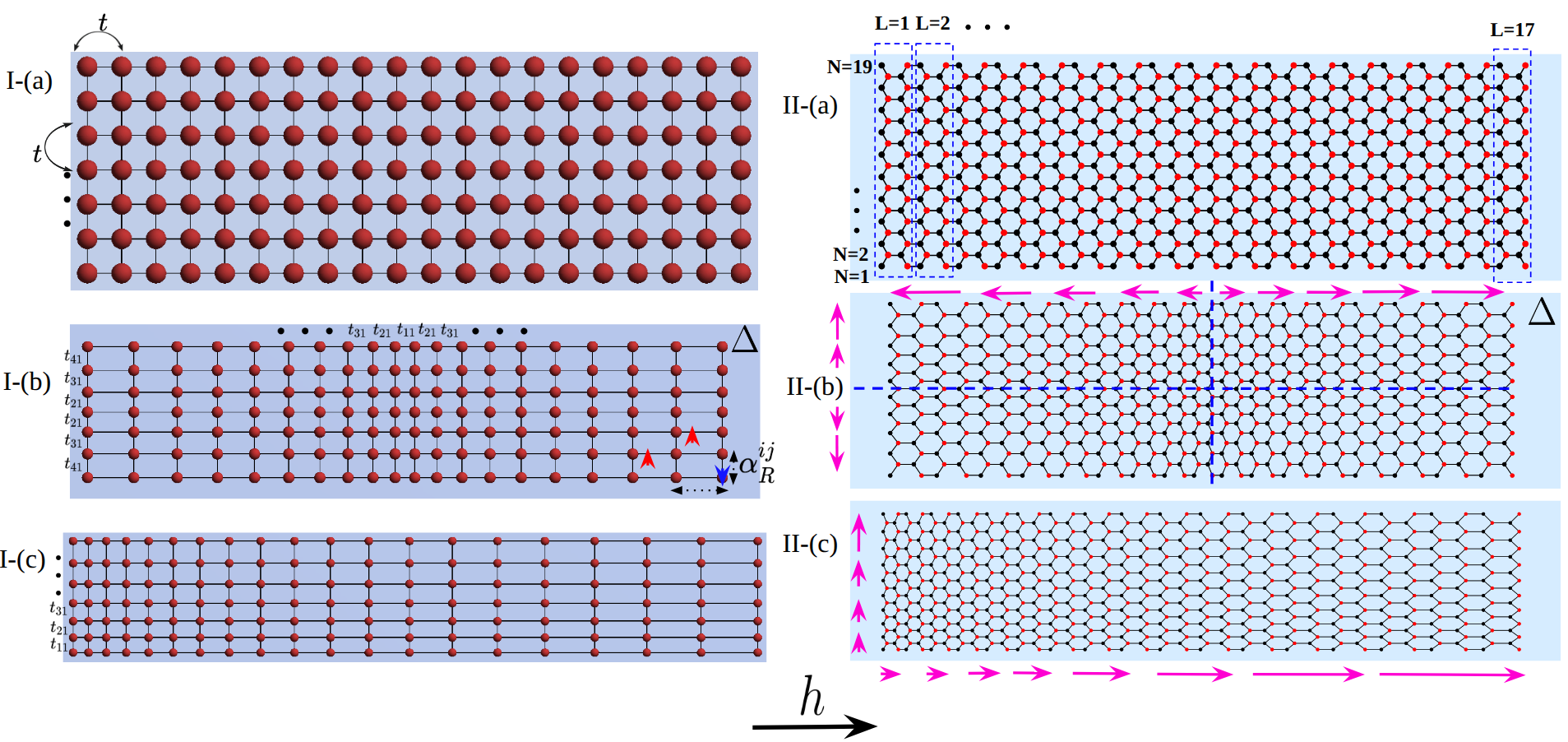}
    \caption{Schematic representation of the strained proximitized systems studied in this work. The light-blue background denotes the parent $s$-wave superconductor with pairing potential $\Delta$. The model includes nearest-neighbor hopping $t$, Rashba spin--orbit coupling $\alpha_{\mathrm{R}}$, and an in-plane Zeeman field $h$ applied along the $x$ direction. Group I corresponds to the quasi-one-dimensional square-lattice geometry: 
(I-a) unstrained chain, 
(I-b) symmetrically strained configuration with the strain centered in the ribbon, and 
(I-c) linearly varying strain along both transverse and longitudinal directions. 
Group II shows the corresponding hexagonal-lattice geometry of graphene: 
(II-a) unstrained ribbon, 
(II-b) symmetrically strained ribbon, and 
(II-c) linearly strained ribbon. 
The magnitude of the arrows represents the spatial variation of strain, i.e., the modulation of local hopping amplitudes. Here, $N$ denotes the number of layers in the transverse direction, and $L$ labels the unit cells along the longitudinal direction.}
    \label{schemetic}
\end{figure*}
\subsection{One-dimensional and quasi-one-dimensional systems}
\label{subsec:1d_model}

We begin with the following effective model that describes the low-energy physics of a one-dimensional semiconductor-superconductor hybrid structure:
\begin{align}
H &= -t\!\!\sum_{\langle i,j\rangle,\sigma}c_{i\sigma}^\dagger c_{j\sigma} + \sum_{i,\sigma}(V_\mathrm{dis}(x_i)-\mu) c^\dagger_{i\sigma}c_{i\sigma} \nonumber\\
&+h\!\sum_{i}c_i^\dagger \sigma_x c_i 
+\frac{\alpha_R}{2}\!\!\sum_{\langle i,j\rangle}\!\left(c_i^\dagger (i\sigma_y) c_j \!+\! \mathrm{h.c.}\right) \nonumber\\
&+\Delta\sum_i\left(c_{i\uparrow}^\dagger c_{i\downarrow}^\dagger \!+\! \mathrm{h.c.}\right),  \label{Eq:Ham1D}
\end{align}
where $\langle i,j\rangle$ refer to the nearest-neighbor sites, the subscript $\sigma$ stands for spin, the column vector $c_i^\dagger = (c_{i\uparrow}^\dagger, c_{i\downarrow}^\dagger)$ is the electron creation operator on site $i$, and $\sigma_\nu$ (with $\nu=x, y, z$) are Pauli matrices. The parameters, $t$, $h$, $\mu$, $\alpha_R$, and $\Delta$ refer to the nearest-neighbor hopping, external Zeeman field strength, chemical potential, Rashba spin-orbit coupling, and the induced superconducting pairing, respectively. The term $V_{\mathrm{dis}}(x_i)$ represents a spatially varying disorder potential. In the absence of disorder and strain, this model undergoes a topological phase transition at $h = \sqrt{\mu^2 + \Delta^2}$, beyond which zero-energy Majorana-bound states emerge at the ends of the wire. This model serves as a minimal reference system for understanding the effects of spatial inhomogeneity introduced by strain and disorder. For numerical calculations, we consider a finite system of $N = 400$ lattice sites, which is sufficiently large to ensure well-separated Majorana modes in the absence of strain and disorder. For the 1D system, unless otherwise specified, we use the following values for the parameters~\cite{zeng2022partially}: $a=5$nm is the lattice constant, $m=0.03 m_e$, where $m_e$ is the bare electronic mass, hopping parameter $t=50$meV ($t=\hbar^2/2ma^2$), $\alpha_R=250$meV$/a$, and $\Delta = 0.3$meV. The length of the wire is chosen to be 2$\mu$m.

To extend this framework to quasi-one-dimensional systems, we consider an array of coupled parallel chains along the transverse ($y$) direction. The transverse coupling introduces additional hopping and spin--orbit terms,
\begin{equation}
H_\perp = -t \sum_{\langle i,j \rangle_y,\sigma} c_{i\sigma}^\dagger c_{j\sigma}
- \frac{\alpha_R}{2} \sum_{\langle i,j \rangle_y} \left( c_i^\dagger i\sigma_x c_j + \mathrm{H.c.} \right),
\label{Ham_q1D_long}
\end{equation}
leading to the full Hamiltonian $H_{\mathrm{q1D}} = H_{\mathrm{1D}} + H_\perp$.
The transverse Rashba term breaks the effective chiral symmetry~\cite{tewari2012topological} present in strictly one-dimensional systems, placing the quasi-one-dimensional system in symmetry class D. This distinction becomes important for defining Majorana polarization and for understanding the robustness of low-energy states in the presence of disorder.
The quasi-one-dimensional system can be constructed by coupling multiple parallel chains. 

\subsection{Graphene nanoribbons}
\label{subsec:graphene_model}
We now turn to graphene nanoribbons, which provide a quasi-one-dimensional platform with a bipartite honeycomb lattice structure and geometry-dependent edge states. 
We describe the system using a tight-binding BdG Hamiltonian in the Nambu basis $\Psi_j = (c_{j\uparrow}, c_{j\downarrow}, c_{j\downarrow}^\dagger, -c_{j\uparrow}^\dagger)^T$. The Hamiltonian can be written as
\begin{equation} \label{Ham_graph}
H_{\mathrm{G}} = H_0 + H_R + H_B,
\end{equation}
where the individual contributions are defined below.
The normal-state and superconducting terms are given by
\begin{equation}
H_0 = \sum_j \Psi_j^\dagger \left[ (V(\mathbf{r}_j) - \mu)\tau^z - \Delta \tau^x \right] \Psi_j
- t \sum_{\langle i,j \rangle} \Psi_i^\dagger \tau^z \Psi_j.
\label{eq:Graphene_H0_long}
\end{equation}
The Zeeman coupling due to an external magnetic field $\vec{h}$ is
\begin{equation}
H_B = \sum_j \Psi_j^\dagger (\vec{h} \cdot \vec{\sigma}) \Psi_j,
\end{equation}
where $\vec{\sigma}$ are Pauli matrices acting in spin space. We neglect orbital magnetic effects, which is justified for in-plane magnetic fields or in the presence of magnetic proximity coupling.
The inversion symmetry-breaking Rashba spin-orbit interaction is modeled as
\begin{equation}
H_R = i \alpha_R \sum_{\langle i,j \rangle}
\Psi_i^\dagger \left[ (\boldsymbol{\delta}_{ij} \times \vec{\sigma}) \cdot \hat{z} \right] \tau^z \Psi_j,
\label{eq:Graphene_HR_long}
\end{equation}
where $\boldsymbol{\delta}_{ij}$ is the unit vector connecting nearest-neighbor sites.
We consider both armchair and zigzag edge terminations by imposing open boundary conditions along the transverse direction. Graphene nanoribbons with shorter edges of zigzag type host localized edge states even in the absence of superconductivity, whereas nanoribbons with shorter edges of armchair type exhibit strong valley mixing and do not support such zero-energy edge modes~\cite{nakada1996edge, castro2009electronic}. These differences play an important role in determining the stability and spatial structure of low-energy bound states in the superconducting regime. In the presence of proximity-induced $s$-wave superconductivity, Rashba spin-orbit coupling, and an external magnetic field, zigzag nanoribbons provide a more favorable platform for realizing and analyzing Majorana physics due to their intrinsic edge-localized states~\cite{karoliya2025majorana}. For this reason, in the main text, we focus on nanoribbons with shorter zigzag edges, where the interplay between superconductivity, spin-orbit coupling, and edge states leads to a richer and more robust low-energy phenomenology. For completeness, results for nanoribbons with shorter armchair edges are presented in the Appendix~\ref{App:zzg}, where the absence of intrinsic edge states leads to comparatively less robust low-energy modes. In our numerical calculations, we consider graphene nanoribbons consisting of $N$ layers in the transverse direction and $L$ unit cells along the longitudinal direction (see Fig.~\ref{schemetic}). Each unit cell contains two lattice sites corresponding to the A and B sublattices, resulting in a total of $2NL$ sites. Unless stated otherwise, we use $N = 3$ and $L = 171$, corresponding to a system size of 1026 lattice sites;
all the energy scales are expressed in terms of the hopping strength $t$, and all the distances are expressed in terms of the nearest-neighbor distance $a$. Also we fixed $\alpha = 0.5t$, $\Delta = 0.2t$.

\subsection{Modeling disorder}
\label{subsec:disorder}
To capture realistic experimental conditions, we introduce disorder through a spatially varying scalar potential $V_{\mathrm{dis}}(\mathbf{r})$. This term enters the BdG Hamiltonian as an onsite energy shift in both nanowire and graphene systems.
We model disorder as a superposition of $N_d$ randomly distributed impurities with finite spatial extent. The potential generated by a single impurity located at $\mathbf{r}_i$ is taken as
\begin{equation}
V_{\mathrm{imp}}^{(i)}(\mathbf{r}) = A_i \exp\left(-\frac{|\mathbf{r} - \mathbf{r}_i|}{\lambda}\right),
\end{equation}
where $\lambda$ is the characteristic decay length and $A_i$ are Gaussian-distributed random amplitudes with zero mean.
The total disorder potential is then
\begin{equation}\label{eq:disorder}
V_{\mathrm{dis}}(\mathbf{r}) = V_0 \sum_{i=1}^{N_d} A_i
\exp\left(-\frac{|\mathbf{r} - \mathbf{r}_i|}{\lambda}\right),
\end{equation}
where $V_0$ controls the overall disorder strength. This correlated disorder suppresses large-momentum scattering processes (e.g., intervalley scattering in graphene) and thus captures the smooth electrostatic inhomogeneity arising from charge puddles and gate fluctuations in realistic nanostructures~\cite{das2011electronic}.

\subsection{Modeling of spatially nonuniform strain}
\label{subsec:strain}
\subsubsection{Strain in one-dimensional systems}
We model strain phenomenologically as a spatial modulation of the effective tight-binding parameters. In particular, strain is incorporated through position-dependent renormalization of the nearest-neighbor hopping amplitude and Rashba spin--orbit coupling,
\begin{equation}
t_i = t_0 \bigl(1 - \varepsilon_i\bigr), \qquad
\alpha_{R,i} = \alpha_{R,0} \bigl(1 - \varepsilon_i\bigr),
\end{equation}
where $i$ labels the lattice sites and $\varepsilon_i$ denotes the local strain profile. This effective description captures the leading influence of strain on the low-energy electronic structure without explicitly solving the underlying elasticity problem. We consider two representative strain configurations. In the first, strain is applied symmetrically from both ends toward the center of the system,
\begin{equation}
\varepsilon_i =
\begin{cases}
\varepsilon_0 \left(\dfrac{N/2 - i}{N/2}\right), & i \leq N/2, \\
\varepsilon_0 \left(\dfrac{i - N/2}{N/2}\right), & i > N/2,
\end{cases}
\end{equation}
which yields an approximately inversion-symmetric strain profile about the center of the wire. This configuration primarily modifies the spatial overlap between low-energy bound states without introducing a directional bias. In the second configuration, strain is applied as a monotonic gradient along the wire,
\begin{equation}
\varepsilon_i = \varepsilon_0 \frac{i}{N},
\end{equation}
which explicitly breaks inversion symmetry and generates a spatial bias across the system. This asymmetric profile enables selective manipulation of edge-localized states by preferentially shifting their localization toward one end of the wire.

\subsubsection{Strain tensor description in graphene}

In graphene, strain is most naturally described within a continuum elasticity framework in terms of a displacement field $\mathbf{u}(\mathbf{r}) = (u_x, u_y)$, which maps the original lattice coordinates $\mathbf{r} = (x,y)$ to deformed coordinates according to
\begin{equation}
\mathbf{r} \rightarrow \mathbf{r}' = \mathbf{r} + \mathbf{u}(\mathbf{r}),
\end{equation}
such that
\begin{equation}
x' = x + u_x(x,y), \qquad y' = y + u_y(x,y).
\end{equation}
The local deformation of the lattice is characterized by the strain tensor
\begin{equation}
\epsilon_{\mu\nu} = \frac{1}{2} \left( \partial_\mu u_\nu + \partial_\nu u_\mu \right),
\end{equation}
which quantifies changes in bond lengths and relative orientations of neighboring sites; see, e.g., Refs.~\cite{guinea2010energy, vozmediano2010gauge, pereira2009strain}. The diagonal components $\epsilon_{xx}$ and $\epsilon_{yy}$ correspond to normal (uniaxial) strain along the $x$ and $y$ directions, respectively, while the off-diagonal component $\epsilon_{xy}$ describes shear deformation. Different physical strain configurations are realized depending on which components of the tensor are nonzero. In particular, strain fields with suitable nonuniform shear components can generate effective gauge fields and pseudomagnetic effects in graphene~\cite{guinea2010energy}. In the present work, however, we focus on diagonal strain profiles to isolate the effect of position-dependent hopping renormalization.

\paragraph{General quadratic displacement and uniaxial strain.}
To model a smooth, spatially varying strain profile effectively, we consider a quadratic displacement field of the form
\begin{align}
u_x(x) = \alpha_x x^2, \qquad u_y(y) = \alpha_y y^2,
\end{align}
where $\alpha_x$ and $\alpha_y$ control the strength of deformation along the longitudinal ($x$) and transverse ($y$) directions, respectively. Substituting this form into the definition of the strain tensor yields
\begin{equation}
\bm{\epsilon} =
\begin{pmatrix}
\partial_x u_x & \frac{1}{2}(\partial_x u_y + \partial_y u_x) \\
\frac{1}{2}(\partial_x u_y + \partial_y u_x) & \partial_y u_y
\end{pmatrix}.
\end{equation}
Since $u_x$ depends only on $x$ and $u_y$ depends only on $y$, the mixed derivatives vanish identically,
\begin{equation}
\partial_x u_y = 0, \qquad \partial_y u_x = 0,
\end{equation}
implying that the shear component $\epsilon_{xy}$ is zero. The strain tensor therefore reduces to a purely diagonal form,
\begin{equation}
\bm{\epsilon} =
\begin{pmatrix}
2\alpha_x x & 0 \\
0 & 2\alpha_y y
\end{pmatrix}.
\end{equation}
This allows us to recover different uniaxial strain configurations as limiting cases. For $\alpha_y = 0$, only $\epsilon_{xx} = 2\alpha_x x$ is nonzero, corresponding to uniaxial strain applied along the longitudinal direction; for $\alpha_x = 0$, only $\epsilon_{yy} = 2\alpha_y y$ is nonzero, corresponding to uniaxial strain along the transverse direction. In both cases, the quadratic displacement field generates a linear strain gradient, $\epsilon_{xx} \propto x$ and $\epsilon_{yy} \propto y$, thus the magnitude of strain increases continuously away from the center of the system toward its boundaries. Such spatially varying strain modifies the local electronic structure by inducing position-dependent changes in bond lengths and hopping amplitudes. 
More general strain configurations, including shear deformation, can be obtained by allowing displacement fields with mixed spatial dependence. However, in the present work, we restrict our analysis to diagonal strain profiles in order to isolate the effects of normal strain and provide a transparent connection between strain gradients and the resulting electronic properties.

\begin{figure}
    \centering
    \includegraphics[width=\columnwidth]{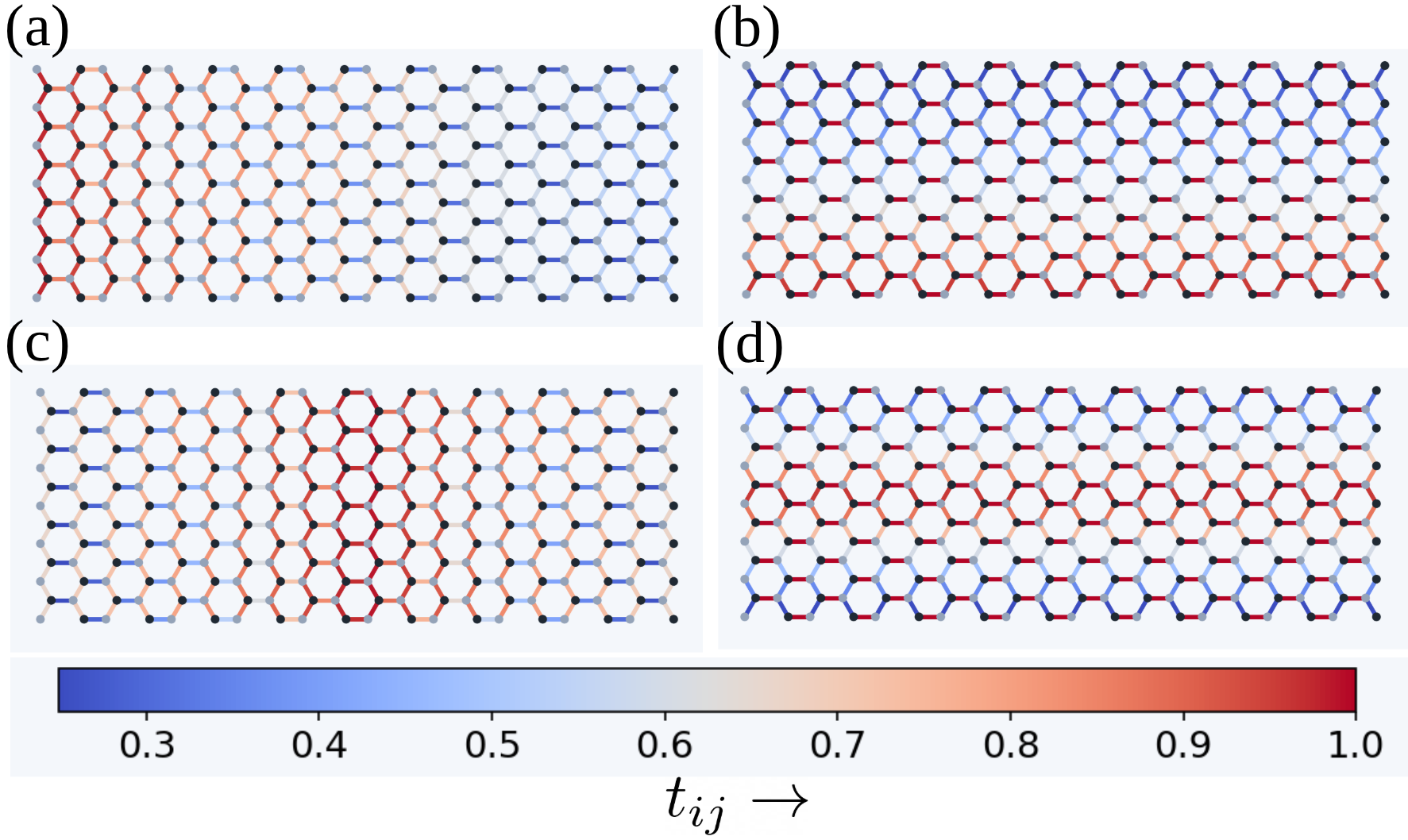}
    \caption{{Spatial modulation of nearest-neighbor hopping amplitudes $t_{ij}$ on the hexagonal lattice for the various strain profiles. The color scale represents the local value of $t_{ij}$, with blue (red) indicating reduced (enhanced) hopping amplitudes. 
(a) Linearly varying strain along the $x$ direction, generated by stretching the right edge.
(b) Linearly varying strain along the $y$ direction, generated by stretching the top edge.
(c) Symmetric strain profile along $y$, generated by stretching both lateral edges.
(d) Symmetric strain profile along $x$, generated by stretching the top and bottom edges. 
These profiles illustrate the resulting bond-dependent hopping inhomogeneity induced by different mechanical deformations of the ribbon.
}}
    \label{fig:hopping_strain}
\end{figure}

\paragraph{Relation to tight-binding parameters.}
Mechanical deformation modifies the graphene lattice geometry and thereby renormalizes the microscopic hopping and spin--orbit parameters entering the tight-binding Hamiltonian. In graphene, each lattice site is connected to three nearest neighbours through the bond vectors:
\begin{equation}
\boldsymbol{\delta}_1=a_0(0,-1), \quad
\boldsymbol{\delta}_2=a_0\left(\frac{\sqrt{3}}{2},\frac{1}{2}\right), \quad
\boldsymbol{\delta}_3=a_0\left(-\frac{\sqrt{3}}{2},\frac{1}{2}\right),
\end{equation}
where $a_0$ is the equilibrium nearest-neighbour bond length. Under a smooth lattice deformation described by the strain tensor $\epsilon$, these bond vectors transform as
\begin{equation}
\boldsymbol{\delta}_n'=(I+\epsilon)\boldsymbol{\delta}_n,
\qquad n=1,2,3,
\end{equation}
where $I$ is the identity matrix. 
The nearest-neighbour hopping amplitudes then acquire bond dependence according to the standard exponential form
\begin{equation}
t_n=t_0\exp\left[-\beta\left(\frac{|\boldsymbol{\delta}_n'|}{a_0}-1\right)\right],
\qquad n=1,2,3,
\label{eq:strained_hopping}
\end{equation}
where $t_0$ is the unstrained hopping parameter and $\beta$ is the Gr\"uneisen parameter ($\beta\approx2$--$3$)~\cite{pereira2009strain}. Thus, strain generally induces anisotropic hoppings, such that the three nearest-neighbour matrix elements are unequal.
For the diagonal strain profiles employed in the present work, the three bond lengths become
\begin{align}
d_1 &= a_0(1+\epsilon_{yy}), \\
d_2 &= a_0\sqrt{\frac{3}{4}(1+\epsilon_{xx})^2+\frac{1}{4}(1+\epsilon_{yy})^2}, \\
d_3 &= a_0\sqrt{\frac{3}{4}(1+\epsilon_{xx})^2+\frac{1}{4}(1+\epsilon_{yy})^2}.
\end{align}
Hence, for diagonal uniaxial strain, one obtains two independent hoppings, with $t_2=t_3\neq t_1$. The relative magnitude of these hoppings depends on whether the deformation is predominantly along the armchair ($x$) or zigzag ($y$) direction.
The Rashba spin--orbit coupling is likewise sensitive to local bond geometry. In a bond-resolved description, one may write
\begin{equation}
\alpha_n=\alpha_0\,f(d_n),
\qquad n=1,2,3,
\end{equation}
where $\alpha_0$ is the unstrained Rashba coupling and $f(d_n)$ encodes the strain-induced renormalization. For weak deformations, both hopping and Rashba terms may be linearized in the local strain amplitude.
In the numerical calculations presented here, we adopt an effective low-strain description in which the dominant effect of strain is captured through a smooth position-dependent renormalization of the local parameters,
\begin{equation}
t_{ij}(\mathbf r)=t_0\left[1-\varepsilon(\mathbf r)\right], \qquad
\alpha_{ij}(\mathbf r)=\alpha_0\left[1-\varepsilon(\mathbf r)\right],
\label{eq:effective_strain_model}
\end{equation}
where $\varepsilon(\mathbf r)$ denotes the local strain magnitude extracted from the continuum strain field. The resulting spatial modulation of the nearest-neighbor hopping amplitudes $t_{ij}$ under the different applied strain profiles is illustrated schematically in Fig.~\ref{fig:hopping_strain}. For uniaxial strain applied along the longitudinal direction of the ribbon, we identify
\begin{equation}
\varepsilon(\mathbf r)\sim \epsilon_{xx}(\mathbf r)
\quad \text{(armchair)}, \qquad
\varepsilon(\mathbf r)\sim \epsilon_{yy}(\mathbf r)
\quad \text{(zigzag)}.
\end{equation}
Equation~(\ref{eq:effective_strain_model}) therefore provides a coarse-grained implementation of the microscopic bond-resolved relations above, while retaining the leading physical consequence relevant for the present work: strain-induced spatial inhomogeneity in the effective hopping and spin--orbit energy scales. This allows us to isolate the influence of strain on spectral redistribution, localization properties, and the stability of Majorana and Andreev bound states without invoking pseudomagnetic-field effects associated with strongly nonuniform deformations. {Elastic relaxation effects such as Poisson-ratio-induced transverse contraction are likewise not treated explicitly, but are absorbed into the effective strain profile $\varepsilon(\mathbf r)$.}

\section{Majorana Polarization}
\label{sec:polarization}
The identification of topological Majorana-bound states in realistic systems remains challenging, as conventional spectroscopic signatures such as zero-bias conductance peaks can be reproduced by trivial low-energy states~\cite{sengupta2001midgap,law2009majorana, flensberg2010tunneling}. Majorana polarization, a wave-function-based diagnostic that probes the local particle-hole structure of Bogoliubov-de Gennes eigenstates~\cite{sticlet2012spin,sedlmayr2015visualizing,sedlmayr2016majorana,bena2017testing,kaladzhyan2017formation,awoga2024identifying}, has been suggested as a probe of topological Majorana-bound states. 
The appropriate definition of MP depends on the symmetry class of the underlying Hamiltonian. In particular, the presence or absence of chiral symmetry determines whether a simplified scalar definition is sufficient or whether a more general particle-hole formulation must be used~\cite{karoliya2025majorana}.
\subsection{Chiral-symmetric systems: one-dimensional nanowires}
\label{subsec:mp_chiral}
For the strictly one-dimensional nanowire described by Eq.~(\ref{Eq:Ham1D}), the system belongs to the BDI symmetry class and preserves chiral symmetry~\cite{tewari2012topological}. In this symmetry class, each low-energy Bogoliubov-de Gennes (BdG) eigenstate admits a decomposition into two Majorana components. This property allows for a natural definition of the Majorana polarization as a measure of the local imbalance between these components. We begin by recasting the Hamiltonian in the BdG formalism using the Nambu basis $(c^\dagger_{i\uparrow}, c^\dagger_{i\downarrow}, c_{i\downarrow}, -c_{i\uparrow})$, and diagonalize it numerically to obtain eigenstates of the form
\begin{equation}
\psi_j(i) = \left(u_{i\uparrow}^j, u_{i\downarrow}^j, v_{i\downarrow}^j, v_{i\uparrow}^j\right)^T,
\end{equation}
with corresponding eigenenergy $\epsilon_j$. Here, $u$ and $v$ represent the particle and hole components of the quasiparticle wavefunction, respectively. Due to particle-hole symmetry, each positive-energy solution at $\epsilon_j$ is accompanied by a corresponding solution at $-\epsilon_j$ with wavefunction components related by complex conjugation. To make the Majorana structure explicit, we introduce a Majorana basis defined by the operators
\begin{align}
\gamma_{i\sigma}^1 &= \frac{1}{\sqrt{2}}\left(c^\dagger_{i\sigma}+c_{i\sigma}\right), \nonumber\\
\gamma_{i\sigma}^2 &= \frac{i}{\sqrt{2}}\left(c^\dagger_{i\sigma} -c_{i\sigma}\right),
\label{Eq_chi_PRB}
\end{align}
which satisfy $\gamma^\dagger = \gamma$ and thus correspond to Majorana fermions. In this representation, each complex fermionic operator is decomposed into two real Majorana components, labeled $1$ and $2$. Expressing the BdG wavefunction in this Majorana basis, one finds that the amplitudes associated with the $1$ and $2$ components are given by linear combinations of the particle and hole amplitudes,
\begin{equation}
\psi_i = \frac{1}{\sqrt{2}}\left(u_{i\uparrow}-v_{i\uparrow}, -i(u_{i\uparrow}+v_{i\uparrow}), u_{i\downarrow}-v_{i\downarrow}, -i(u_{i\downarrow}+v_{i\downarrow})\right).
\end{equation}
The Majorana polarization is defined as the difference between the local probabilities of occupying the two Majorana components. Physically, this quantity measures the extent to which a given state is dominated by one Majorana component over the other. A fully localized Majorana-bound state corresponds to a maximal imbalance, whereas a conventional fermionic state corresponds to equal contributions from both components. For a BdG eigenstate $\psi_j(i)$, the local Majorana polarization is given by~\cite{sticlet2012spin}
\begin{equation}
\mathcal{P}_i = 2\,\mathrm{Re}\!\left[u_{i\downarrow}^j v^{j*}_{i\downarrow} - u_{i\uparrow}^j v^{j*}_{i\uparrow}\right],
\label{eq:MP_chiral_local_PRB}
\end{equation}
which follows directly from evaluating the difference in probabilities associated with $\gamma^1$ and $\gamma^2$. Importantly, in the presence of chiral symmetry, the particle and hole components of each eigenstate are not independent but are related by a fixed phase constraint. As a result, one can choose a representation in which the relative phase between $u$ and $v$ is real, ensuring that the local Majorana polarization is purely real. Consequently, in this symmetry class, the Majorana polarization is fully characterized by a single real component. To probe the low-energy sector relevant for Majorana physics, we define the energy-resolved Majorana polarization as
\begin{equation}
\mathcal{P}_i(\omega) = 2\sum_j \delta(\omega-\epsilon_j)\,
\mathrm{Re}\!\left[u^j_{i\downarrow}v^{j*}_{i\downarrow} - u^j_{i\uparrow}v^{j*}_{i\uparrow}\right],
\label{eq:MP_chiral_energy_PRB}
\end{equation}
where the Dirac delta function is approximated numerically by a Gaussian of width $\sigma$. In practice, we focus on $\omega = 0$ to isolate the contribution from low-energy states. To further characterize the spatial structure of Majorana modes, we define the polarization integrated over the left and right halves of the system,
\begin{align}
\mathcal{P}_\mathrm{left}(\omega) &= \sum_{i \leq N/2} \mathcal{P}_i(\omega), \\ \nonumber
\mathcal{P}_\mathrm{right}(\omega) &= \sum_{i > N/2} \mathcal{P}_i(\omega).
\end{align}
For well-separated Majorana-bound states, these quantities are equal in magnitude and opposite in sign, reflecting the nonlocal nature of the Majorana components. Consequently, the product $\mathcal{P}_\mathrm{left} \times \mathcal{P}_\mathrm{right}$ provides a useful measure of nonlocal correlations. We emphasize that the Majorana polarization captures the internal structure of BdG eigenstates in terms of their Majorana components. In particular, states with $|\mathcal{P}| \approx 1$ and vanishing energy splitting correspond to well-separated Majorana-bound states, whereas deviations from unity indicate increasing overlap between Majorana components, as occurs in psABSs. This makes the Majorana polarization a particularly powerful diagnostic for distinguishing between topological and trivial low-energy states in one-dimensional systems.

\begin{figure*}
    \centering
    \includegraphics[width=1.9\columnwidth]{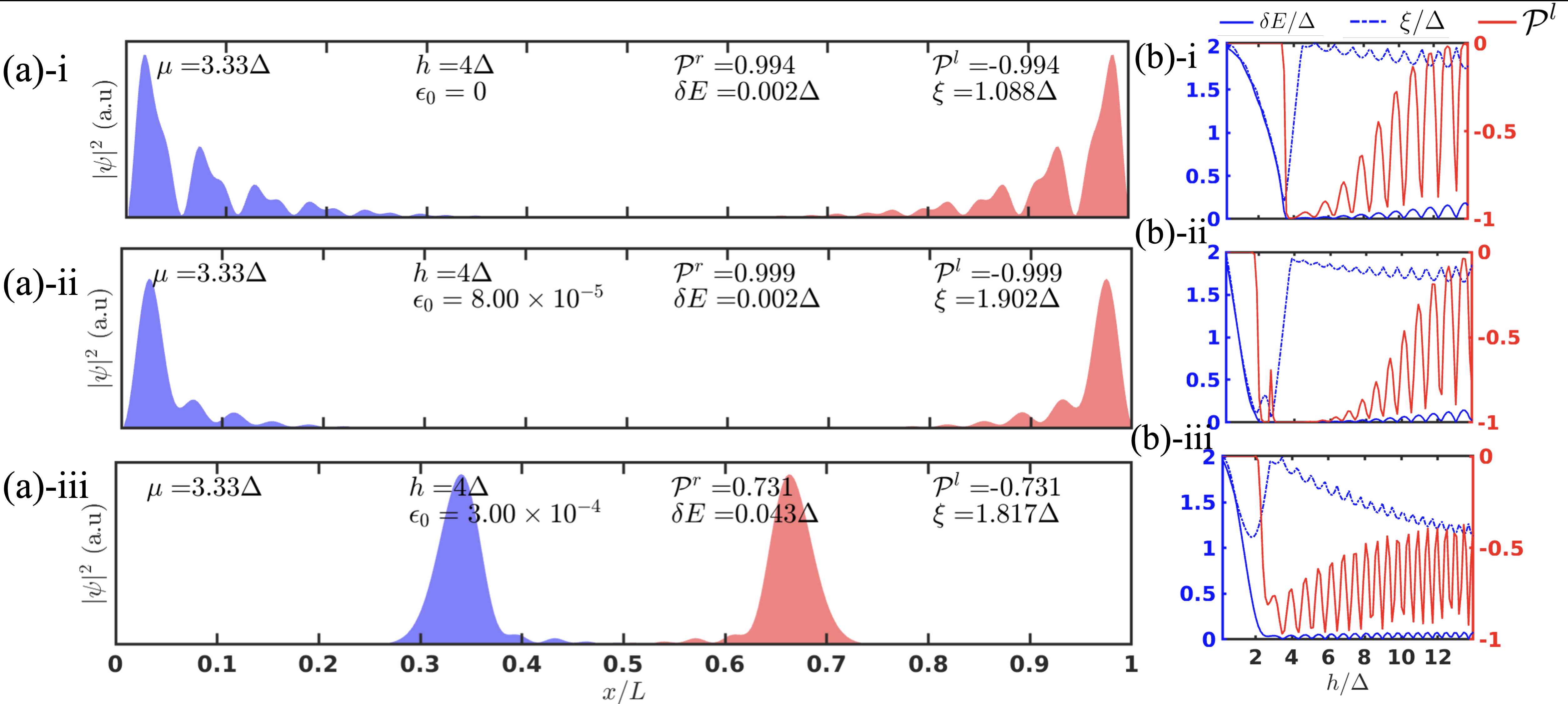}
    \caption{\textit{Clean one-dimensional proximitized nanowire.} 
(a) Spatial profiles of the Majorana components $\chi_A$ (blue) and $\chi_B$ (red) of the lowest-energy eigenstate for different strain strengths: 
(a-i) unstrained case ($\epsilon_0=0$), showing well-localized Majorana-bound states at opposite ends of the wire; 
(a-ii) weak strain, where the zero-energy modes remain robust with negligible change in localization; 
(a-iii) stronger strain, leading to the emergence of Andreev bound states with reduced spatial separation and overlap in the bulk. 
Relevant parameters ($\mu$, $h$, $\delta E$, $\xi$, and Majorana polarization $\mathcal{P}^{l}$) are indicated in each panel.
(b) Evolution with Zeeman field $h/\Delta$ corresponding to (a-i)–(a-iii): 
(b-i) unstrained, (b-ii) weak strain, and (b-iii) stronger strain. 
Shown are the excitation gap $\xi$ (blue dashed), the energy splitting between the two lowest modes $\delta E$ (blue solid), and the Majorana polarization $\mathcal{P}^{l}$ (red), with $\mathcal{P}^{r}=-\mathcal{P}^{l}$. 
}
    \label{Clean}
\end{figure*}
\subsection{General particle--hole definition: quasi-one-dimensional and graphene systems}
\label{subsec:mp_general} 

The Hamiltonian in Eq.~\ref{Ham_q1D_long}, \ref{Ham_graph} breaks chiral symmetry due to the presence of transverse Rashba spin--orbit coupling~\cite{tewari2012topological}, thereby placing the system in symmetry class D. In this regime, the Majorana polarization is no longer constrained to be purely real, and a more general formulation based on particle-hole symmetry is required. To this end, we employ the definition of Majorana polarization introduced by Sedlmayr \textit{et al.}~\cite{sedlmayr2015visualizing}, which is applicable irrespective of the presence or absence of chiral symmetry. The Bogoliubov--de Gennes Hamiltonian ($H_\mathrm{BdG}$), which describes superconducting quasiparticles, possesses an intrinsic particle--hole symmetry. This symmetry is represented by an antiunitary operator $\mathcal{C}$ that anticommutes with the Hamiltonian, $\{H_\mathrm{BdG}, \mathcal{C}\} = 0,$ ensuring that for every eigenstate at energy $\epsilon$, there exists a corresponding state at $-\epsilon$. A true Majorana bound state is self-conjugate under this symmetry, satisfying $\mathcal{C}|\psi\rangle = |\psi\rangle,$ and therefore must occur at zero energy. Physically, this condition implies that the state is an equal-weight superposition of particle and hole components, making the quasiparticle operator Hermitian. To evaluate the Majorana character of eigenstates, we recast the Hamiltonian in the Nambu basis $(c^\dagger_{\mathbf{r}_i\uparrow}, c^\dagger_{\mathbf{r}_i\downarrow}, c_{\mathbf{r}_i\downarrow}, -c_{\mathbf{r}_i\uparrow})$ and diagonalize it numerically. The resulting eigenstates take the form
\begin{equation}
\psi(\mathbf{r}_i,\epsilon_j) = \left(u_{\mathbf{r}_i\uparrow}, u_{\mathbf{r}_i\downarrow}, v_{\mathbf{r}_i\downarrow}, v_{\mathbf{r}_i\uparrow}\right),
\end{equation}
where $u$ and $v$ denote the particle and hole amplitudes, respectively. In this representation, the particle--hole operator is given by $\mathcal{C} = \sigma_y \tau_y \mathcal{K},$ where $\sigma_y$ and $\tau_y$ act in spin and particle--hole spaces, and $\mathcal{K}$ denotes complex conjugation. The local Majorana polarization at site $\mathbf{r}_i$ and energy $\epsilon_j$ is then defined as the expectation value of $\mathcal{C}$,
\begin{equation}
\mathcal{P}(\mathbf{r}_i,\epsilon_j) =
\langle\psi(\mathbf{r}_i,\epsilon_j) | \mathcal{C}|\psi(\mathbf{r}_i,\epsilon_j)\rangle.
\label{eq:MP_general_local_PRB}
\end{equation}

Unlike the chiral case, $\mathcal{P}(\mathbf{r}_i,\epsilon_j)$ is, in general, a complex quantity. Its magnitude quantifies the degree of particle--hole symmetry of the state, while its phase encodes the relative phase between particle and hole components. For a perfect Majorana-bound state, the polarization satisfies
\begin{equation}
|\mathcal{P}_\mathcal{R}| = 1,
\end{equation}
within the region $\mathcal{R}$ where the state is localized. To characterize spatially extended modes, we define the normalized Majorana polarization integrated over a region $\mathcal{R}$,
\begin{align}
\mathcal{P}_\mathcal{R}(\epsilon_j)=
\frac{\sum_{i\in\mathcal{R}}\langle\psi(\mathbf{r}_i,\epsilon_j) | \mathcal{C}|\psi(\mathbf{r}_i,\epsilon_j)\rangle}
{\sum_{i\in\mathcal{R}}\langle\psi(\mathbf{r}_i,\epsilon_j) | \psi(\mathbf{r}_i,\epsilon_j)\rangle}.
\label{Eq:Pregion1_PRB}
\end{align}
For an ideal Majorana mode, this quantity takes the form
$\mathcal{P}_\mathcal{R}(\epsilon_j) = e^{i\theta},$
with unit magnitude and an arbitrary phase $\theta$, reflecting the gauge freedom in defining the Majorana operator. To probe the low-energy sector, we further define the frequency-resolved polarization as
\begin{align}
\mathcal{P}_\mathcal{R}(\omega)=
\frac{\sum_{i\in\mathcal{R}}\sum_j\langle\psi(\mathbf{r}_i,\epsilon_j) | \mathcal{C}|\psi(\mathbf{r}_i,\epsilon_j)\rangle \delta(\epsilon_j-\omega)}
{\sum_{i\in\mathcal{R}}\sum_j\langle\psi(\mathbf{r}_i,\epsilon_j) | \psi(\mathbf{r}_i,\epsilon_j)\rangle \delta(\epsilon_j-\omega)}.
\label{Eq:P_region2_PRB}
\end{align}
Here, the summation over $j$ runs over all eigenstates, and the Dirac delta function is implemented numerically using a Gaussian broadening of width $\sigma$ (we typically use $\sigma=\Delta/25$). Finally, we note that the polarization $\mathcal{P}_\mathcal{R}$ can be written in terms of its real and imaginary components, $\mathcal{P}_\mathcal{R} = \mathcal{P}_\mathcal{R}^x + i \mathcal{P}_\mathcal{R}^y$.

In contrast to the strictly one-dimensional case, both components are generally nonzero in quasi-one-dimensional and graphene systems. This reflects the absence of chiral symmetry and the presence of additional phase degrees of freedom in the particle--hole superposition. Physically, this complex structure encodes the richer internal composition of low-energy states in multiband systems, where hybridization between different channels leads to a nontrivial phase structure. Consequently, both the magnitude and phase of $\mathcal{P}_\mathcal{R}$ must be considered when characterizing Majorana-like states in these systems.

\subsection{Nonlocal correlations and identification criteria}
\label{subsec:mp_criteria}
While a finite magnitude of Majorana polarization indicates the presence of Majorana-like correlations, the defining property of topological Majorana-bound states is their nonlocality, i.e., the spatial separation of the two Majorana components.
For one-dimensional and quasi-one-dimensional systems, we partition the system into left ($l$) and right ($r$) halves and compute the integrated polarizations $\mathcal{P}_l$ and $\mathcal{P}_r$. The nonlocal correlation is quantified through the product $\mathcal{P}_l^* \mathcal{P}_r,$ which approaches $\mathcal{P}_l^* \mathcal{P}_r \approx -1$ for well-separated Majorana modes localized at opposite ends of the system~\cite{awoga2024identifying}.
In finite graphene nanoribbons, however, this correlation metric becomes less reliable. Due to the bipartite lattice structure and the complex nature of the polarization, $\mathcal{P}_l^* \mathcal{P}_r$ generally acquires a complex value and no longer serves as a robust indicator of topological phase transitions. In this case, we instead analyze the magnitude of the polarization in different spatial regions, $|\mathcal{P}^\nu|,$ 
in which $\nu = l, r, u, d,$ and $l$, $r$, $u$, and $d$ denote left, right, upper, and lower halves of the system.
It is important to emphasize that Majorana polarization alone is not sufficient to establish the topological nature of a state~\cite{karoliya2025majorana}. In particular, partially separated Andreev bound states can exhibit large polarization values while lacking topological protection. Therefore, throughout this work, we identify Majorana-bound states based on a combined set of criteria: (i) the presence of near-zero-energy modes within a finite bulk gap, (ii) nonzero Majorana polarization, and (iii) spatial separation of the Majorana components as revealed by real-space wave-function profiles.

\section{Results and Discussion}\label{sec:results}
\subsection{Strain-induced control of bound states in one-dimensional systems}
We begin by analyzing the effect of symmetric strain on a clean one-dimensional proximitized nanowire. To unambiguously characterize the nature of low-energy states, we employ three complementary diagnostics: (i) the Majorana polarization $\mathcal{P}$, (ii) the energy splitting $\delta E$ of the lowest-energy states, and (iii) the spatial separation (or overlap) of the corresponding wave functions. As established in our previous work~\cite{karoliya2025majorana}, true Majorana-bound states are identified by (i) $|\mathcal{P}| \approx 1$, (ii) $\delta E \approx 0$, and (iii) well-separated wave functions localized at opposite ends of the system.

\paragraph*{(i) Clean system: Majorana to ABSs crossover.}
We first consider the clean system in the absence of strain. As shown in Fig.~\ref{Clean}(a-i), in the topological regime, the system hosts well-localized Majorana-bound states at the two ends of the wire. This identification is also supported by all three diagnostics: the Majorana polarization approaches unity, the energy splitting $\delta E$ is negligible, and the corresponding wave functions are spatially well-separated. Upon introducing symmetric strain, a qualitative modification of the spatial structure of these modes is observed. As the strain strength increases [Fig.~\ref{Clean}(a-ii) and (a-iii)], the Majorana wave functions are progressively displaced toward the interior of the system, leading to an enhanced spatial overlap between their components. Although the energy of the lowest states remains close to zero, the increase in overlap is reflected in a reduction of the polarization. 

The evolution of the system with the Zeeman field is shown in Fig.~\ref{Clean}(b). In the absence of strain, a clear topological phase transition occurs at $h \approx 4\Delta$, characterized by a closing and reopening of the bulk gap, simultaneously accompanied by near-zero energy modes and polarization approaching unity. In contrast, with increasing strain, this gap-closing feature is progressively suppressed. In particular, for sufficiently large strain [Fig.~\ref{Clean}(b-iii)], the bulk gap no longer closes but acquires a finite minima, indicating the absence of a topological phase transition. This behavior demonstrates that symmetric strain drives a crossover from true Majorana-bound states to Andreev-bound states. Importantly, this transition occurs without a significant change in the low-energy spectrum, but is instead governed by a strain-induced redistribution of the wave-function overlap and the resulting loss of topological protection.

\begin{figure*}
    \centering
    \includegraphics[width=1.99\columnwidth]{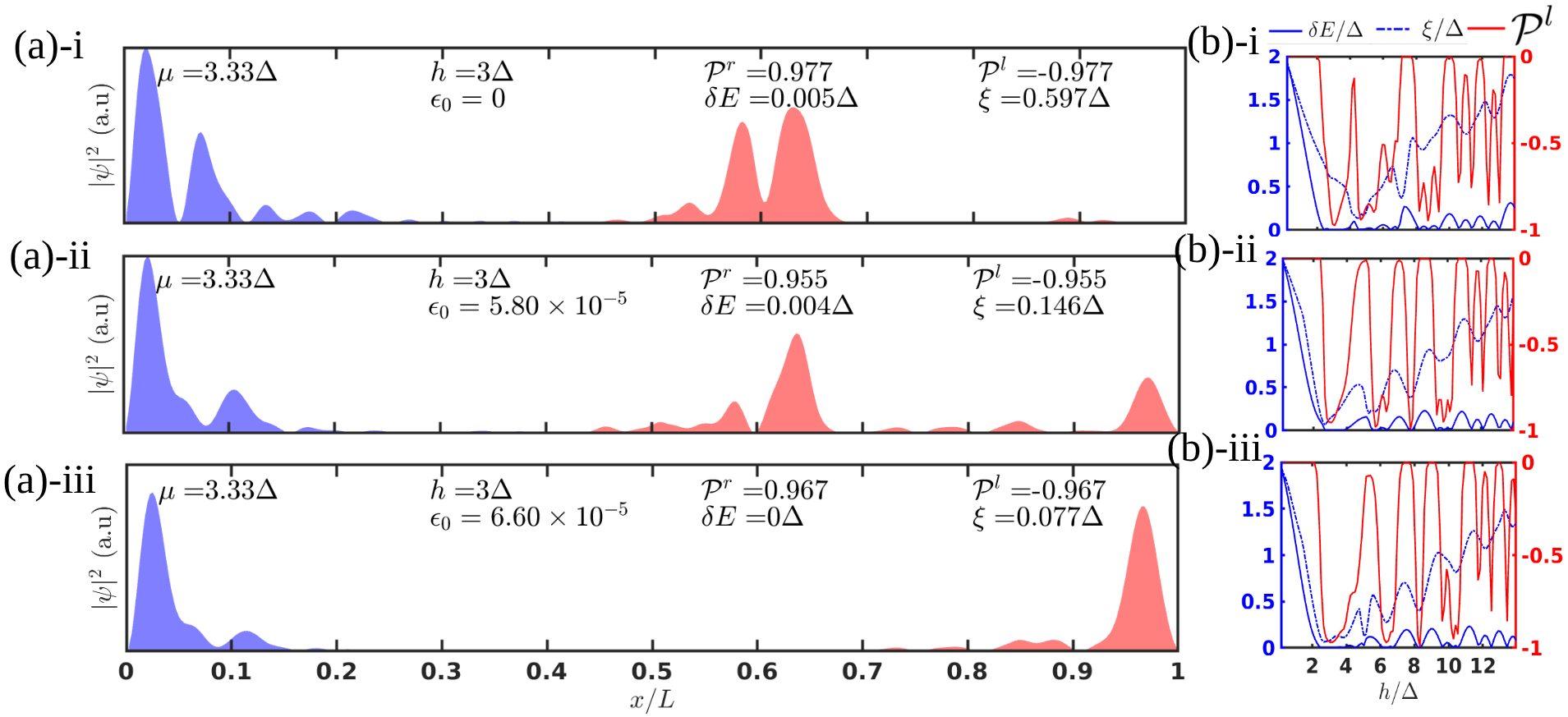}
    \caption{Disordered proximitized nanowire corresponding to the disorder profile shown in Fig.~\ref{trivial}(d). 
(a) Spatial profiles of the Majorana components $\chi_A$ (blue) and $\chi_B$ (red) of the lowest-energy state for increasing strain strength: 
(a-i) in the absence of strain ($\epsilon_0=0$), psABSs are clearly visible with significant overlap in the bulk; 
(a-ii) intermediate strain, where partial separation persists but localization improves; 
(a-iii) stronger strain, where the psABS evolves into well-separated modes, consistent with topological Majorana modes localized at opposite ends of the wire. 
The corresponding parameters ($\mu$, $h$, $\delta E$, $\xi$, and Majorana polarization $\mathcal{P}^{l}$) are indicated in each panel.
(b) Evolution with Zeeman field $h/\Delta$ for the same disorder realization: 
(b-i) no strain, (b-ii) intermediate strain, and (b-iii) stronger strain. 
Shown are the excitation gap $\xi$ (blue dashed), the energy splitting $\delta E$ (blue solid), and the Majorana polarization $\mathcal{P}^{l}$ (red), with $\mathcal{P}^{r}=-\mathcal{P}^{l}$. 
With increasing strain, the topological phase transition is identified by gap closing and reopening in $\xi$-shifts to lower Zeeman fields, indicating strain-assisted stabilization of Majorana modes in the disordered system.
 }
    \label{psABS}
\end{figure*}

\begin{figure*}
    \centering
    \includegraphics[width=1.99\columnwidth]{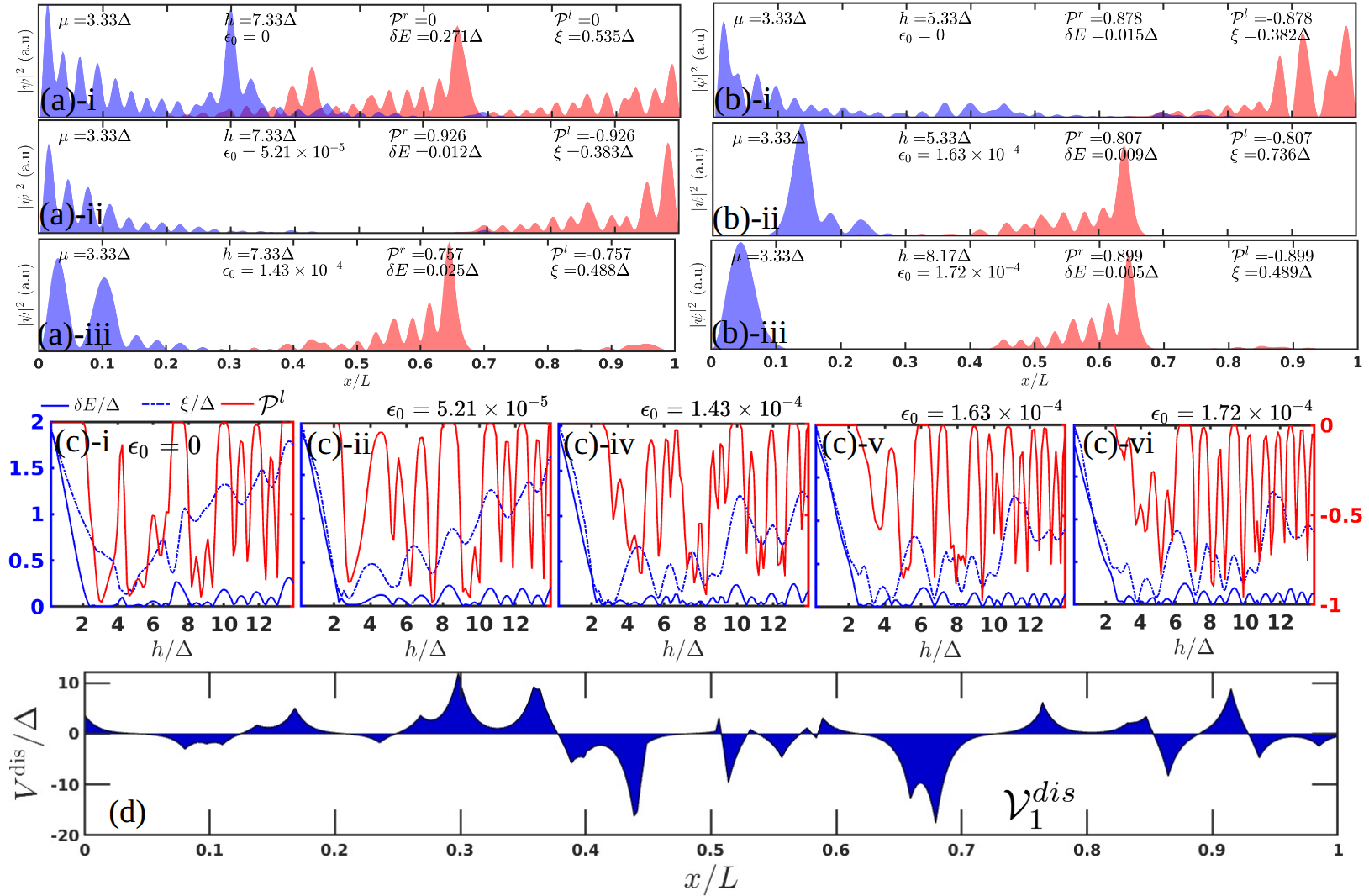}
    \caption{Disordered proximitized nanowire illustrating strain-induced transitions between trivial states, Majorana modes (MMs), and psABSs. The disorder profile used is shown in (d).
(a) Evolution starting from a trivial phase at fixed Zeeman field: 
(a-i) trivial state with no strain ($\epsilon_0=0$), exhibiting no Majorana polarization; 
(a-ii) with increasing strain, a topological phase emerges with well-localized Majorana modes at the wire ends; 
(a-iii) beyond a critical strain, the system transitions into a psABS regime characterized by partially overlapping modes and reduced topological protection.
(b) Complementary evolution for a different Zeeman field: 
(b-i) topological phase with Majorana modes; 
(b-ii) intermediate regime with reduced localization; 
(b-iii) psABS regime at higher strain, showing significant spatial overlap.
(c) Corresponding evolution as a function of Zeeman field $h/\Delta$ for increasing strain strengths mentioned above the figures. 
Shown are the excitation gap $\xi$ (blue dashed), the energy splitting $\delta E$ (blue solid), and the Majorana polarization $\mathcal{P}^{l}$ (red), with $\mathcal{P}^{r}=-\mathcal{P}^{l}$.
(d) Spatial profile of the disorder potential $V_{\mathrm{dis}}(x)$ used throughout the figure.
These results demonstrate that strain can both induce topological Majorana modes from a trivial phase and, beyond a threshold, drive a crossover into psABSs, highlighting the nonmonotonic role of strain in disordered systems.
}
    \label{trivial}
\end{figure*}

\paragraph*{(ii) psABSs to Majorana transition induced by strain.}
Next, we consider the effect of symmetric strain in the presence of disorder. The corresponding disorder profile, denoted by $\mathcal{V}^\mathrm{dis}_1$, is shown in Fig.~\ref{trivial}(d), which was randomly generated using the parameters $V_0=2$meV, $n_d=20/\mu$m, $\lambda=20$nm. As shown in Fig.~\ref{psABS}(a-i), we start with the scenario when in the absence of strain, the system hosts low-energy states that are not topological Majorana modes but partially separated Andreev bound states. These states exhibit a finite spatial overlap, while the energy splitting $\delta E$ remains close to zero. Notably, the Majorana polarization approaches unity despite the trivial nature of these states, illustrating that polarization alone is insufficient to distinguish between psABSs and true Majorana-bound states~\cite{karoliya2025majorana}. Upon introducing a symmetric strain, a qualitative transition in the nature of the low-energy states is observed. As the strain strength increases [Fig.~\ref{psABS}(a-ii) and (a-iii)], the wave functions become progressively more localized at opposite ends of the system, indicating a reduction in their spatial overlap. Concurrently, the polarization remains close to unity, and the energy splitting is further suppressed, consistent with the emergence of well-separated Majorana-bound states. This demonstrates that strain can convert initially trivial psABSs into topological Majorana modes by enhancing their nonlocal character. 

The Zeeman-field dependence of this transition is shown in Fig.~\ref{psABS}(b). In the absence of strain [Fig.~\ref{psABS}(b-i)], the bulk gap closes and reopens at $h \approx 4\Delta$, indicating a topological phase transition. However, the Majorana polarization approaches unity already at lower fields ($h \approx 2\Delta$), leading to a misleading identification of psABSs as Majorana modes in this regime. With increasing strain, the topological phase boundary shifts to lower Zeeman fields. In particular, for sufficiently large strain [Fig.~\ref{psABS}(b-iii)], the bulk gap closes and reopens near $h \approx 2.2\Delta$, precisely where the polarization approaches unity.  This alignment between the gap-closing transition and polarization provides clear evidence of a strain-induced transition into a true topological phase. Therefore, symmetric strain not only modifies the spatial structure of low-energy states but also shifts the topological phase boundary, enabling the conversion of trivial psABSs into robust Majorana-bound states.

\paragraph*{(iii) Trivial states to Majorana and psABSs: re-entrant behavior.}

Finally, we consider a regime where, under the same disorder profile [Fig.~\ref{trivial}(d)], the low-energy states are neither Majorana-bound states nor psABSs, but correspond to trivial subgap states. As shown in Fig.~\ref{trivial}(a-i), these states are characterized by a finite energy splitting $\delta E$, negligible Majorana polarization, and the absence of spatial separation. Upon introducing symmetric strain, the system undergoes a sequence of transitions. At moderate strain [Fig.~\ref{trivial}(a-ii)], the wave functions become localized near opposite ends of the system, the polarization approaches unity, and the energy splitting is strongly suppressed, which are features consistent with the emergence of well-separated Majorana-bound states. Importantly, these modes remain robust over a finite range of strain values, indicating that the induced topological phase is stable within this window. However, upon further increasing the strain [Fig.~\ref{trivial}(a-iii)], the wave functions begin to overlap again, while the energy remains close to zero. Although the polarization does not vanish and remains relatively large, the increasing overlap signals a transition from Majorana-bound states to psABSs. This evolution is further illustrated in Fig.~\ref{trivial}(b), which shows how initially robust Majorana modes in the disordered system are progressively distorted by strain. In the absence of strain [Fig.~\ref{trivial}(b-i)], the system hosts well-defined Majorana-bound states. As strain is increased [Fig.~\ref{trivial}(b-ii) and (b-iii)], these modes lose their spatial separation and evolve into psABSs. The dependence on the Zeeman field, shown in Fig.~\ref{trivial}(c). These results demonstrate that symmetric strain can induce a re-entrant sequence of phases, driving the system from trivial subgap states to Majorana-bound states and subsequently to psABSs. This highlights the \textit{dual role of strain in both stabilizing and destabilizing topological phases}. 

\paragraph*{Summary.}
The above results establish that symmetric strain in one-dimensional systems provides a controlled mechanism for manipulating low-energy bound states. It enables continuous and reversible tuning between trivial states, psABSs, and true Majorana-bound states. 
We have focused only on the symmetric strain profile applied from both ends of the system. For completeness, we note that applying strain from a single end leads to a qualitatively similar behavior, with the primary difference being that only the bound state localized near the strained edge is significantly affected, while the opposite-end mode remains largely unchanged. In contrast, symmetric strain modifies both end-localized modes simultaneously by shifting their spatial profiles toward the interior of the system. A direct comparison between single-end and symmetric strain configurations is provided in the Appendix~\ref{A1}. Before closing this subsection, we note that the results for a quasi-one-dimensional system remain qualitatively the same, hence they are relegated to Appendix \ref{A1_2}.

\subsection{Multiband effects and ambiguity in graphene nanoribbons}\label{graphene_section}

We now turn to graphene nanoribbons with zigzag short edges, where the magnetic field is applied along the longitudinal ($x$) direction. In contrast to the single-channel nanowire, graphene hosts a multiband spectrum originating from its bipartite lattice, transverse confinement, and edge-dependent states. In addition, the presence of disorder further enhances mode mixing, leading to a dense and highly hybridized low-energy spectrum. To characterize the low-energy states, we employ the generalized definition of Majorana polarization introduced in Sec.~\ref{subsec:mp_general}. Unlike in one-dimensional systems, where left and right polarizations are equal in magnitude and opposite in sign, in graphene these quantities are generally unequal. For this reason, we analyze the absolute values of polarizations on the left-, right-, upper-, and lower half of the system. The low-energy states cannot be unambiguously identified as Majorana-bound states despite their proximity to zero energy because graphene nanoribbons support multiple competing channels that generate a complex and highly hybridized low-energy landscape. Consequently, in the presence of strong multiband mixing, standard diagnostics such as polarization and energy splitting can become ambiguous, unless additional mechanisms, such as strain, act to suppress hybridization and isolate the low-energy modes.

\begin{figure*}
    \centering
    \includegraphics[width=1.99\columnwidth]{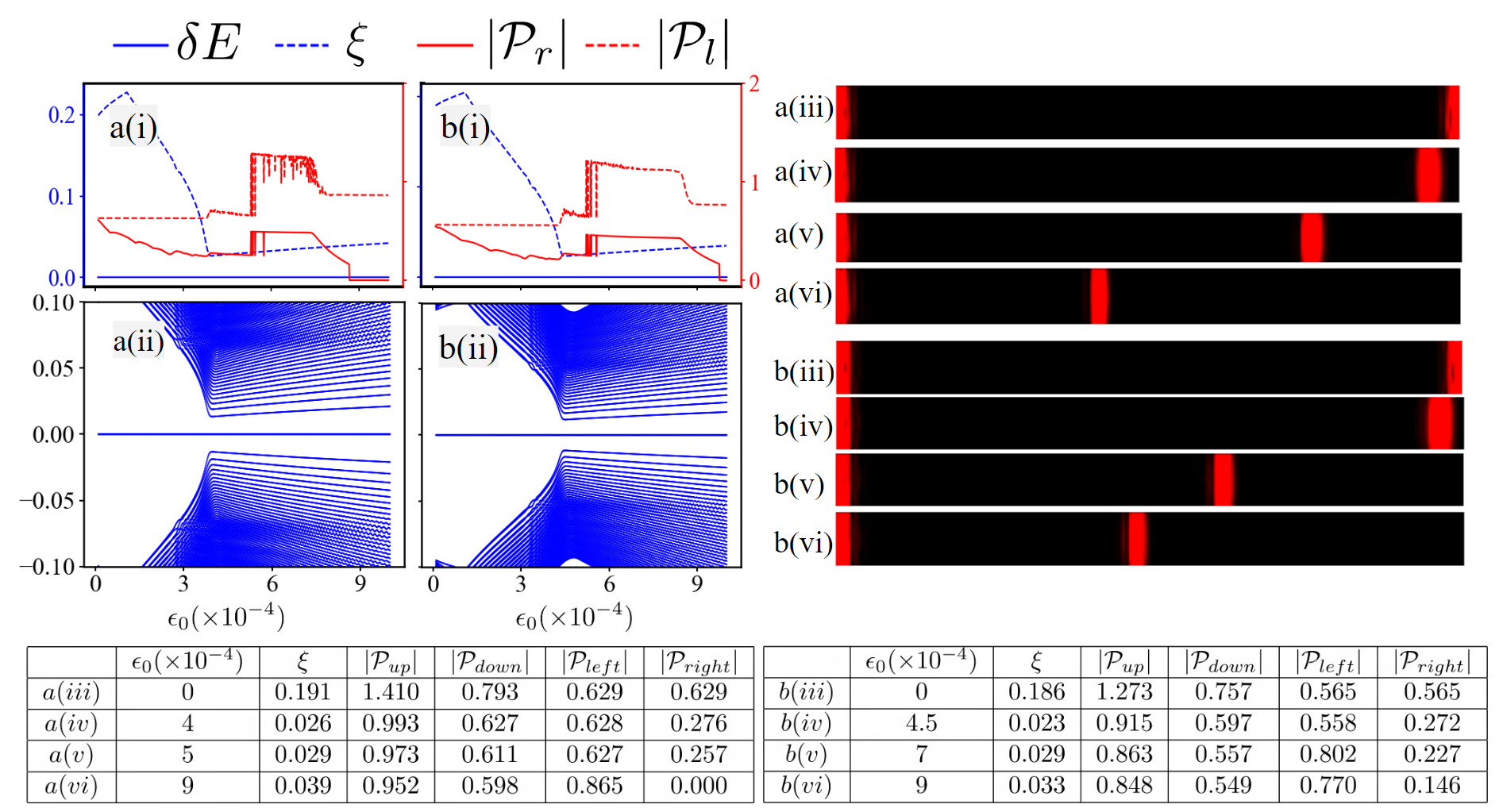}
    \caption{Majorana polarization and spectral evolution in a finite-size \textbf{armchair} graphene strip under nonuniform strain in longitudinal (\(x\)) direction for two Zeeman fields: (a) $B_x = 0.7t$ and (b) $B_x = 0.8t$ at fixed chemical potential $\mu = 2.0t$. Panels (a-i) and (b-i) show the mode splitting $\delta E$ (solid blue), bulk gap $\xi$ (dashed blue), and the absolute values of Majorana polarization at the right ($|P_r|$, solid red) and left ($|P_l|$, dashed red) ends as a function of strain $\epsilon_0$. Panels (a-ii) and (b-ii) display the low-energy spectrum, showing the $100$ eigenvalues closest to zero energy. Panels (a-iii)--(a-vi) and (b-iii)--(b-vi) present the spatial profile of the local density of states (LDOS) near zero energy for representative strain values listed in the tables below.}
    \label{cleang}
\end{figure*}

\paragraph*{(i) Clean system: Majorana to psABSs crossover.}
Graphene nanoribbons proximitized by an $s$-wave superconductor and subject to Rashba spin-orbit coupling and a Zeeman field can realize an effective topological superconducting phase supporting Majorana-bound states at their boundaries~\cite{ma2025graphene,kaladzhyan2017formation,kaladzhyan2017majorana,PhysRevX.5.041042, laubscher2020majorana,wang2018strain, manesco2019effective}. In this regime, the system exhibits a finite bulk gap along with finite absolute values of the left and right Majorana polarizations, as defined in Eq.~\ref{Eq:P_region2_PRB}. To investigate the effect of strain, we introduce a \textit{non-uniform uniaxial deformation} in clean graphene by fixing the left end of the ribbon while stretching the right end. This deformation modifies the bond lengths and induces a spatial gradient in the hopping amplitudes along the direction of deformation. As observed in the energy spectrum  Fig.~\ref{cleang}(a-ii) and ~\ref{cleang}(b-ii), the applied strain drives the bulk states toward zero energy, leading to a significant reduction of the band gap, {which closes at sufficiently large strain. From the figure ~\ref{cleang}a(iii)-a(vi) and b(iii)-b(vi), it is well observed that increasing the uniaxial strain moves the mode from the right towards the left, and after a certain value of strain, the right mode completely moves near the left edge. The overlap of the two modes leads to closing of the band gap at that critical strain, which is quite large for our system}. This redistribution of bulk spectral weight is accompanied by a corresponding modification of the polarization, as shown in Fig.~\ref{cleang}(a-i) and ~\ref{cleang}(b-i). Notably, while the localization and polarization of the left-end mode remain largely unaffected (Fig.~~\ref{cleang} table), the right-end mode progressively coalesces with the bulk, with its polarization decreasing rapidly as the strain strength increases. The gradual delocalization of the near-zero-energy density from the right boundary into the bulk is clearly observed in Fig.~\ref{cleang}(a-iii)–~\ref{cleang}(a-vi) for $B_x = 0.7t$, and Fig.~\ref{cleang}(b-iii)–~\ref{cleang}(b-vi) for $B_x = 0.8t$. Figures~\ref{cleang}(a-i) and ~\ref{cleang}(b-i) show the variation of the left and right polarizations, together with the reduction in the band gap, for the two representative magnetic field regimes.

This behavior signals a crossover from well-localized, nonlocal Majorana-bound states to psABSs, analogous to the mechanism observed in clean nanowire systems under strain. In particular, although the low-energy states remain near zero energy, the increasing spatial overlap of their wave functions and the loss of boundary localization indicate a breakdown of their topological character. As in the clean nanowire case, where strain primarily enhances wave-function overlap, the non-uniform strain in clean graphene also drives bulk states toward low energies, thereby accelerating the loss of topological protection. 

Under the application of symmetric strain where the center of the wire is fixed while both ends are stretched, as shown in Fig.~\ref{fig:hopping_strain}(c)-\ref{fig:hopping_strain}(d). In panel (c), the ends are stretched along the $x$-direction (i.e., along the length of the wire), whereas in panel (d), the stretching is done along the $y$-direction (i.e., along the width). 
Such deformation induces a non-uniform strain profile, which effectively shifts the low-energy modes toward the center of the system, similar to what is observed in nanowire setups. This approach is not particularly effective in tuning the band gap or driving a topological phase transition in disordered systems. In contrast, uniaxial strain can achieve this more efficiently.
In the presence of disorder, symmetric strain modifies the spatial profile of the modes, causing them to acquire a finite probability distribution across the entire strip. It also fails to satisfy additional necessary conditions, such as opening a finite band gap, and leads to an overcrowding of bulk states near zero energy (as discussed in Appendix~\ref{App:zzg}). Consequently, in disordered systems, unidirectional strain is employed to facilitate the transition from psABS to Majorana, as discussed below.
 
\paragraph*{(ii) psABSs to Majorana transition induced by strain.}

\begin{figure*}
    \centering
    \includegraphics[width=1.99\columnwidth]{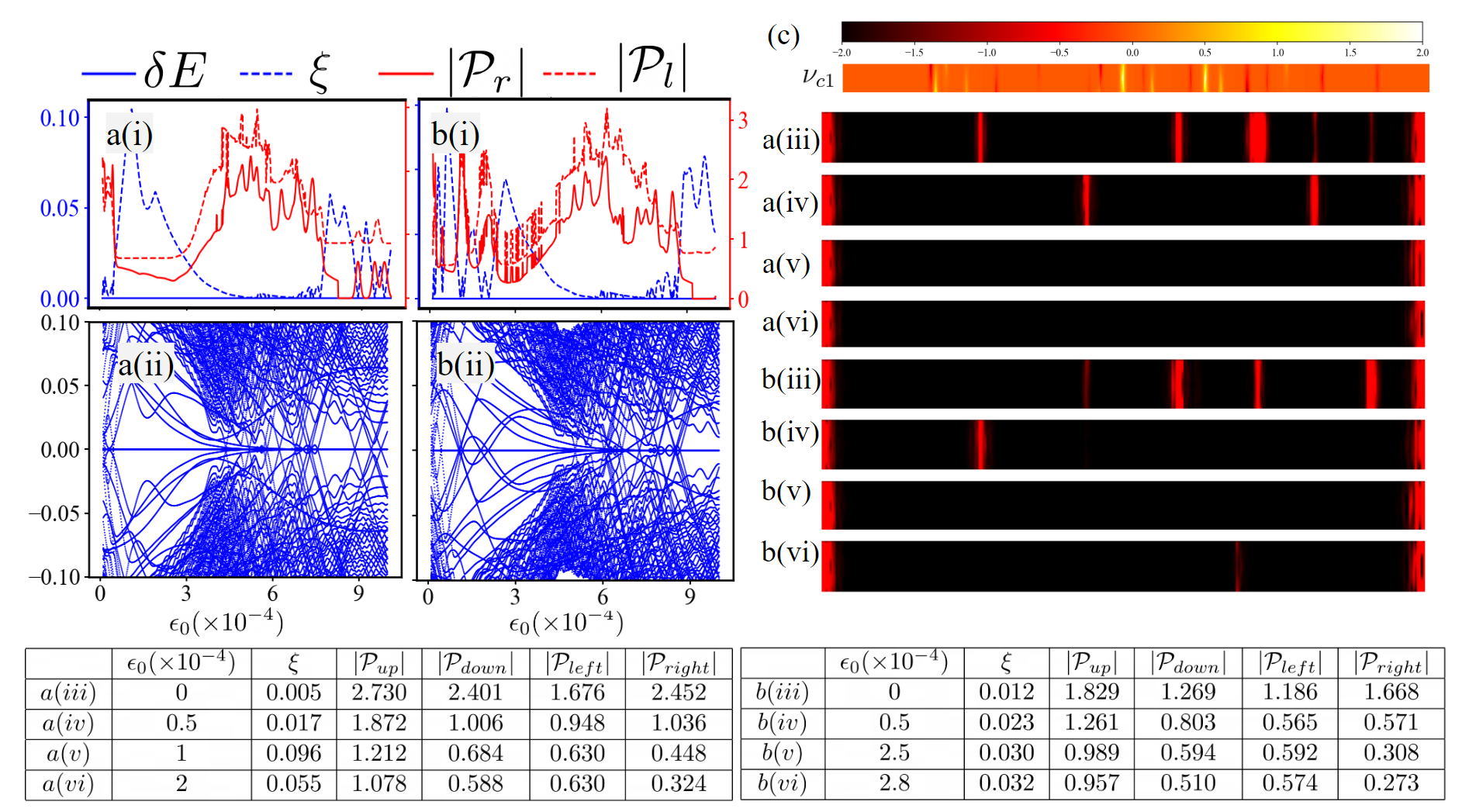}
    \caption{Majorana polarization and spectral evolution in a finite-size \textbf{armchair} graphene strip in the presence of disorder for two Zeeman fields: (a) $B_x = 0.7t$ and (b) $B_x = 0.8t$, at fixed chemical potential $\mu = 2.0t$. The disorder configuration $\nu_{c1}$ is shown in panel (c) and is generated using Eq.~\ref{eq:disorder} with parameters $V_0 = 1.0t$, $\lambda = 1.0t$, and on-site disorder amplitudes $A_i \in [-2.0,\,2.0]t$ applied to $N_d = 20$ lattice sites. Panels (a-i) and (b-i) show the mode splitting $\delta E$ (solid blue), bulk gap $\xi$ (dashed blue), and the absolute values of Majorana polarization at the right ($|P_r|$, solid red) and left ($|P_l|$, dashed red) ends as a function of strain $\epsilon_0$. Panels (a-ii) and (b-ii) display the low-energy spectrum, showing the $100$ eigenvalues closest to zero energy. Panels (a-iii)--(a-vi) and (b-iii)--(b-vi) show the spatial profile of the LDOS near zero energy for representative strain values listed in the tables below. }
    \label{disorderg}
\end{figure*}

We now investigate the effect of spatially nonuniform strain on a disordered graphene nanoribbon. The disorder configuration, denoted by $\nu_{c1}$, and shown in Fig.~\ref{disorderg}(c), is generated using Eq.~\ref{eq:disorder} with parameters $V_0 = 1.0\,\mathrm{meV}$, $\lambda = t$, and on-site disorder potential amplitudes are randomly chosen from $A_i \in [-2.0,\,2.0]\,t$ for \(N_d=20\) lattice sites.

We consider an asymmetric strain profile that induces a spatial gradient across the system, where the left end is held fixed, and strain is applied at the right end. In the presence of disorder, the low-energy local density of states exhibits multiple spatial maxima distributed across both the edges and the bulk. Instead of forming well-localized edge modes, these states display a fragmented spatial structure with significant weight throughout the system [Fig.~\ref{disorderg}(a-iii) and (b-iii)]. This behavior, together with strongly nonuniform polarization, indicates that the low-energy states originate from multiband hybridization. Consequently, the system hosts trivial low-energy states or psABSs, characterized by a dense accumulation of states near zero energy.

For finite strain, the spectral evolution [Fig.~\ref{disorderg}(a-ii) and Fig.~\ref{disorderg}(b-ii)] shows a reduction in spectral crowding near zero energy, driven by the lifting of accidental degeneracies and suppression of inter-subband hybridization. As the strain is further increased, the low-energy spectrum undergoes a qualitative reorganization, leading to the emergence of a gap-like feature. At a critical strain, the spectrum exhibits a near closing and subsequent reopening of the gap, consistent with a transition into a topological regime. In this regime, well-defined zero-energy modes appear and become strongly localized at the boundaries, as confirmed by the LDOS and finite polarization [Fig.~\ref{disorderg}(a-iii)--(a-v)]. These modes remain robust over a finite range of strain, provided the gap remains open [Fig.~\ref{disorderg}(a-vi) and Fig.~\ref{disorderg}(b-vi)].

This evolution indicates a crossover from disorder-induced partially separated Andreev bound states to boundary-localized modes consistent with Majorana-bound states. 
Overall, nonuniform strain acts as an effective tuning parameter that suppresses trivial low-energy states and stabilizes boundary-localized Majorana-like modes in the disordered system. \label{B-ii}
\paragraph*{Summary.} These results establish that non-uniform strain acts as an effective tuning parameter in graphene nanoribbon platforms hosting Majorana fermions. In the presence of disorder, the multiband spectrum leads to intertwined low-energy states and degeneracies near zero energy. We demonstrate that asymmetric strain lifts these degeneracies and reorganizes the spectrum, resulting in a finite band gap and well-localized boundary modes with finite Majorana polarization, consistent with topologically protected Majorana-bound states. In contrast, in the clean system, strain drives a crossover from well-localized Majorana modes to partially separated states. 

\subsection{Unified physical interpretation}
The contrasting behavior observed in one-dimensional and graphene systems can be understood in terms of their underlying band structure and the spatial profile of the applied strain. In the nanowire, where the low-energy physics is governed by a single effective channel and strain is applied symmetrically, the primary effect is a controlled redistribution of Majorana components. This enables continuous and reversible tuning between trivial states, partially separated Andreev bound states, and true Majorana-bound states due to strain-induced shifts in the topological phase boundary. In graphene nanoribbons, by contrast, the presence of multiple dispersive subbands leads to strong hybridization and a dense low-energy spectrum. Strain lifts accidental degeneracies and suppresses interband mixing, resulting in a reorganization of the low-energy spectrum characterized by reduced spectral density near zero energy and enhanced spatial localization of individual modes. 

Although strain modifies the local electronic structure in both systems, its macroscopic manifestation depends fundamentally on the number of active low-energy channels. In effectively single-channel systems, strain can shift the topological boundary and enable controlled phase transitions. In multiband systems, however, strain acts nonuniformly across different subbands, preventing a direct correspondence between strain and topology. Instead, it primarily serves to suppress multiband hybridization and enhance the distinguishability of low-energy states.
Taken together, these results demonstrate that the role of strain is dictated by both dimensionality and band structure: in single-channel systems, it functions as a direct tuning knob for Majorana physics, while in multiband systems, it acts as a mechanism for spectral reorganization that improves the identification of localized bound states. To further understand the results of our numerical simulations, in the next section, we develop an analytical understanding of the same. 

\section{Analytical theory of strain-driven crossovers}\label{sec:analytical}
In this section, we develop an analytical theory of strain-driven crossovers that captures the interplay of topology, disorder, and spatial inhomogeneity in real space in a one-dimensional proximitized semiconductor nanowire and the effects of strain in graphene nanoribbons. 
\subsection{One-dimensional nanowire}
In the continuum limit, the Hamiltonian (Eq.~\ref{Eq:Ham1D}) including the strain and disorder terms in the Nambu basis $[\psi^\dagger_\uparrow(x),\psi^\dagger_\downarrow(x),\psi_\downarrow(x),-\psi_\uparrow(x)]$
becomes 
\begin{align}
H_{\rm BdG}(x)
&=
\left[
-\frac{\hbar^2}{2m(x)}\partial^2_x
-\mu_{\rm eff}(x)
\right]\tau_z\nonumber\\
&-i\alpha(x)\partial_x\sigma_y\tau_z
+h\sigma_x
+\Delta\tau_x,
\label{eq:continuum_bdg}
\end{align}
where we have introduced a continuous electron field $\psi_\sigma(x)$. The renormalization of hopping enters through a position-dependent mass term ($m(x)$), and the spin-orbit coupling also becomes position dependent ($\alpha(x)$).
The Pauli matrices $\sigma_a$ act in spin space and $\tau_a$ act in
particle-hole space. 
In the presence of strain and disorder, the local band bottom is shifted, and the effective position-dependent chemical potential is
\begin{align}
\mu_{\rm eff}(x)
=
\mu
-
V_{\rm dis}(x)
+
\lambda_{\epsilon}\epsilon(x),
\label{eq:mueff}
\end{align}
where $\lambda_{\epsilon}$ is a phenomenological deformation-potential
coefficient.

We now examine how the topological condition $h^2=\Delta^2+\mu^2$ is modified in the presence of position-dependent parameters, assuming that the spatial variations occur on length scales large compared to the superconducting coherence length and Fermi wavelength. At a given point $x$, we first neglect the gradient term and consider the local zeroth-order BdG matrix:
\begin{align}
H_0(x)
=
-\mu_{\rm eff}(x)\tau_z
+
h\sigma_x
+
\Delta\tau_x .
\label{eq:local_matrix}
\end{align}
The local eigenvalues are
\begin{align}
E_{\sigma,\pm}(x)
=
\sigma h
\pm
\sqrt{\mu_{\rm eff}^2(x)+\Delta^2}.
\label{eq:local_eigs}
\end{align}
where $\sigma=\pm 1$. The inner two bands encounter a gap-closing condition when $h^2=\Delta^2+\mu_{\rm eff}^2(x)$. This motivates us to define a real-space topological mass:
\begin{align}
M(x)
=
h^2-\Delta^2-\mu_{\rm eff}^2(x).
\label{eq:mass_def}
\end{align}
Regions with $M(x)>0$, 
are locally topological, while regions with
$M(x)<0$ are locally trivial. A Majorana component can appear at a domain wall when 
$M(x)=0$. Equivalently, this domain wall condition can be written as 
\begin{align}
V_{\rm eff}(x)=V_c,
\qquad
V_c=\mu\mp\sqrt{h^2-\Delta^2},
\label{eq:crossing_geometry}
\end{align}
where $V_{\rm eff}(x) = V_{\rm dis}(x)-\lambda_{\epsilon}\epsilon(x)$. Therefore, disorder and strain enter the local topological criterion only through the effective local potential $V_{\rm eff}(x)$.

We now examine how the domain walls move as a function of strain amplitude. We write the strain profile as $\epsilon(x)=\epsilon_0 f(x)$, where $\epsilon_0$ is the strain amplitude and $f(x)$ is a smooth function. The domain-wall position $x_a(\epsilon_0)$ is defined by
\begin{align}
F(x_a,\epsilon_0)=0, \qquad
F(x,\epsilon_0)
=
h^2-\Delta^2-\mu_{\rm eff}^2(x,\epsilon_0).
\label{eq:F_def}
\end{align}
The total derivative of $F$ must vanish, since the domain-wall position $x_a(\epsilon_0)$ is implicitly defined by the condition $F(x_a,\epsilon_0)=0$, which holds identically for all $\epsilon_0$. Therefore,
\begin{align}
\frac{dx_a}{d\epsilon_0}
=
-
\frac{\partial_{\epsilon_0}F}
{\partial_x F},
\label{eq:implicit_shift_general}
\end{align}
which is evaluated to be
\begin{align}
\frac{dx_a}{d\epsilon_0}
=
\frac{\lambda_{\epsilon}f(x_a)}
{
\partial_xV_{\rm dis}(x_a)
-
\lambda_{\epsilon}\epsilon_0 f'(x_a)
}.
\label{eq:domain_wall_velocity}
\end{align}
Eq.~\ref{eq:domain_wall_velocity} gives the motion of a Majorana component in real space as the strain amplitude is varied. Note that if the strain gradient nearly cancels the disorder gradient, the domain wall moves very rapidly with strain. This provides a natural mechanism for large strain-induced displacement of Majorana components. 
In the absence of strong disorder gradients, a symmetric strain profile, which does not have a definite sign throughout the wire, generates an inward-directed motion of the domain walls from both ends. However, in the presence of spatially varying disorder, the direction of motion is determined by the competition between strain and disorder gradients. It can locally reverse, leading to either inward or outward displacement. This is responsible for the psABS to edge-localized Majorana modes as observed in our numerical simulations. For asymmetric strain profiles, the gradient $f'(x)$ retains a definite sign across the wire, leading to a directional displacement of the domain wall and enabling selective motion of one Majorana component relative to the other.

We now derive the low-energy effective Hamiltonian near a real-space topological domain by projecting the BdG Hamiltonian onto its gap-closing low-energy subspace.
From Eq.~\ref{eq:local_matrix}, the relevant low-energy states are
\begin{align}
    |1\rangle &= |\sigma_x;+\rangle\otimes|u;+\rangle\nonumber\\
    |2\rangle &= |\sigma_x;-\rangle\otimes|u;-\rangle,
\end{align}
where $(-\mu(x_a)\tau_z+\Delta\tau_x) |u;\pm\rangle = \pm h|\pm u\rangle$. Projecting $H_{\rm BdG}$ to the subspace $(|1\rangle, |2\rangle)$ gives 
\begin{align}
H_{\rm eff}^{(a)}
&=
-\frac{i}{2}
\left\{
v_a(x), \partial_x
\right\}
\rho_y
+
m_a(x)\rho_z,
\end{align}
where  $v_a(x)={\alpha(x)\Delta}/{h}$, 
\begin{align}
m_a(x)
&=
-\frac{s_a\mu_c}{h}
(
\mu
-
V_{\rm dis}(x)
+
\lambda_{\epsilon}\epsilon(x)
-
s_a\mu_c
),
\end{align}
$\mu_c =
\sqrt{h^2-\Delta^2}$ with $
s_a=\pm 1$, 
and $\rho$'s indicate Pauli-spin matrices in the projected subspace. The zero-energy domain wall state can be derived by the condition $H_{\rm eff}^{(a)}\Psi_a = 0$, akin to the solution of the Jackiw-Rebbi bound state~\cite{jackiw1976solitons}:
\begin{align}
\Psi_a(x)
\propto
\frac{1}{\sqrt{v_a(x)}}
\exp\left[
-\int_{x_a}^{x} dx'\,
\frac{1}{\xi(x')}
\right]\chi_a,
\label{eq:majorana_solution}
\end{align}
$\chi_a$ being a $\rho_x$-eigenstate, and $\xi(x)={v_a(x')}/{|m_a(x')|}$ is the Majorana decay length. Since Majorana modes come in pairs, let us assume that the two components are localized near $x_1$ and $x_2$, with their wavefunctions given by Eq.~\ref{eq:majorana_solution}. Their overlap gives rise to a nonzero splitting, which can be substantial if either the decay length is large or their spatial separation is small. The distinction between an ordinary ABS, a psABS, and a topological Majorana mode is governed by the separation between the two Majorana components. If 
\begin{align}
d_{12}\lesssim \bar{\xi}
&\quad\Rightarrow\quad
\text{ordinary ABS},
\nonumber\\
\bar{\xi}\ll d_{12}\ll L
&\quad\Rightarrow\quad
\text{partially separated ABS},
\nonumber\\
d_{12}\sim L
&\quad\Rightarrow\quad
\text{Majorana-bound states}.
\nonumber
\end{align}

We can also define the following quantity 
\begin{align}
\mathcal{S}_{12}
=
\int_{x_1}^{x_2}\frac{dx}{\xi(x)}
\label{eq:nonlocality_action}
\end{align}
that encodes the nonlocality of the two modes. Equivalently, in terms of $\mathcal{S}_{12}$ 
\begin{align}
\mathcal{S}_{12}\lesssim 1
&\quad\Rightarrow\quad
\text{ordinary ABS},\nonumber\\
\mathcal{S}_{12}\sim O(1)
&\quad\Rightarrow\quad
\text{partially separated ABS},\nonumber\\
\mathcal{S}_{12}\gg1
&\quad\Rightarrow\quad
\text{Majorana-bound states}.
\nonumber
\end{align}
The numerical boundaries between these regimes are not universal, but the robust conclusion is that exponentially increasing $\mathcal{S}_{12}$ suppresses the splitting exponentially.
The derivative of $\mathcal{S}_{12}$ is evaluated to be:
\begin{align}
\frac{d\mathcal{S}_{12}}{d\epsilon_0}
&=
\frac{1}{\xi(x_2)}
\frac{dx_2}{d\epsilon_0}
-
\frac{1}{\xi(x_1)}
\frac{dx_1}{d\epsilon_0}
+
\int_{x_1}^{x_2}dx\,
\frac{\partial}{\partial\epsilon_0}
\left[
\frac{1}{\xi(x;\epsilon_0)}
\right].
\label{eq:dSdeps}
\end{align}
The first two terms in Eq.~\ref{eq:dSdeps} describe the motion of the domain walls, and the third term describes the change in the decay length due to strain-induced changes in
$v(x)$ and $m(x)$.
Substituting Eq.~\eqref{eq:domain_wall_velocity}, we obtain a fully real-space criterion for the crossover. Therefore, a strain-induced psABS-to-MBS crossover occurs when strain increases the nonlocal action,
\begin{align}
\frac{d\mathcal{S}_{12}}{d\epsilon_0}>0, \qquad 
\textrm{psABS-to-MBS}.
\label{eq:psabs_to_mbs_condition}
\end{align}
Physically, this means that strain either pushes the two Majorana components
farther apart or reduces their decay length. Conversely, an MBS-to-psABS
crossover occurs when
\begin{align}
\frac{d\mathcal{S}_{12}}{d\epsilon_0}<0, \qquad 
\textrm{MBS-to-psABS}.
\label{eq:mbs_to_psabs_condition}
\end{align}
which means that strain pulls the components closer together or increases
their overlap.

\subsection{Graphene nanoribbons}
In the absence of superconductivity, the low-energy electronic structure of graphene is described by Dirac fermions near the two valleys. 
Here, we employ a single-valley continuum Hamiltonian as a minimal local description. The effects of finite-width confinement and intervalley mixing are incorporated through the projected subband parameters and inter-subband couplings.
Expanding around a valley point, the continuum Hamiltonian takes the form
\begin{align}
h_0({\bf r})
=
\hbar v_F({\bf r})
\left[
(-i\partial_x)\rho_x
+
(-i\partial_y)\rho_y
\right]
+
V_{\rm dis}({\bf r})
-
\mu ,
\label{eq:h0_graphene}
\end{align}
where $\rho_a$ are Pauli matrices in sublattice space, and $\sigma$ act on spin-space. The Fermi velocity inherits the spatial dependence of the hopping, $v_F(\mathbf{r}) = v_{F} \left[1-\epsilon(\mathbf{r})\right]$.
Thus, strain enters as a position-dependent velocity, leading to a spatial modulation of the band structure~\cite{de2012space}. Similarly, the Rashba spin-orbit hopping is  also modified as:
$h_R({\bf r})
=
\alpha({\bf r})
\left[
\rho_x\sigma_y
-
\rho_y\sigma_x
\right]$, 
and the Zeeman term is $
h_Z=h\sigma_x$. 
The normal-state Hamiltonian is therefore
\begin{align}
h({\bf r})
=
h_0({\bf r})
+
h_R({\bf r})
+
h_Z .
\label{eq:normal_eta_graphene}
\end{align}
Including superconductivity, the continuum BdG Hamiltonian in the Nambu basis $[
\psi_{\uparrow}({\bf r}),
\psi_{\downarrow}({\bf r}),
\psi^\dagger_{\downarrow}({\bf r}),
-\psi^\dagger_{\uparrow}({\bf r})
]^T$ can be written as: 
\begin{align}
\mathcal{H}_{\rm BdG}({\bf r})
=
\left[
h_0({\bf r})
+
h_R({\bf r})
\right]\tau_z
+
h\sigma_x
+
\Delta\tau_x .
\label{eq:bdg_eta_full_revised}
\end{align}
Note that we no longer write the sublattice dependence explicitly in the Nambu spinor. This is because even though in the continuum Hamiltonian $h_0({\bf r})$ acts on a sublattice-spinor through the Pauli matrices $\rho_a$, for a finite-width ribbon, the physically relevant low-energy degrees of freedom are the discrete eigenmodes obtained by solving the full transverse problem including the sublattice structure.

We now reduce the two-dimensional ribbon to an effective quasi-one-dimensional problem. Since the system is finite in the transverse direction, the transverse momentum is quantized. We expand the field operator as:
\begin{align}
\psi(x,y)
=
\sum_n
\varphi_{n}(y)\,
\chi_{n}(x),
\label{eq:transverse_expansion}
\end{align}
where $\varphi_{n}(y)$, and $n$ is the subband index, which already contains the microscopic information about the
finite-width ribbon, including boundary condition, edge termination, and sublattice structure. Projecting onto the subband basis gives
\begin{align}
H_{\rm BdG}
=
\sum_{n,m}
\int dx\,
\chi_{n}^\dagger(x)
\,
\mathcal{H}_{nm}(x)
\,
\chi_{m}(x),
\label{eq:projected_hamiltonian}
\end{align}
with
\begin{align}
\mathcal{H}_{nm}(x)
=
\int dy\,
\varphi_{n}^\dagger(y)
\mathcal{H}_{\rm BdG}(x,y)
\varphi_{m}(y).
\label{eq:projected_matrix_element}
\end{align}
The diagonal component is the effective Hamiltonian for the subband $n$, i.e., 
\begin{align}
\mathcal{H}_{n}(x)
&=
\left[
-\frac{\hbar^2}{2m_{n}^*(x)}\partial_x^2
-
\mu_{n}^{\rm eff}(x)
\right]\tau_z\nonumber\\
&-i\alpha_{n}(x)\partial_x\sigma_y\tau_z
+
h\sigma_x
+
\Delta\tau_x .
\label{eq:effective_1d_clean}
\end{align}
The Hamiltonian thus assumes the form of the standard proximitized Rashba-wire structure, with parameters inherited from the finite graphene ribbon.
The effective chemical potential is
\begin{align}\mu_{n}^{\rm eff}(x)=\mu - V_{n}(x) - \varepsilon_{n}(x),\end{align}
where
\begin{align}V_{n}(x)=\int dy\, \varphi_{n}^\dagger(y)\,V_{\rm dis}(x,y)\,\varphi_{n}(y)\end{align}
is the projected disorder potential, and $\varepsilon_{n}(x)$ is the
strain-renormalized subband energy. For weak, slowly varying strain,
$\varepsilon_{n}(x)\simeq \varepsilon_{n}^{0}+\lambda_{n}\epsilon_{n}(x)$,
with $\epsilon_{n}(x)=\int dy\, \varphi_{n}^\dagger(y)\epsilon(x,y)\varphi_{n}(y)$.
Thus strain shifts the local subband bottom and enters as a subband-dependent
contribution to the chemical potential. The projected Rashba couplings behave
similarly.

The off-diagonal terms with $n\neq m$ describe inter-subband mixing,
$\mathcal{U}_{nm}(x)=\mathcal{H}_{nm}(x)$, arising from disorder, strain
gradients, Rashba modulation, and boundary effects; e.g.,
$\mathcal{U}_{nm}^{\rm dis}(x)=\int dy\,\varphi_n^\dagger(y)V_{\rm dis}(x,y)\varphi_m(y)\,\tau_z$.
The full effective Hamiltonian is
\begin{align}
H_{\rm GNR}^{\rm BdG}
&=
\sum_{n}\int dx\,\chi_{n}^\dagger(x)\mathcal{H}_{n}(x)\chi_{n}(x)\nonumber\\
&+
\sum_{n\neq m}\int dx\,\chi_{n}^\dagger(x)\mathcal{U}_{nm}(x)\chi_{m}(x).
\end{align}

For an (approximately) isolated subband, the local topological mass is $M_{n}(x)=h^2-\Delta^2-\mu_{n}^{\rm eff}(x)^2$. Regions with $M_n(x)>0(<0)$ are locally topological (trivial), while $M_n(x)=0$ defines domain walls supporting low-energy bound states.
Now, crucially, different subbands experience different projected strain and disorder, so in general $M_n(x)\neq M_m(x)$ for $n\neq m$. Strain can thus separate the effective topological transitions of different channels and reduce their hybridization. On the other hand, if several $M_n(x)$ vanish in the same region, the inter-subband couplings $\mathcal{U}_{nm}(x)$ generate a dense set of hybridized near-zero-energy states, leading naturally to partially
separated Andreev bound states rather than isolated Majorana modes. This is in close agreement with our earlier numerical findings.

\section{Conclusion}\label{conclude}
In this work, we have investigated the role of spatially nonuniform strain in controlling low-energy bound states in proximitized low-dimensional systems, focusing on both one-dimensional semiconductor nanowires and graphene nanoribbons. By combining superconductivity, Rashba spin--orbit coupling, Zeeman fields, and disorder within a unified Bogoliubov--de Gennes framework, we have demonstrated that strain acts as a versatile and experimentally accessible tuning parameter for manipulating the nature of subgap excitations.

Our results establish that strain enables continuous and reversible evolution between trivial bound states, partially separated Andreev bound states, and topological Majorana-bound states. In one-dimensional nanowires, strain primarily modifies the spatial overlap of Majorana components and shifts the effective topological phase boundary, thereby inducing transitions between these regimes. In contrast, in graphene nanoribbons, the multiband structure leads to strong hybridization and spectral crowding near zero energy; here, strain plays a qualitatively different role by lifting degeneracies, suppressing interband mixing, and reorganizing the low-energy spectrum to favor boundary-localized modes.

A central outcome of this work is that strain-induced evolution provides a robust framework for distinguishing psABSs from true topological MBSs. Since strain systematically modifies the spatial structure and spectral properties of low-energy states, psABSs and MBSs exhibit qualitatively distinct responses under strain: trivial psABSs can be driven into well-separated Majorana modes through enhanced nonlocality, whereas initially topological states may lose their nonlocal character and evolve into psABSs due to increased overlap. This contrasting behavior, when analyzed in conjunction with the low-energy spectra, Majorana polarization, and real-space wave-function profiles, enables a consistent identification of the underlying nature of the states. Furthermore, in this work, we developed an analytical theory of strain-driven crossovers, which provides a unified real-space picture in which strain and disorder enter through an effective local potential, shifting the zeros of a position-dependent topological mass and thereby controlling the motion and overlap of Majorana components. In nanowires, this directly governs the transition between trivial states, psABSs, and topological Majorana modes. In graphene nanoribbons, the same framework generalizes to multiple projected subbands, explaining how strain lifts degeneracies, suppresses hybridization, and reorganizes the low-energy spectrum to stabilize boundary-localized modes.

Our findings demonstrate that strain is not merely an irrelevant perturbation, but an important control parameter that can both stabilize and destabilize topological phases depending on the system and regime. In this sense, strain provides a physically transparent and experimentally relevant route for probing the interplay between topology and nonlocality in realistic platforms.
We expect our results to be directly relevant for ongoing experiments in hybrid nanowire and graphene-based superconducting systems, where controlled strain engineering is becoming increasingly feasible. Lastly, strain-driven manipulation of low-energy bound states offers a promising pathway toward better control, identification, and implementation of Majorana modes in complex, disordered environments for topological quantum computation.

\textit{Acknowledgment:} G.S. and Ekta were funded by ANRF-SERB Core Research Grant CRG/2023/005628. S.K. was funded by the IIT Mandi HTRA fellowship. The authors thank IIT Delhi’s HPC facility for computational resources. The authors acknowledge useful discussions with Karsten Held, Anna Kauch, and Sumanta Tewari.

\bibliography{biblio.bib}

\begin{thebibliography}{86}%
\makeatletter
\providecommand \@ifxundefined [1]{%
 \@ifx{#1\undefined}
}%
\providecommand \@ifnum [1]{%
 \ifnum #1\expandafter \@firstoftwo
 \else \expandafter \@secondoftwo
 \fi
}%
\providecommand \@ifx [1]{%
 \ifx #1\expandafter \@firstoftwo
 \else \expandafter \@secondoftwo
 \fi
}%
\providecommand \natexlab [1]{#1}%
\providecommand \enquote  [1]{``#1''}%
\providecommand \bibnamefont  [1]{#1}%
\providecommand \bibfnamefont [1]{#1}%
\providecommand \citenamefont [1]{#1}%
\providecommand \href@noop [0]{\@secondoftwo}%
\providecommand \href [0]{\begingroup \@sanitize@url \@href}%
\providecommand \@href[1]{\@@startlink{#1}\@@href}%
\providecommand \@@href[1]{\endgroup#1\@@endlink}%
\providecommand \@sanitize@url [0]{\catcode `\\12\catcode `\$12\catcode `\&12\catcode `\#12\catcode `\^12\catcode `\_12\catcode `\%12\relax}%
\providecommand \@@startlink[1]{}%
\providecommand \@@endlink[0]{}%
\providecommand \url  [0]{\begingroup\@sanitize@url \@url }%
\providecommand \@url [1]{\endgroup\@href {#1}{\urlprefix }}%
\providecommand \urlprefix  [0]{URL }%
\providecommand \Eprint [0]{\href }%
\providecommand \doibase [0]{https://doi.org/}%
\providecommand \selectlanguage [0]{\@gobble}%
\providecommand \bibinfo  [0]{\@secondoftwo}%
\providecommand \bibfield  [0]{\@secondoftwo}%
\providecommand \translation [1]{[#1]}%
\providecommand \BibitemOpen [0]{}%
\providecommand \bibitemStop [0]{}%
\providecommand \bibitemNoStop [0]{.\EOS\space}%
\providecommand \EOS [0]{\spacefactor3000\relax}%
\providecommand \BibitemShut  [1]{\csname bibitem#1\endcsname}%
\let\auto@bib@innerbib\@empty
\bibitem [{\citenamefont {Majorana}(1937)}]{Majorana1937}%
  \BibitemOpen
  \bibfield  {author} {\bibinfo {author} {\bibfnamefont {E.}~\bibnamefont {Majorana}},\ }\href@noop {} {\bibfield  {journal} {\bibinfo  {journal} {Nuovo Cimento}\ }\textbf {\bibinfo {volume} {14}},\ \bibinfo {pages} {171} (\bibinfo {year} {1937})}\BibitemShut {NoStop}%
\bibitem [{\citenamefont {Moore}\ and\ \citenamefont {Read}(1991)}]{moore1991nonabelions}%
  \BibitemOpen
  \bibfield  {author} {\bibinfo {author} {\bibfnamefont {G.}~\bibnamefont {Moore}}\ and\ \bibinfo {author} {\bibfnamefont {N.}~\bibnamefont {Read}},\ }\bibfield  {title} {\bibinfo {title} {Nonabelions in the fractional quantum hall effect},\ }\href@noop {} {\bibfield  {journal} {\bibinfo  {journal} {Nuclear Physics B}\ }\textbf {\bibinfo {volume} {360}},\ \bibinfo {pages} {362} (\bibinfo {year} {1991})}\BibitemShut {NoStop}%
\bibitem [{\citenamefont {Read}\ and\ \citenamefont {Green}(2000)}]{read2000paired}%
  \BibitemOpen
  \bibfield  {author} {\bibinfo {author} {\bibfnamefont {N.}~\bibnamefont {Read}}\ and\ \bibinfo {author} {\bibfnamefont {D.}~\bibnamefont {Green}},\ }\bibfield  {title} {\bibinfo {title} {Paired states of fermions in two dimensions with breaking of parity and time-reversal symmetries and the fractional quantum hall effect},\ }\href@noop {} {\bibfield  {journal} {\bibinfo  {journal} {Physical Review B}\ }\textbf {\bibinfo {volume} {61}},\ \bibinfo {pages} {10267} (\bibinfo {year} {2000})}\BibitemShut {NoStop}%
\bibitem [{\citenamefont {Nayak}\ and\ \citenamefont {Wilczek}(1996)}]{nayak19962n}%
  \BibitemOpen
  \bibfield  {author} {\bibinfo {author} {\bibfnamefont {C.}~\bibnamefont {Nayak}}\ and\ \bibinfo {author} {\bibfnamefont {F.}~\bibnamefont {Wilczek}},\ }\bibfield  {title} {\bibinfo {title} {2n-quasihole states realize 2n- 1-dimensional spinor braiding statistics in paired quantum hall states},\ }\href@noop {} {\bibfield  {journal} {\bibinfo  {journal} {Nuclear Physics B}\ }\textbf {\bibinfo {volume} {479}},\ \bibinfo {pages} {529} (\bibinfo {year} {1996})}\BibitemShut {NoStop}%
\bibitem [{\citenamefont {Kitaev}(2003)}]{Kitaev2003}%
  \BibitemOpen
  \bibfield  {author} {\bibinfo {author} {\bibfnamefont {A.~Y.}\ \bibnamefont {Kitaev}},\ }\href@noop {} {\bibfield  {journal} {\bibinfo  {journal} {Annals. Phys.}\ }\textbf {\bibinfo {volume} {303}},\ \bibinfo {pages} {2} (\bibinfo {year} {2003})}\BibitemShut {NoStop}%
\bibitem [{\citenamefont {Nayak}\ \emph {et~al.}(2008)\citenamefont {Nayak}, \citenamefont {Simon}, \citenamefont {Stern}, \citenamefont {Freedman},\ and\ \citenamefont {Das~Sarma}}]{Nayak2008}%
  \BibitemOpen
  \bibfield  {author} {\bibinfo {author} {\bibfnamefont {C.}~\bibnamefont {Nayak}}, \bibinfo {author} {\bibfnamefont {S.~H.}\ \bibnamefont {Simon}}, \bibinfo {author} {\bibfnamefont {A.}~\bibnamefont {Stern}}, \bibinfo {author} {\bibfnamefont {M.}~\bibnamefont {Freedman}},\ and\ \bibinfo {author} {\bibfnamefont {S.}~\bibnamefont {Das~Sarma}},\ }\bibfield  {title} {\bibinfo {title} {Non-abelian anyons and topological quantum computation},\ }\href@noop {} {\bibfield  {journal} {\bibinfo  {journal} {Rev. Mod. Phys.}\ }\textbf {\bibinfo {volume} {80}},\ \bibinfo {pages} {1083} (\bibinfo {year} {2008})}\BibitemShut {NoStop}%
\bibitem [{\citenamefont {Sau}\ \emph {et~al.}(2010{\natexlab{a}})\citenamefont {Sau}, \citenamefont {Lutchyn}, \citenamefont {Tewari},\ and\ \citenamefont {Sarma}}]{sau2010generic}%
  \BibitemOpen
  \bibfield  {author} {\bibinfo {author} {\bibfnamefont {J.~D.}\ \bibnamefont {Sau}}, \bibinfo {author} {\bibfnamefont {R.~M.}\ \bibnamefont {Lutchyn}}, \bibinfo {author} {\bibfnamefont {S.}~\bibnamefont {Tewari}},\ and\ \bibinfo {author} {\bibfnamefont {S.~D.}\ \bibnamefont {Sarma}},\ }\bibfield  {title} {\bibinfo {title} {Generic new platform for topological quantum computation using semiconductor heterostructures},\ }\href@noop {} {\bibfield  {journal} {\bibinfo  {journal} {Physical review letters}\ }\textbf {\bibinfo {volume} {104}},\ \bibinfo {pages} {040502} (\bibinfo {year} {2010}{\natexlab{a}})}\BibitemShut {NoStop}%
\bibitem [{\citenamefont {Sau}\ \emph {et~al.}(2010{\natexlab{b}})\citenamefont {Sau}, \citenamefont {Tewari}, \citenamefont {Lutchyn}, \citenamefont {Stanescu},\ and\ \citenamefont {Sarma}}]{sau2010non}%
  \BibitemOpen
  \bibfield  {author} {\bibinfo {author} {\bibfnamefont {J.~D.}\ \bibnamefont {Sau}}, \bibinfo {author} {\bibfnamefont {S.}~\bibnamefont {Tewari}}, \bibinfo {author} {\bibfnamefont {R.~M.}\ \bibnamefont {Lutchyn}}, \bibinfo {author} {\bibfnamefont {T.~D.}\ \bibnamefont {Stanescu}},\ and\ \bibinfo {author} {\bibfnamefont {S.~D.}\ \bibnamefont {Sarma}},\ }\bibfield  {title} {\bibinfo {title} {Non-abelian quantum order in spin-orbit-coupled semiconductors: Search for topological majorana particles in solid-state systems},\ }\href@noop {} {\bibfield  {journal} {\bibinfo  {journal} {Physical Review B}\ }\textbf {\bibinfo {volume} {82}},\ \bibinfo {pages} {214509} (\bibinfo {year} {2010}{\natexlab{b}})}\BibitemShut {NoStop}%
\bibitem [{\citenamefont {Oreg}\ \emph {et~al.}(2010)\citenamefont {Oreg}, \citenamefont {Refael},\ and\ \citenamefont {von Oppen}}]{oreg2010helical}%
  \BibitemOpen
  \bibfield  {author} {\bibinfo {author} {\bibfnamefont {Y.}~\bibnamefont {Oreg}}, \bibinfo {author} {\bibfnamefont {G.}~\bibnamefont {Refael}},\ and\ \bibinfo {author} {\bibfnamefont {F.}~\bibnamefont {von Oppen}},\ }\bibfield  {title} {\bibinfo {title} {Helical liquids and majorana bound states in quantum wires},\ }\href@noop {} {\bibfield  {journal} {\bibinfo  {journal} {Physical review letters}\ }\textbf {\bibinfo {volume} {105}},\ \bibinfo {pages} {177002} (\bibinfo {year} {2010})}\BibitemShut {NoStop}%
\bibitem [{\citenamefont {Lutchyn}\ \emph {et~al.}(2010)\citenamefont {Lutchyn}, \citenamefont {Sau},\ and\ \citenamefont {Sarma}}]{lutchyn2010majorana}%
  \BibitemOpen
  \bibfield  {author} {\bibinfo {author} {\bibfnamefont {R.~M.}\ \bibnamefont {Lutchyn}}, \bibinfo {author} {\bibfnamefont {J.~D.}\ \bibnamefont {Sau}},\ and\ \bibinfo {author} {\bibfnamefont {S.~D.}\ \bibnamefont {Sarma}},\ }\bibfield  {title} {\bibinfo {title} {Majorana fermions and a topological phase transition in semiconductor-superconductor heterostructures},\ }\href@noop {} {\bibfield  {journal} {\bibinfo  {journal} {Physical review letters}\ }\textbf {\bibinfo {volume} {105}},\ \bibinfo {pages} {077001} (\bibinfo {year} {2010})}\BibitemShut {NoStop}%
\bibitem [{\citenamefont {Sengupta}\ \emph {et~al.}(2001)\citenamefont {Sengupta}, \citenamefont {{\v{Z}}uti{\'c}}, \citenamefont {Kwon}, \citenamefont {Yakovenko},\ and\ \citenamefont {Sarma}}]{sengupta2001midgap}%
  \BibitemOpen
  \bibfield  {author} {\bibinfo {author} {\bibfnamefont {K.}~\bibnamefont {Sengupta}}, \bibinfo {author} {\bibfnamefont {I.}~\bibnamefont {{\v{Z}}uti{\'c}}}, \bibinfo {author} {\bibfnamefont {H.-J.}\ \bibnamefont {Kwon}}, \bibinfo {author} {\bibfnamefont {V.~M.}\ \bibnamefont {Yakovenko}},\ and\ \bibinfo {author} {\bibfnamefont {S.~D.}\ \bibnamefont {Sarma}},\ }\bibfield  {title} {\bibinfo {title} {Midgap edge states and pairing symmetry of quasi-one-dimensional organic superconductors},\ }\href@noop {} {\bibfield  {journal} {\bibinfo  {journal} {Physical Review B}\ }\textbf {\bibinfo {volume} {63}},\ \bibinfo {pages} {144531} (\bibinfo {year} {2001})}\BibitemShut {NoStop}%
\bibitem [{\citenamefont {Law}\ \emph {et~al.}(2009)\citenamefont {Law}, \citenamefont {Lee},\ and\ \citenamefont {Ng}}]{law2009majorana}%
  \BibitemOpen
  \bibfield  {author} {\bibinfo {author} {\bibfnamefont {K.}~\bibnamefont {Law}}, \bibinfo {author} {\bibfnamefont {P.~A.}\ \bibnamefont {Lee}},\ and\ \bibinfo {author} {\bibfnamefont {T.}~\bibnamefont {Ng}},\ }\bibfield  {title} {\bibinfo {title} {Majorana fermion induced resonant andreev reflection},\ }\href@noop {} {\bibfield  {journal} {\bibinfo  {journal} {Physical review letters}\ }\textbf {\bibinfo {volume} {103}},\ \bibinfo {pages} {237001} (\bibinfo {year} {2009})}\BibitemShut {NoStop}%
\bibitem [{\citenamefont {Flensberg}(2010)}]{flensberg2010tunneling}%
  \BibitemOpen
  \bibfield  {author} {\bibinfo {author} {\bibfnamefont {K.}~\bibnamefont {Flensberg}},\ }\bibfield  {title} {\bibinfo {title} {Tunneling characteristics of a chain of majorana bound states},\ }\href@noop {} {\bibfield  {journal} {\bibinfo  {journal} {Physical Review B}\ }\textbf {\bibinfo {volume} {82}},\ \bibinfo {pages} {180516} (\bibinfo {year} {2010})}\BibitemShut {NoStop}%
\bibitem [{\citenamefont {Mourik}\ \emph {et~al.}(2012)\citenamefont {Mourik}, \citenamefont {Zuo}, \citenamefont {Frolov}, \citenamefont {Plissard}, \citenamefont {Bakkers},\ and\ \citenamefont {Kouwenhoven}}]{mourik2012signatures}%
  \BibitemOpen
  \bibfield  {author} {\bibinfo {author} {\bibfnamefont {V.}~\bibnamefont {Mourik}}, \bibinfo {author} {\bibfnamefont {K.}~\bibnamefont {Zuo}}, \bibinfo {author} {\bibfnamefont {S.~M.}\ \bibnamefont {Frolov}}, \bibinfo {author} {\bibfnamefont {S.}~\bibnamefont {Plissard}}, \bibinfo {author} {\bibfnamefont {E.~P.}\ \bibnamefont {Bakkers}},\ and\ \bibinfo {author} {\bibfnamefont {L.~P.}\ \bibnamefont {Kouwenhoven}},\ }\bibfield  {title} {\bibinfo {title} {Signatures of majorana fermions in hybrid superconductor-semiconductor nanowire devices},\ }\href@noop {} {\bibfield  {journal} {\bibinfo  {journal} {Science}\ }\textbf {\bibinfo {volume} {336}},\ \bibinfo {pages} {1003} (\bibinfo {year} {2012})}\BibitemShut {NoStop}%
\bibitem [{\citenamefont {Deng}\ \emph {et~al.}(2012)\citenamefont {Deng}, \citenamefont {Yu}, \citenamefont {Huang}, \citenamefont {Larsson}, \citenamefont {Caroff},\ and\ \citenamefont {Xu}}]{deng2012anomalous}%
  \BibitemOpen
  \bibfield  {author} {\bibinfo {author} {\bibfnamefont {M.}~\bibnamefont {Deng}}, \bibinfo {author} {\bibfnamefont {C.}~\bibnamefont {Yu}}, \bibinfo {author} {\bibfnamefont {G.}~\bibnamefont {Huang}}, \bibinfo {author} {\bibfnamefont {M.}~\bibnamefont {Larsson}}, \bibinfo {author} {\bibfnamefont {P.}~\bibnamefont {Caroff}},\ and\ \bibinfo {author} {\bibfnamefont {H.}~\bibnamefont {Xu}},\ }\bibfield  {title} {\bibinfo {title} {Anomalous zero-bias conductance peak in a nb--insb nanowire--nb hybrid device},\ }\href@noop {} {\bibfield  {journal} {\bibinfo  {journal} {Nano letters}\ }\textbf {\bibinfo {volume} {12}},\ \bibinfo {pages} {6414} (\bibinfo {year} {2012})}\BibitemShut {NoStop}%
\bibitem [{\citenamefont {Das}\ \emph {et~al.}(2012)\citenamefont {Das}, \citenamefont {Ronen}, \citenamefont {Most}, \citenamefont {Oreg}, \citenamefont {Heiblum},\ and\ \citenamefont {Shtrikman}}]{das2012zero}%
  \BibitemOpen
  \bibfield  {author} {\bibinfo {author} {\bibfnamefont {A.}~\bibnamefont {Das}}, \bibinfo {author} {\bibfnamefont {Y.}~\bibnamefont {Ronen}}, \bibinfo {author} {\bibfnamefont {Y.}~\bibnamefont {Most}}, \bibinfo {author} {\bibfnamefont {Y.}~\bibnamefont {Oreg}}, \bibinfo {author} {\bibfnamefont {M.}~\bibnamefont {Heiblum}},\ and\ \bibinfo {author} {\bibfnamefont {H.}~\bibnamefont {Shtrikman}},\ }\bibfield  {title} {\bibinfo {title} {Zero-bias peaks and splitting in an al--inas nanowire topological superconductor as a signature of majorana fermions},\ }\href@noop {} {\bibfield  {journal} {\bibinfo  {journal} {Nature Physics}\ }\textbf {\bibinfo {volume} {8}},\ \bibinfo {pages} {887} (\bibinfo {year} {2012})}\BibitemShut {NoStop}%
\bibitem [{\citenamefont {Rokhinson}\ \emph {et~al.}(2012)\citenamefont {Rokhinson}, \citenamefont {Liu},\ and\ \citenamefont {Furdyna}}]{rokhinson2012fractional}%
  \BibitemOpen
  \bibfield  {author} {\bibinfo {author} {\bibfnamefont {L.~P.}\ \bibnamefont {Rokhinson}}, \bibinfo {author} {\bibfnamefont {X.}~\bibnamefont {Liu}},\ and\ \bibinfo {author} {\bibfnamefont {J.~K.}\ \bibnamefont {Furdyna}},\ }\bibfield  {title} {\bibinfo {title} {The fractional ac josephson effect in a semiconductor--superconductor nanowire as a signature of majorana particles},\ }\href@noop {} {\bibfield  {journal} {\bibinfo  {journal} {Nature Physics}\ }\textbf {\bibinfo {volume} {8}},\ \bibinfo {pages} {795} (\bibinfo {year} {2012})}\BibitemShut {NoStop}%
\bibitem [{\citenamefont {Churchill}\ \emph {et~al.}(2013)\citenamefont {Churchill}, \citenamefont {Fatemi}, \citenamefont {Grove-Rasmussen}, \citenamefont {Deng}, \citenamefont {Caroff}, \citenamefont {Xu},\ and\ \citenamefont {Marcus}}]{churchill2013superconductor}%
  \BibitemOpen
  \bibfield  {author} {\bibinfo {author} {\bibfnamefont {H.}~\bibnamefont {Churchill}}, \bibinfo {author} {\bibfnamefont {V.}~\bibnamefont {Fatemi}}, \bibinfo {author} {\bibfnamefont {K.}~\bibnamefont {Grove-Rasmussen}}, \bibinfo {author} {\bibfnamefont {M.}~\bibnamefont {Deng}}, \bibinfo {author} {\bibfnamefont {P.}~\bibnamefont {Caroff}}, \bibinfo {author} {\bibfnamefont {H.}~\bibnamefont {Xu}},\ and\ \bibinfo {author} {\bibfnamefont {C.~M.}\ \bibnamefont {Marcus}},\ }\bibfield  {title} {\bibinfo {title} {Superconductor-nanowire devices from tunneling to the multichannel regime: Zero-bias oscillations and magnetoconductance crossover},\ }\href@noop {} {\bibfield  {journal} {\bibinfo  {journal} {Physical Review B}\ }\textbf {\bibinfo {volume} {87}},\ \bibinfo {pages} {241401} (\bibinfo {year} {2013})}\BibitemShut {NoStop}%
\bibitem [{\citenamefont {Finck}\ \emph {et~al.}(2013)\citenamefont {Finck}, \citenamefont {Van~Harlingen}, \citenamefont {Mohseni}, \citenamefont {Jung},\ and\ \citenamefont {Li}}]{finck2013anomalous}%
  \BibitemOpen
  \bibfield  {author} {\bibinfo {author} {\bibfnamefont {A.}~\bibnamefont {Finck}}, \bibinfo {author} {\bibfnamefont {D.~J.}\ \bibnamefont {Van~Harlingen}}, \bibinfo {author} {\bibfnamefont {P.}~\bibnamefont {Mohseni}}, \bibinfo {author} {\bibfnamefont {K.}~\bibnamefont {Jung}},\ and\ \bibinfo {author} {\bibfnamefont {X.}~\bibnamefont {Li}},\ }\bibfield  {title} {\bibinfo {title} {Anomalous modulation of a zero-bias peak in a hybrid nanowire-superconductor device},\ }\href@noop {} {\bibfield  {journal} {\bibinfo  {journal} {Physical review letters}\ }\textbf {\bibinfo {volume} {110}},\ \bibinfo {pages} {126406} (\bibinfo {year} {2013})}\BibitemShut {NoStop}%
\bibitem [{\citenamefont {Deng}\ \emph {et~al.}(2016)\citenamefont {Deng}, \citenamefont {Vaitiek{\.e}nas}, \citenamefont {Hansen}, \citenamefont {Danon}, \citenamefont {Leijnse}, \citenamefont {Flensberg}, \citenamefont {Nyg{\aa}rd}, \citenamefont {Krogstrup},\ and\ \citenamefont {Marcus}}]{deng2016majorana}%
  \BibitemOpen
  \bibfield  {author} {\bibinfo {author} {\bibfnamefont {M.}~\bibnamefont {Deng}}, \bibinfo {author} {\bibfnamefont {S.}~\bibnamefont {Vaitiek{\.e}nas}}, \bibinfo {author} {\bibfnamefont {E.~B.}\ \bibnamefont {Hansen}}, \bibinfo {author} {\bibfnamefont {J.}~\bibnamefont {Danon}}, \bibinfo {author} {\bibfnamefont {M.}~\bibnamefont {Leijnse}}, \bibinfo {author} {\bibfnamefont {K.}~\bibnamefont {Flensberg}}, \bibinfo {author} {\bibfnamefont {J.}~\bibnamefont {Nyg{\aa}rd}}, \bibinfo {author} {\bibfnamefont {P.}~\bibnamefont {Krogstrup}},\ and\ \bibinfo {author} {\bibfnamefont {C.~M.}\ \bibnamefont {Marcus}},\ }\bibfield  {title} {\bibinfo {title} {Majorana bound state in a coupled quantum-dot hybrid-nanowire system},\ }\href@noop {} {\bibfield  {journal} {\bibinfo  {journal} {Science}\ }\textbf {\bibinfo {volume} {354}},\ \bibinfo {pages} {1557} (\bibinfo {year} {2016})}\BibitemShut {NoStop}%
\bibitem [{\citenamefont {Zhang}\ \emph {et~al.}(2017)\citenamefont {Zhang}, \citenamefont {G{\"u}l}, \citenamefont {Conesa-Boj}, \citenamefont {Nowak}, \citenamefont {Wimmer}, \citenamefont {Zuo}, \citenamefont {Mourik}, \citenamefont {De~Vries}, \citenamefont {Van~Veen}, \citenamefont {De~Moor} \emph {et~al.}}]{zhang2017ballistic}%
  \BibitemOpen
  \bibfield  {author} {\bibinfo {author} {\bibfnamefont {H.}~\bibnamefont {Zhang}}, \bibinfo {author} {\bibfnamefont {{\"O}.}~\bibnamefont {G{\"u}l}}, \bibinfo {author} {\bibfnamefont {S.}~\bibnamefont {Conesa-Boj}}, \bibinfo {author} {\bibfnamefont {M.~P.}\ \bibnamefont {Nowak}}, \bibinfo {author} {\bibfnamefont {M.}~\bibnamefont {Wimmer}}, \bibinfo {author} {\bibfnamefont {K.}~\bibnamefont {Zuo}}, \bibinfo {author} {\bibfnamefont {V.}~\bibnamefont {Mourik}}, \bibinfo {author} {\bibfnamefont {F.~K.}\ \bibnamefont {De~Vries}}, \bibinfo {author} {\bibfnamefont {J.}~\bibnamefont {Van~Veen}}, \bibinfo {author} {\bibfnamefont {M.~W.}\ \bibnamefont {De~Moor}}, \emph {et~al.},\ }\bibfield  {title} {\bibinfo {title} {Ballistic superconductivity in semiconductor nanowires},\ }\href@noop {} {\bibfield  {journal} {\bibinfo  {journal} {Nature communications}\ }\textbf {\bibinfo {volume} {8}},\ \bibinfo {pages} {16025} (\bibinfo {year} {2017})}\BibitemShut {NoStop}%
\bibitem [{\citenamefont {Chen}\ \emph {et~al.}(2017)\citenamefont {Chen}, \citenamefont {Yu}, \citenamefont {Stenger}, \citenamefont {Hocevar}, \citenamefont {Car}, \citenamefont {Plissard}, \citenamefont {Bakkers}, \citenamefont {Stanescu},\ and\ \citenamefont {Frolov}}]{chen2017experimental}%
  \BibitemOpen
  \bibfield  {author} {\bibinfo {author} {\bibfnamefont {J.}~\bibnamefont {Chen}}, \bibinfo {author} {\bibfnamefont {P.}~\bibnamefont {Yu}}, \bibinfo {author} {\bibfnamefont {J.}~\bibnamefont {Stenger}}, \bibinfo {author} {\bibfnamefont {M.}~\bibnamefont {Hocevar}}, \bibinfo {author} {\bibfnamefont {D.}~\bibnamefont {Car}}, \bibinfo {author} {\bibfnamefont {S.~R.}\ \bibnamefont {Plissard}}, \bibinfo {author} {\bibfnamefont {E.~P.}\ \bibnamefont {Bakkers}}, \bibinfo {author} {\bibfnamefont {T.~D.}\ \bibnamefont {Stanescu}},\ and\ \bibinfo {author} {\bibfnamefont {S.~M.}\ \bibnamefont {Frolov}},\ }\bibfield  {title} {\bibinfo {title} {Experimental phase diagram of zero-bias conductance peaks in superconductor/semiconductor nanowire devices},\ }\href@noop {} {\bibfield  {journal} {\bibinfo  {journal} {Science advances}\ }\textbf {\bibinfo {volume} {3}},\ \bibinfo {pages} {e1701476} (\bibinfo {year} {2017})}\BibitemShut {NoStop}%
\bibitem [{\citenamefont {Nichele}\ \emph {et~al.}(2017)\citenamefont {Nichele}, \citenamefont {Drachmann}, \citenamefont {Whiticar}, \citenamefont {O’Farrell}, \citenamefont {Suominen}, \citenamefont {Fornieri}, \citenamefont {Wang}, \citenamefont {Gardner}, \citenamefont {Thomas}, \citenamefont {Hatke} \emph {et~al.}}]{nichele2017scaling}%
  \BibitemOpen
  \bibfield  {author} {\bibinfo {author} {\bibfnamefont {F.}~\bibnamefont {Nichele}}, \bibinfo {author} {\bibfnamefont {A.~C.}\ \bibnamefont {Drachmann}}, \bibinfo {author} {\bibfnamefont {A.~M.}\ \bibnamefont {Whiticar}}, \bibinfo {author} {\bibfnamefont {E.~C.}\ \bibnamefont {O’Farrell}}, \bibinfo {author} {\bibfnamefont {H.~J.}\ \bibnamefont {Suominen}}, \bibinfo {author} {\bibfnamefont {A.}~\bibnamefont {Fornieri}}, \bibinfo {author} {\bibfnamefont {T.}~\bibnamefont {Wang}}, \bibinfo {author} {\bibfnamefont {G.~C.}\ \bibnamefont {Gardner}}, \bibinfo {author} {\bibfnamefont {C.}~\bibnamefont {Thomas}}, \bibinfo {author} {\bibfnamefont {A.~T.}\ \bibnamefont {Hatke}}, \emph {et~al.},\ }\bibfield  {title} {\bibinfo {title} {Scaling of majorana zero-bias conductance peaks},\ }\href@noop {} {\bibfield  {journal} {\bibinfo  {journal} {Physical review letters}\ }\textbf {\bibinfo {volume} {119}},\ \bibinfo {pages} {136803} (\bibinfo {year} {2017})}\BibitemShut {NoStop}%
\bibitem [{\citenamefont {Albrecht}\ \emph {et~al.}(2017)\citenamefont {Albrecht}, \citenamefont {Hansen}, \citenamefont {Higginbotham}, \citenamefont {Kuemmeth}, \citenamefont {Jespersen}, \citenamefont {Nyg{\aa}rd}, \citenamefont {Krogstrup}, \citenamefont {Danon}, \citenamefont {Flensberg},\ and\ \citenamefont {Marcus}}]{albrecht2017transport}%
  \BibitemOpen
  \bibfield  {author} {\bibinfo {author} {\bibfnamefont {S.}~\bibnamefont {Albrecht}}, \bibinfo {author} {\bibfnamefont {E.}~\bibnamefont {Hansen}}, \bibinfo {author} {\bibfnamefont {A.~P.}\ \bibnamefont {Higginbotham}}, \bibinfo {author} {\bibfnamefont {F.}~\bibnamefont {Kuemmeth}}, \bibinfo {author} {\bibfnamefont {T.}~\bibnamefont {Jespersen}}, \bibinfo {author} {\bibfnamefont {J.}~\bibnamefont {Nyg{\aa}rd}}, \bibinfo {author} {\bibfnamefont {P.}~\bibnamefont {Krogstrup}}, \bibinfo {author} {\bibfnamefont {J.}~\bibnamefont {Danon}}, \bibinfo {author} {\bibfnamefont {K.}~\bibnamefont {Flensberg}},\ and\ \bibinfo {author} {\bibfnamefont {C.}~\bibnamefont {Marcus}},\ }\bibfield  {title} {\bibinfo {title} {Transport signatures of quasiparticle poisoning in a majorana island},\ }\href@noop {} {\bibfield  {journal} {\bibinfo  {journal} {Physical review letters}\ }\textbf {\bibinfo {volume} {118}},\ \bibinfo {pages} {137701} (\bibinfo {year} {2017})}\BibitemShut {NoStop}%
\bibitem [{\citenamefont {O~Farrell}\ \emph {et~al.}(2018)\citenamefont {O~Farrell}, \citenamefont {Drachmann}, \citenamefont {Hell}, \citenamefont {Fornieri}, \citenamefont {Whiticar}, \citenamefont {Hansen}, \citenamefont {Gronin}, \citenamefont {Gardner}, \citenamefont {Thomas}, \citenamefont {Manfra} \emph {et~al.}}]{o2018hybridization}%
  \BibitemOpen
  \bibfield  {author} {\bibinfo {author} {\bibfnamefont {E.}~\bibnamefont {O~Farrell}}, \bibinfo {author} {\bibfnamefont {A.}~\bibnamefont {Drachmann}}, \bibinfo {author} {\bibfnamefont {M.}~\bibnamefont {Hell}}, \bibinfo {author} {\bibfnamefont {A.}~\bibnamefont {Fornieri}}, \bibinfo {author} {\bibfnamefont {A.}~\bibnamefont {Whiticar}}, \bibinfo {author} {\bibfnamefont {E.}~\bibnamefont {Hansen}}, \bibinfo {author} {\bibfnamefont {S.}~\bibnamefont {Gronin}}, \bibinfo {author} {\bibfnamefont {G.}~\bibnamefont {Gardner}}, \bibinfo {author} {\bibfnamefont {C.}~\bibnamefont {Thomas}}, \bibinfo {author} {\bibfnamefont {M.}~\bibnamefont {Manfra}}, \emph {et~al.},\ }\bibfield  {title} {\bibinfo {title} {Hybridization of subgap states in one-dimensional superconductor-semiconductor coulomb islands},\ }\href@noop {} {\bibfield  {journal} {\bibinfo  {journal} {Physical review letters}\ }\textbf {\bibinfo {volume} {121}},\ \bibinfo {pages} {256803} (\bibinfo {year} {2018})}\BibitemShut {NoStop}%
\bibitem [{\citenamefont {Shen}\ \emph {et~al.}(2018)\citenamefont {Shen}, \citenamefont {Heedt}, \citenamefont {Borsoi}, \citenamefont {Van~Heck}, \citenamefont {Gazibegovic}, \citenamefont {het Veld}, \citenamefont {Car}, \citenamefont {Logan}, \citenamefont {Pendharkar}, \citenamefont {Ramakers} \emph {et~al.}}]{shen2018parity}%
  \BibitemOpen
  \bibfield  {author} {\bibinfo {author} {\bibfnamefont {J.}~\bibnamefont {Shen}}, \bibinfo {author} {\bibfnamefont {S.}~\bibnamefont {Heedt}}, \bibinfo {author} {\bibfnamefont {F.}~\bibnamefont {Borsoi}}, \bibinfo {author} {\bibfnamefont {B.}~\bibnamefont {Van~Heck}}, \bibinfo {author} {\bibfnamefont {S.}~\bibnamefont {Gazibegovic}}, \bibinfo {author} {\bibfnamefont {R.~L.~O.}\ \bibnamefont {het Veld}}, \bibinfo {author} {\bibfnamefont {D.}~\bibnamefont {Car}}, \bibinfo {author} {\bibfnamefont {J.~A.}\ \bibnamefont {Logan}}, \bibinfo {author} {\bibfnamefont {M.}~\bibnamefont {Pendharkar}}, \bibinfo {author} {\bibfnamefont {S.~J.}\ \bibnamefont {Ramakers}}, \emph {et~al.},\ }\bibfield  {title} {\bibinfo {title} {Parity transitions in the superconducting ground state of hybrid insb--al coulomb islands},\ }\href@noop {} {\bibfield  {journal} {\bibinfo  {journal} {Nature communications}\ }\textbf {\bibinfo {volume} {9}},\ \bibinfo {pages} {4801} (\bibinfo {year} {2018})}\BibitemShut {NoStop}%
\bibitem [{\citenamefont {Sherman}\ \emph {et~al.}(2017)\citenamefont {Sherman}, \citenamefont {Yodh}, \citenamefont {Albrecht}, \citenamefont {Nyg{\aa}rd}, \citenamefont {Krogstrup},\ and\ \citenamefont {Marcus}}]{sherman2017normal}%
  \BibitemOpen
  \bibfield  {author} {\bibinfo {author} {\bibfnamefont {D.}~\bibnamefont {Sherman}}, \bibinfo {author} {\bibfnamefont {J.}~\bibnamefont {Yodh}}, \bibinfo {author} {\bibfnamefont {S.~M.}\ \bibnamefont {Albrecht}}, \bibinfo {author} {\bibfnamefont {J.}~\bibnamefont {Nyg{\aa}rd}}, \bibinfo {author} {\bibfnamefont {P.}~\bibnamefont {Krogstrup}},\ and\ \bibinfo {author} {\bibfnamefont {C.~M.}\ \bibnamefont {Marcus}},\ }\bibfield  {title} {\bibinfo {title} {Normal, superconducting and topological regimes of hybrid double quantum dots},\ }\href@noop {} {\bibfield  {journal} {\bibinfo  {journal} {Nature nanotechnology}\ }\textbf {\bibinfo {volume} {12}},\ \bibinfo {pages} {212} (\bibinfo {year} {2017})}\BibitemShut {NoStop}%
\bibitem [{\citenamefont {Vaitiek{\.e}nas}\ \emph {et~al.}(2018)\citenamefont {Vaitiek{\.e}nas}, \citenamefont {Whiticar}, \citenamefont {Deng}, \citenamefont {Krizek}, \citenamefont {Sestoft}, \citenamefont {Palmstr{\o}m}, \citenamefont {Marti-Sanchez}, \citenamefont {Arbiol}, \citenamefont {Krogstrup}, \citenamefont {Casparis} \emph {et~al.}}]{vaitiekenas2018selective}%
  \BibitemOpen
  \bibfield  {author} {\bibinfo {author} {\bibfnamefont {S.}~\bibnamefont {Vaitiek{\.e}nas}}, \bibinfo {author} {\bibfnamefont {A.}~\bibnamefont {Whiticar}}, \bibinfo {author} {\bibfnamefont {M.-T.}\ \bibnamefont {Deng}}, \bibinfo {author} {\bibfnamefont {F.}~\bibnamefont {Krizek}}, \bibinfo {author} {\bibfnamefont {J.}~\bibnamefont {Sestoft}}, \bibinfo {author} {\bibfnamefont {C.}~\bibnamefont {Palmstr{\o}m}}, \bibinfo {author} {\bibfnamefont {S.}~\bibnamefont {Marti-Sanchez}}, \bibinfo {author} {\bibfnamefont {J.}~\bibnamefont {Arbiol}}, \bibinfo {author} {\bibfnamefont {P.}~\bibnamefont {Krogstrup}}, \bibinfo {author} {\bibfnamefont {L.}~\bibnamefont {Casparis}}, \emph {et~al.},\ }\bibfield  {title} {\bibinfo {title} {Selective-area-grown semiconductor-superconductor hybrids: A basis for topological networks},\ }\href@noop {} {\bibfield  {journal} {\bibinfo  {journal} {Physical review letters}\ }\textbf {\bibinfo {volume} {121}},\ \bibinfo {pages} {147701} (\bibinfo {year} {2018})}\BibitemShut {NoStop}%
\bibitem [{\citenamefont {Albrecht}\ \emph {et~al.}(2016)\citenamefont {Albrecht}, \citenamefont {Higginbotham}, \citenamefont {Madsen}, \citenamefont {Kuemmeth}, \citenamefont {Jespersen}, \citenamefont {Nyg{\aa}rd}, \citenamefont {Krogstrup},\ and\ \citenamefont {Marcus}}]{albrecht2016exponential}%
  \BibitemOpen
  \bibfield  {author} {\bibinfo {author} {\bibfnamefont {S.~M.}\ \bibnamefont {Albrecht}}, \bibinfo {author} {\bibfnamefont {A.~P.}\ \bibnamefont {Higginbotham}}, \bibinfo {author} {\bibfnamefont {M.}~\bibnamefont {Madsen}}, \bibinfo {author} {\bibfnamefont {F.}~\bibnamefont {Kuemmeth}}, \bibinfo {author} {\bibfnamefont {T.~S.}\ \bibnamefont {Jespersen}}, \bibinfo {author} {\bibfnamefont {J.}~\bibnamefont {Nyg{\aa}rd}}, \bibinfo {author} {\bibfnamefont {P.}~\bibnamefont {Krogstrup}},\ and\ \bibinfo {author} {\bibfnamefont {C.}~\bibnamefont {Marcus}},\ }\bibfield  {title} {\bibinfo {title} {Exponential protection of zero modes in majorana islands},\ }\href@noop {} {\bibfield  {journal} {\bibinfo  {journal} {Nature}\ }\textbf {\bibinfo {volume} {531}},\ \bibinfo {pages} {206} (\bibinfo {year} {2016})}\BibitemShut {NoStop}%
\bibitem [{\citenamefont {Yu}\ \emph {et~al.}(2021)\citenamefont {Yu}, \citenamefont {Chen}, \citenamefont {Gomanko}, \citenamefont {Badawy}, \citenamefont {Bakkers}, \citenamefont {Zuo}, \citenamefont {Mourik},\ and\ \citenamefont {Frolov}}]{Yu_2021}%
  \BibitemOpen
  \bibfield  {author} {\bibinfo {author} {\bibfnamefont {P.}~\bibnamefont {Yu}}, \bibinfo {author} {\bibfnamefont {J.}~\bibnamefont {Chen}}, \bibinfo {author} {\bibfnamefont {M.}~\bibnamefont {Gomanko}}, \bibinfo {author} {\bibfnamefont {G.}~\bibnamefont {Badawy}}, \bibinfo {author} {\bibfnamefont {E.~P. A.~M.}\ \bibnamefont {Bakkers}}, \bibinfo {author} {\bibfnamefont {K.}~\bibnamefont {Zuo}}, \bibinfo {author} {\bibfnamefont {V.}~\bibnamefont {Mourik}},\ and\ \bibinfo {author} {\bibfnamefont {S.~M.}\ \bibnamefont {Frolov}},\ }\bibfield  {title} {\bibinfo {title} {Non-majorana states yield nearly quantized conductance in proximatized nanowires},\ }\href@noop {} {\bibfield  {journal} {\bibinfo  {journal} {Nature Physics}\ }\textbf {\bibinfo {volume} {17}},\ \bibinfo {pages} {482–488} (\bibinfo {year} {2021})}\BibitemShut {NoStop}%
\bibitem [{\citenamefont {Quantum}\ \emph {et~al.}(2025)\citenamefont {Quantum}, \citenamefont {Aghaee}, \citenamefont {Alcaraz~Ramirez}, \citenamefont {Alam}, \citenamefont {Ali}, \citenamefont {Andrzejczuk}, \citenamefont {Antipov}, \citenamefont {Astafev}, \citenamefont {Barzegar}, \citenamefont {Bauer} \emph {et~al.}}]{microsoft2025interferometric}%
  \BibitemOpen
  \bibfield  {author} {\bibinfo {author} {\bibfnamefont {M.~A.}\ \bibnamefont {Quantum}}, \bibinfo {author} {\bibfnamefont {M.}~\bibnamefont {Aghaee}}, \bibinfo {author} {\bibfnamefont {A.}~\bibnamefont {Alcaraz~Ramirez}}, \bibinfo {author} {\bibfnamefont {Z.}~\bibnamefont {Alam}}, \bibinfo {author} {\bibfnamefont {R.}~\bibnamefont {Ali}}, \bibinfo {author} {\bibfnamefont {M.}~\bibnamefont {Andrzejczuk}}, \bibinfo {author} {\bibfnamefont {A.}~\bibnamefont {Antipov}}, \bibinfo {author} {\bibfnamefont {M.}~\bibnamefont {Astafev}}, \bibinfo {author} {\bibfnamefont {A.}~\bibnamefont {Barzegar}}, \bibinfo {author} {\bibfnamefont {B.}~\bibnamefont {Bauer}}, \emph {et~al.},\ }\bibfield  {title} {\bibinfo {title} {Interferometric single-shot parity measurement in inas--al hybrid devices},\ }\href@noop {} {\bibfield  {journal} {\bibinfo  {journal} {Nature}\ }\textbf {\bibinfo {volume} {638}},\ \bibinfo {pages} {651} (\bibinfo {year} {2025})}\BibitemShut {NoStop}%
\bibitem [{\citenamefont {Mondal}\ \emph {et~al.}(2025)\citenamefont {Mondal}, \citenamefont {Pal}, \citenamefont {Saha},\ and\ \citenamefont {Nag}}]{mondal2025distinguishing}%
  \BibitemOpen
  \bibfield  {author} {\bibinfo {author} {\bibfnamefont {D.}~\bibnamefont {Mondal}}, \bibinfo {author} {\bibfnamefont {A.}~\bibnamefont {Pal}}, \bibinfo {author} {\bibfnamefont {A.}~\bibnamefont {Saha}},\ and\ \bibinfo {author} {\bibfnamefont {T.}~\bibnamefont {Nag}},\ }\bibfield  {title} {\bibinfo {title} {Distinguishing between topological majorana and trivial zero modes via transport and shot noise study in an altermagnet heterostructure},\ }\href@noop {} {\bibfield  {journal} {\bibinfo  {journal} {Physical Review B}\ }\textbf {\bibinfo {volume} {111}},\ \bibinfo {pages} {L121401} (\bibinfo {year} {2025})}\BibitemShut {NoStop}%
\bibitem [{\citenamefont {G{\l}odzik}\ \emph {et~al.}(2020)\citenamefont {G{\l}odzik}, \citenamefont {Sedlmayr},\ and\ \citenamefont {Doma{\'n}ski}}]{glodzik2020measure}%
  \BibitemOpen
  \bibfield  {author} {\bibinfo {author} {\bibfnamefont {S.}~\bibnamefont {G{\l}odzik}}, \bibinfo {author} {\bibfnamefont {N.}~\bibnamefont {Sedlmayr}},\ and\ \bibinfo {author} {\bibfnamefont {T.}~\bibnamefont {Doma{\'n}ski}},\ }\bibfield  {title} {\bibinfo {title} {How to measure the majorana polarization of a topological planar josephson junction},\ }\href@noop {} {\bibfield  {journal} {\bibinfo  {journal} {Physical Review B}\ }\textbf {\bibinfo {volume} {102}},\ \bibinfo {pages} {085411} (\bibinfo {year} {2020})}\BibitemShut {NoStop}%
\bibitem [{\citenamefont {Rossi}\ \emph {et~al.}(2020)\citenamefont {Rossi}, \citenamefont {Dolcini},\ and\ \citenamefont {Rossi}}]{rossi2020majorana}%
  \BibitemOpen
  \bibfield  {author} {\bibinfo {author} {\bibfnamefont {L.}~\bibnamefont {Rossi}}, \bibinfo {author} {\bibfnamefont {F.}~\bibnamefont {Dolcini}},\ and\ \bibinfo {author} {\bibfnamefont {F.}~\bibnamefont {Rossi}},\ }\bibfield  {title} {\bibinfo {title} {Majorana-like localized spin density without bound states in topologically trivial spin-orbit coupled nanowires},\ }\href@noop {} {\bibfield  {journal} {\bibinfo  {journal} {Physical Review B}\ }\textbf {\bibinfo {volume} {101}},\ \bibinfo {pages} {195421} (\bibinfo {year} {2020})}\BibitemShut {NoStop}%
\bibitem [{\citenamefont {Tanaka}\ \emph {et~al.}(2024)\citenamefont {Tanaka}, \citenamefont {Tamura},\ and\ \citenamefont {Cayao}}]{tanaka2024theory}%
  \BibitemOpen
  \bibfield  {author} {\bibinfo {author} {\bibfnamefont {Y.}~\bibnamefont {Tanaka}}, \bibinfo {author} {\bibfnamefont {S.}~\bibnamefont {Tamura}},\ and\ \bibinfo {author} {\bibfnamefont {J.}~\bibnamefont {Cayao}},\ }\bibfield  {title} {\bibinfo {title} {Theory of majorana zero modes in unconventional superconductors},\ }\href@noop {} {\bibfield  {journal} {\bibinfo  {journal} {Progress of Theoretical and Experimental Physics}\ }\textbf {\bibinfo {volume} {2024}},\ \bibinfo {pages} {08C105} (\bibinfo {year} {2024})}\BibitemShut {NoStop}%
\bibitem [{\citenamefont {Tanaka}\ \emph {et~al.}(2011)\citenamefont {Tanaka}, \citenamefont {Sato},\ and\ \citenamefont {Nagaosa}}]{tanaka2011symmetry}%
  \BibitemOpen
  \bibfield  {author} {\bibinfo {author} {\bibfnamefont {Y.}~\bibnamefont {Tanaka}}, \bibinfo {author} {\bibfnamefont {M.}~\bibnamefont {Sato}},\ and\ \bibinfo {author} {\bibfnamefont {N.}~\bibnamefont {Nagaosa}},\ }\bibfield  {title} {\bibinfo {title} {Symmetry and topology in superconductors--odd-frequency pairing and edge states--},\ }\href@noop {} {\bibfield  {journal} {\bibinfo  {journal} {Journal of the Physical Society of Japan}\ }\textbf {\bibinfo {volume} {81}},\ \bibinfo {pages} {011013} (\bibinfo {year} {2011})}\BibitemShut {NoStop}%
\bibitem [{\citenamefont {Sharma}\ and\ \citenamefont {Tewari}(2016)}]{sharma2016tunneling}%
  \BibitemOpen
  \bibfield  {author} {\bibinfo {author} {\bibfnamefont {G.}~\bibnamefont {Sharma}}\ and\ \bibinfo {author} {\bibfnamefont {S.}~\bibnamefont {Tewari}},\ }\bibfield  {title} {\bibinfo {title} {Tunneling conductance for majorana fermions in spin-orbit coupled semiconductor-superconductor heterostructures using superconducting leads},\ }\href@noop {} {\bibfield  {journal} {\bibinfo  {journal} {Physical Review B}\ }\textbf {\bibinfo {volume} {93}},\ \bibinfo {pages} {195161} (\bibinfo {year} {2016})}\BibitemShut {NoStop}%
\bibitem [{\citenamefont {Tanaka}\ \emph {et~al.}(2009)\citenamefont {Tanaka}, \citenamefont {Yokoyama},\ and\ \citenamefont {Nagaosa}}]{tanaka2009manipulation}%
  \BibitemOpen
  \bibfield  {author} {\bibinfo {author} {\bibfnamefont {Y.}~\bibnamefont {Tanaka}}, \bibinfo {author} {\bibfnamefont {T.}~\bibnamefont {Yokoyama}},\ and\ \bibinfo {author} {\bibfnamefont {N.}~\bibnamefont {Nagaosa}},\ }\bibfield  {title} {\bibinfo {title} {Manipulation of the majorana fermion, andreev reflection, and josephson current on topological insulators},\ }\href@noop {} {\bibfield  {journal} {\bibinfo  {journal} {Physical review letters}\ }\textbf {\bibinfo {volume} {103}},\ \bibinfo {pages} {107002} (\bibinfo {year} {2009})}\BibitemShut {NoStop}%
\bibitem [{\citenamefont {Kells}\ \emph {et~al.}(2012)\citenamefont {Kells}, \citenamefont {Meidan},\ and\ \citenamefont {Brouwer}}]{Brouwer2012}%
  \BibitemOpen
  \bibfield  {author} {\bibinfo {author} {\bibfnamefont {G.}~\bibnamefont {Kells}}, \bibinfo {author} {\bibfnamefont {D.}~\bibnamefont {Meidan}},\ and\ \bibinfo {author} {\bibfnamefont {P.~W.}\ \bibnamefont {Brouwer}},\ }\bibfield  {title} {\bibinfo {title} {Near-zero-energy end states in topologically trivial spin-orbit coupled superconducting nanowires with a smooth confinement},\ }\href@noop {} {\bibfield  {journal} {\bibinfo  {journal} {Phys. Rev. B}\ }\textbf {\bibinfo {volume} {86}},\ \bibinfo {pages} {100503} (\bibinfo {year} {2012})}\BibitemShut {NoStop}%
\bibitem [{\citenamefont {Mi}\ \emph {et~al.}(2014)\citenamefont {Mi}, \citenamefont {Pikulin}, \citenamefont {Marciani},\ and\ \citenamefont {Beenakker}}]{Mi2014}%
  \BibitemOpen
  \bibfield  {author} {\bibinfo {author} {\bibfnamefont {S.}~\bibnamefont {Mi}}, \bibinfo {author} {\bibfnamefont {D.~I.}\ \bibnamefont {Pikulin}}, \bibinfo {author} {\bibfnamefont {M.}~\bibnamefont {Marciani}},\ and\ \bibinfo {author} {\bibfnamefont {C.~W.~J.}\ \bibnamefont {Beenakker}},\ }\bibfield  {title} {\bibinfo {title} {X-shaped and y-shaped andreev resonance profiles in a superconducting quantum dot},\ }\href@noop {} {\bibfield  {journal} {\bibinfo  {journal} {Journal of Experimental and Theoretical Physics}\ }\textbf {\bibinfo {volume} {119}},\ \bibinfo {pages} {1018} (\bibinfo {year} {2014})}\BibitemShut {NoStop}%
\bibitem [{\citenamefont {Bagrets}\ and\ \citenamefont {Altland}(2012)}]{Bagrets2012}%
  \BibitemOpen
  \bibfield  {author} {\bibinfo {author} {\bibfnamefont {D.}~\bibnamefont {Bagrets}}\ and\ \bibinfo {author} {\bibfnamefont {A.}~\bibnamefont {Altland}},\ }\bibfield  {title} {\bibinfo {title} {Class $d$ spectral peak in majorana quantum wires},\ }\href@noop {} {\bibfield  {journal} {\bibinfo  {journal} {Phys. Rev. Lett.}\ }\textbf {\bibinfo {volume} {109}},\ \bibinfo {pages} {227005} (\bibinfo {year} {2012})}\BibitemShut {NoStop}%
\bibitem [{\citenamefont {Pikulin}\ \emph {et~al.}(2012)\citenamefont {Pikulin}, \citenamefont {Dahlhaus}, \citenamefont {Wimmer}, \citenamefont {Schomerus},\ and\ \citenamefont {Beenakker}}]{pikulin2012zero}%
  \BibitemOpen
  \bibfield  {author} {\bibinfo {author} {\bibfnamefont {D.~I.}\ \bibnamefont {Pikulin}}, \bibinfo {author} {\bibfnamefont {J.}~\bibnamefont {Dahlhaus}}, \bibinfo {author} {\bibfnamefont {M.}~\bibnamefont {Wimmer}}, \bibinfo {author} {\bibfnamefont {H.}~\bibnamefont {Schomerus}},\ and\ \bibinfo {author} {\bibfnamefont {C.}~\bibnamefont {Beenakker}},\ }\bibfield  {title} {\bibinfo {title} {A zero-voltage conductance peak from weak antilocalization in a majorana nanowire},\ }\href@noop {} {\bibfield  {journal} {\bibinfo  {journal} {New Journal of Physics}\ }\textbf {\bibinfo {volume} {14}},\ \bibinfo {pages} {125011} (\bibinfo {year} {2012})}\BibitemShut {NoStop}%
\bibitem [{\citenamefont {Prada}\ \emph {et~al.}(2012)\citenamefont {Prada}, \citenamefont {San-Jose},\ and\ \citenamefont {Aguado}}]{ramon2012transport}%
  \BibitemOpen
  \bibfield  {author} {\bibinfo {author} {\bibfnamefont {E.}~\bibnamefont {Prada}}, \bibinfo {author} {\bibfnamefont {P.}~\bibnamefont {San-Jose}},\ and\ \bibinfo {author} {\bibfnamefont {R.}~\bibnamefont {Aguado}},\ }\bibfield  {title} {\bibinfo {title} {Transport spectroscopy of $ns$ nanowire junctions with majorana fermions},\ }\href@noop {} {\bibfield  {journal} {\bibinfo  {journal} {Phys. Rev. B}\ }\textbf {\bibinfo {volume} {86}},\ \bibinfo {pages} {180503} (\bibinfo {year} {2012})}\BibitemShut {NoStop}%
\bibitem [{\citenamefont {Pan}\ and\ \citenamefont {Sarma}(2020)}]{pan2020physical}%
  \BibitemOpen
  \bibfield  {author} {\bibinfo {author} {\bibfnamefont {H.}~\bibnamefont {Pan}}\ and\ \bibinfo {author} {\bibfnamefont {S.~D.}\ \bibnamefont {Sarma}},\ }\bibfield  {title} {\bibinfo {title} {Physical mechanisms for zero-bias conductance peaks in majorana nanowires},\ }\href@noop {} {\bibfield  {journal} {\bibinfo  {journal} {Physical Review Research}\ }\textbf {\bibinfo {volume} {2}},\ \bibinfo {pages} {013377} (\bibinfo {year} {2020})}\BibitemShut {NoStop}%
\bibitem [{\citenamefont {Moore}\ \emph {et~al.}(2018{\natexlab{a}})\citenamefont {Moore}, \citenamefont {Stanescu},\ and\ \citenamefont {Tewari}}]{moore2018two}%
  \BibitemOpen
  \bibfield  {author} {\bibinfo {author} {\bibfnamefont {C.}~\bibnamefont {Moore}}, \bibinfo {author} {\bibfnamefont {T.~D.}\ \bibnamefont {Stanescu}},\ and\ \bibinfo {author} {\bibfnamefont {S.}~\bibnamefont {Tewari}},\ }\bibfield  {title} {\bibinfo {title} {Two-terminal charge tunneling: Disentangling majorana zero modes from partially separated andreev bound states in semiconductor-superconductor heterostructures},\ }\href@noop {} {\bibfield  {journal} {\bibinfo  {journal} {Physical Review B}\ }\textbf {\bibinfo {volume} {97}},\ \bibinfo {pages} {165302} (\bibinfo {year} {2018}{\natexlab{a}})}\BibitemShut {NoStop}%
\bibitem [{\citenamefont {Moore}\ \emph {et~al.}(2018{\natexlab{b}})\citenamefont {Moore}, \citenamefont {Zeng}, \citenamefont {Stanescu},\ and\ \citenamefont {Tewari}}]{moore2018quantized}%
  \BibitemOpen
  \bibfield  {author} {\bibinfo {author} {\bibfnamefont {C.}~\bibnamefont {Moore}}, \bibinfo {author} {\bibfnamefont {C.}~\bibnamefont {Zeng}}, \bibinfo {author} {\bibfnamefont {T.~D.}\ \bibnamefont {Stanescu}},\ and\ \bibinfo {author} {\bibfnamefont {S.}~\bibnamefont {Tewari}},\ }\bibfield  {title} {\bibinfo {title} {Quantized zero-bias conductance plateau in semiconductor-superconductor heterostructures without topological majorana zero modes},\ }\href@noop {} {\bibfield  {journal} {\bibinfo  {journal} {Physical Review B}\ }\textbf {\bibinfo {volume} {98}},\ \bibinfo {pages} {155314} (\bibinfo {year} {2018}{\natexlab{b}})}\BibitemShut {NoStop}%
\bibitem [{\citenamefont {Vuik}\ \emph {et~al.}(2018)\citenamefont {Vuik}, \citenamefont {Nijholt}, \citenamefont {Akhmerov},\ and\ \citenamefont {Wimmer}}]{vuik2018reproducing}%
  \BibitemOpen
  \bibfield  {author} {\bibinfo {author} {\bibfnamefont {A.}~\bibnamefont {Vuik}}, \bibinfo {author} {\bibfnamefont {B.}~\bibnamefont {Nijholt}}, \bibinfo {author} {\bibfnamefont {A.~R.}\ \bibnamefont {Akhmerov}},\ and\ \bibinfo {author} {\bibfnamefont {M.}~\bibnamefont {Wimmer}},\ }\bibfield  {title} {\bibinfo {title} {Reproducing topological properties with quasi-majorana states},\ }\href@noop {} {\bibfield  {journal} {\bibinfo  {journal} {arXiv preprint arXiv:1806.02801}\ } (\bibinfo {year} {2018})}\BibitemShut {NoStop}%
\bibitem [{\citenamefont {Stanescu}\ and\ \citenamefont {Tewari}(2019)}]{Stanescu_Robust}%
  \BibitemOpen
  \bibfield  {author} {\bibinfo {author} {\bibfnamefont {T.~D.}\ \bibnamefont {Stanescu}}\ and\ \bibinfo {author} {\bibfnamefont {S.}~\bibnamefont {Tewari}},\ }\bibfield  {title} {\bibinfo {title} {Robust low-energy andreev bound states in semiconductor-superconductor structures: Importance of partial separation of component majorana bound states},\ }\href@noop {} {\bibfield  {journal} {\bibinfo  {journal} {Phys. Rev. B}\ }\textbf {\bibinfo {volume} {100}},\ \bibinfo {pages} {155429} (\bibinfo {year} {2019})}\BibitemShut {NoStop}%
\bibitem [{\citenamefont {{San-Jose}}\ \emph {et~al.}(2016)\citenamefont {{San-Jose}}, \citenamefont {{Cayao}}, \citenamefont {{Prada}},\ and\ \citenamefont {{Aguado}}}]{ramon_Jorge2106exceptional}%
  \BibitemOpen
  \bibfield  {author} {\bibinfo {author} {\bibfnamefont {P.}~\bibnamefont {{San-Jose}}}, \bibinfo {author} {\bibfnamefont {J.}~\bibnamefont {{Cayao}}}, \bibinfo {author} {\bibfnamefont {E.}~\bibnamefont {{Prada}}},\ and\ \bibinfo {author} {\bibfnamefont {R.}~\bibnamefont {{Aguado}}},\ }\bibfield  {title} {\bibinfo {title} {{Majorana bound states from exceptional points in non-topological superconductors}},\ }\href@noop {} {\bibfield  {journal} {\bibinfo  {journal} {Scientific Reports}\ }\textbf {\bibinfo {volume} {6}},\ \bibinfo {eid} {21427} (\bibinfo {year} {2016})}\BibitemShut {NoStop}%
\bibitem [{\citenamefont {Avila}\ \emph {et~al.}(2019)\citenamefont {Avila}, \citenamefont {Peñaranda}, \citenamefont {Prada}, \citenamefont {San-Jose},\ and\ \citenamefont {Aguado}}]{ramon2019nonhermitian}%
  \BibitemOpen
  \bibfield  {author} {\bibinfo {author} {\bibfnamefont {J.}~\bibnamefont {Avila}}, \bibinfo {author} {\bibfnamefont {F.}~\bibnamefont {Peñaranda}}, \bibinfo {author} {\bibfnamefont {E.}~\bibnamefont {Prada}}, \bibinfo {author} {\bibfnamefont {P.}~\bibnamefont {San-Jose}},\ and\ \bibinfo {author} {\bibfnamefont {R.}~\bibnamefont {Aguado}},\ }\bibfield  {title} {\bibinfo {title} {Non-hermitian topology as a unifying framework for the andreev versus majorana states controversy},\ }\href@noop {} {\bibfield  {journal} {\bibinfo  {journal} {Communications Physics}\ }\textbf {\bibinfo {volume} {2}} (\bibinfo {year} {2019})}\BibitemShut {NoStop}%
\bibitem [{\citenamefont {Awoga}\ \emph {et~al.}(2019)\citenamefont {Awoga}, \citenamefont {Cayao},\ and\ \citenamefont {Black-Schaffer}}]{Jorge2019supercurrent}%
  \BibitemOpen
  \bibfield  {author} {\bibinfo {author} {\bibfnamefont {O.~A.}\ \bibnamefont {Awoga}}, \bibinfo {author} {\bibfnamefont {J.}~\bibnamefont {Cayao}},\ and\ \bibinfo {author} {\bibfnamefont {A.~M.}\ \bibnamefont {Black-Schaffer}},\ }\bibfield  {title} {\bibinfo {title} {Supercurrent detection of topologically trivial zero-energy states in nanowire junctions},\ }\href@noop {} {\bibfield  {journal} {\bibinfo  {journal} {Phys. Rev. Lett.}\ }\textbf {\bibinfo {volume} {123}},\ \bibinfo {pages} {117001} (\bibinfo {year} {2019})}\BibitemShut {NoStop}%
\bibitem [{\citenamefont {Sharma}\ \emph {et~al.}(2020)\citenamefont {Sharma}, \citenamefont {Zeng}, \citenamefont {Stanescu},\ and\ \citenamefont {Tewari}}]{sharma2020hybridization}%
  \BibitemOpen
  \bibfield  {author} {\bibinfo {author} {\bibfnamefont {G.}~\bibnamefont {Sharma}}, \bibinfo {author} {\bibfnamefont {C.}~\bibnamefont {Zeng}}, \bibinfo {author} {\bibfnamefont {T.~D.}\ \bibnamefont {Stanescu}},\ and\ \bibinfo {author} {\bibfnamefont {S.}~\bibnamefont {Tewari}},\ }\bibfield  {title} {\bibinfo {title} {Hybridization energy oscillations of majorana and andreev bound states in semiconductor-superconductor nanowire heterostructures},\ }\href@noop {} {\bibfield  {journal} {\bibinfo  {journal} {Physical Review B}\ }\textbf {\bibinfo {volume} {101}},\ \bibinfo {pages} {245405} (\bibinfo {year} {2020})}\BibitemShut {NoStop}%
\bibitem [{\citenamefont {Zeng}\ \emph {et~al.}(2020)\citenamefont {Zeng}, \citenamefont {Sharma}, \citenamefont {Stanescu},\ and\ \citenamefont {Tewari}}]{zeng2020feasibility}%
  \BibitemOpen
  \bibfield  {author} {\bibinfo {author} {\bibfnamefont {C.}~\bibnamefont {Zeng}}, \bibinfo {author} {\bibfnamefont {G.}~\bibnamefont {Sharma}}, \bibinfo {author} {\bibfnamefont {T.~D.}\ \bibnamefont {Stanescu}},\ and\ \bibinfo {author} {\bibfnamefont {S.}~\bibnamefont {Tewari}},\ }\bibfield  {title} {\bibinfo {title} {Feasibility of measurement-based braiding in the quasi-majorana regime of semiconductor-superconductor heterostructures},\ }\href@noop {} {\bibfield  {journal} {\bibinfo  {journal} {Physical Review B}\ }\textbf {\bibinfo {volume} {102}},\ \bibinfo {pages} {205101} (\bibinfo {year} {2020})}\BibitemShut {NoStop}%
\bibitem [{\citenamefont {Zeng}\ \emph {et~al.}(2022)\citenamefont {Zeng}, \citenamefont {Sharma}, \citenamefont {Tewari},\ and\ \citenamefont {Stanescu}}]{zeng2022partially}%
  \BibitemOpen
  \bibfield  {author} {\bibinfo {author} {\bibfnamefont {C.}~\bibnamefont {Zeng}}, \bibinfo {author} {\bibfnamefont {G.}~\bibnamefont {Sharma}}, \bibinfo {author} {\bibfnamefont {S.}~\bibnamefont {Tewari}},\ and\ \bibinfo {author} {\bibfnamefont {T.}~\bibnamefont {Stanescu}},\ }\bibfield  {title} {\bibinfo {title} {Partially separated majorana modes in a disordered medium},\ }\href@noop {} {\bibfield  {journal} {\bibinfo  {journal} {Physical Review B}\ }\textbf {\bibinfo {volume} {105}},\ \bibinfo {pages} {205122} (\bibinfo {year} {2022})}\BibitemShut {NoStop}%
\bibitem [{\citenamefont {Prada}\ \emph {et~al.}(2020)\citenamefont {Prada}, \citenamefont {San-Jose}, \citenamefont {de~Moor}, \citenamefont {Geresdi}, \citenamefont {Lee}, \citenamefont {Klinovaja}, \citenamefont {Loss}, \citenamefont {Nygård}, \citenamefont {Aguado},\ and\ \citenamefont {Kouwenhoven}}]{ramon2020from}%
  \BibitemOpen
  \bibfield  {author} {\bibinfo {author} {\bibfnamefont {E.}~\bibnamefont {Prada}}, \bibinfo {author} {\bibfnamefont {P.}~\bibnamefont {San-Jose}}, \bibinfo {author} {\bibfnamefont {M.~W.~A.}\ \bibnamefont {de~Moor}}, \bibinfo {author} {\bibfnamefont {A.}~\bibnamefont {Geresdi}}, \bibinfo {author} {\bibfnamefont {E.~J.~H.}\ \bibnamefont {Lee}}, \bibinfo {author} {\bibfnamefont {J.}~\bibnamefont {Klinovaja}}, \bibinfo {author} {\bibfnamefont {D.}~\bibnamefont {Loss}}, \bibinfo {author} {\bibfnamefont {J.}~\bibnamefont {Nygård}}, \bibinfo {author} {\bibfnamefont {R.}~\bibnamefont {Aguado}},\ and\ \bibinfo {author} {\bibfnamefont {L.~P.}\ \bibnamefont {Kouwenhoven}},\ }\bibfield  {title} {\bibinfo {title} {From andreev to majorana bound states in hybrid superconductor–semiconductor nanowires},\ }\href@noop {} {\bibfield  {journal} {\bibinfo  {journal} {Nature Reviews Physics}\ } (\bibinfo {year} {2020})}\BibitemShut {NoStop}%
\bibitem [{\citenamefont {Cayao}\ and\ \citenamefont {Black-Schaffer}(2021)}]{Jorge2021distinguishing}%
  \BibitemOpen
  \bibfield  {author} {\bibinfo {author} {\bibfnamefont {J.}~\bibnamefont {Cayao}}\ and\ \bibinfo {author} {\bibfnamefont {A.~M.}\ \bibnamefont {Black-Schaffer}},\ }\bibfield  {title} {\bibinfo {title} {Distinguishing trivial and topological zero-energy states in long nanowire junctions},\ }\href@noop {} {\bibfield  {journal} {\bibinfo  {journal} {Physical Review B}\ }\textbf {\bibinfo {volume} {104}} (\bibinfo {year} {2021})}\BibitemShut {NoStop}%
\bibitem [{\citenamefont {Zhang}\ \emph {et~al.}(2021)\citenamefont {Zhang}, \citenamefont {de~Moor}, \citenamefont {Bommer}, \citenamefont {Xu}, \citenamefont {Wang}, \citenamefont {van Loo}, \citenamefont {Liu}, \citenamefont {Gazibegovic}, \citenamefont {Logan}, \citenamefont {Car}, \citenamefont {het Veld}, \citenamefont {van Veldhoven}, \citenamefont {Koelling}, \citenamefont {Verheijen}, \citenamefont {Pendharkar}, \citenamefont {Pennachio}, \citenamefont {Shojaei}, \citenamefont {Lee}, \citenamefont {Palmstrøm}, \citenamefont {Bakkers}, \citenamefont {Sarma},\ and\ \citenamefont {Kouwenhoven}}]{zhang2021}%
  \BibitemOpen
  \bibfield  {author} {\bibinfo {author} {\bibfnamefont {H.}~\bibnamefont {Zhang}}, \bibinfo {author} {\bibfnamefont {M.~W.~A.}\ \bibnamefont {de~Moor}}, \bibinfo {author} {\bibfnamefont {J.~D.~S.}\ \bibnamefont {Bommer}}, \bibinfo {author} {\bibfnamefont {D.}~\bibnamefont {Xu}}, \bibinfo {author} {\bibfnamefont {G.}~\bibnamefont {Wang}}, \bibinfo {author} {\bibfnamefont {N.}~\bibnamefont {van Loo}}, \bibinfo {author} {\bibfnamefont {C.-X.}\ \bibnamefont {Liu}}, \bibinfo {author} {\bibfnamefont {S.}~\bibnamefont {Gazibegovic}}, \bibinfo {author} {\bibfnamefont {J.~A.}\ \bibnamefont {Logan}}, \bibinfo {author} {\bibfnamefont {D.}~\bibnamefont {Car}}, \bibinfo {author} {\bibfnamefont {R.~L. M.~O.}\ \bibnamefont {het Veld}}, \bibinfo {author} {\bibfnamefont {P.~J.}\ \bibnamefont {van Veldhoven}}, \bibinfo {author} {\bibfnamefont {S.}~\bibnamefont {Koelling}}, \bibinfo {author} {\bibfnamefont {M.~A.}\ \bibnamefont {Verheijen}}, \bibinfo {author} {\bibfnamefont {M.}~\bibnamefont {Pendharkar}}, \bibinfo {author}
  {\bibfnamefont {D.~J.}\ \bibnamefont {Pennachio}}, \bibinfo {author} {\bibfnamefont {B.}~\bibnamefont {Shojaei}}, \bibinfo {author} {\bibfnamefont {J.~S.}\ \bibnamefont {Lee}}, \bibinfo {author} {\bibfnamefont {C.~J.}\ \bibnamefont {Palmstrøm}}, \bibinfo {author} {\bibfnamefont {E.~P. A.~M.}\ \bibnamefont {Bakkers}}, \bibinfo {author} {\bibfnamefont {S.~D.}\ \bibnamefont {Sarma}},\ and\ \bibinfo {author} {\bibfnamefont {L.~P.}\ \bibnamefont {Kouwenhoven}},\ }\href@noop {} {\bibinfo {title} {Large zero-bias peaks in insb-al hybrid semiconductor-superconductor nanowire devices}} (\bibinfo {year} {2021})\BibitemShut {NoStop}%
\bibitem [{\citenamefont {Sarma}\ and\ \citenamefont {Pan}(2021)}]{DasSarma2021}%
  \BibitemOpen
  \bibfield  {author} {\bibinfo {author} {\bibfnamefont {S.~D.}\ \bibnamefont {Sarma}}\ and\ \bibinfo {author} {\bibfnamefont {H.}~\bibnamefont {Pan}},\ }\href@noop {} {\bibinfo {title} {Disorder-induced zero-bias peaks in majorana nanowires}} (\bibinfo {year} {2021})\BibitemShut {NoStop}%
\bibitem [{\citenamefont {Frolov}(2021)}]{Frolov2021}%
  \BibitemOpen
  \bibfield  {author} {\bibinfo {author} {\bibfnamefont {S.}~\bibnamefont {Frolov}},\ }\bibfield  {title} {\bibinfo {title} {Quantum computing’s reproducibility crisis: Majorana fermions},\ }\href@noop {} {\bibfield  {journal} {\bibinfo  {journal} {Nature}\ }\textbf {\bibinfo {volume} {592}},\ \bibinfo {pages} {350} (\bibinfo {year} {2021})}\BibitemShut {NoStop}%
\bibitem [{\citenamefont {Nakada}\ \emph {et~al.}(1996)\citenamefont {Nakada}, \citenamefont {Fujita}, \citenamefont {Dresselhaus},\ and\ \citenamefont {Dresselhaus}}]{nakada1996edge}%
  \BibitemOpen
  \bibfield  {author} {\bibinfo {author} {\bibfnamefont {K.}~\bibnamefont {Nakada}}, \bibinfo {author} {\bibfnamefont {M.}~\bibnamefont {Fujita}}, \bibinfo {author} {\bibfnamefont {G.}~\bibnamefont {Dresselhaus}},\ and\ \bibinfo {author} {\bibfnamefont {M.~S.}\ \bibnamefont {Dresselhaus}},\ }\bibfield  {title} {\bibinfo {title} {Edge state in graphene ribbons: Nanometer size effect and edge shape dependence},\ }\href@noop {} {\bibfield  {journal} {\bibinfo  {journal} {Physical Review B}\ }\textbf {\bibinfo {volume} {54}},\ \bibinfo {pages} {17954} (\bibinfo {year} {1996})}\BibitemShut {NoStop}%
\bibitem [{\citenamefont {Castro~Neto}\ \emph {et~al.}(2009)\citenamefont {Castro~Neto}, \citenamefont {Guinea}, \citenamefont {Peres}, \citenamefont {Novoselov},\ and\ \citenamefont {Geim}}]{castro2009electronic}%
  \BibitemOpen
  \bibfield  {author} {\bibinfo {author} {\bibfnamefont {A.~H.}\ \bibnamefont {Castro~Neto}}, \bibinfo {author} {\bibfnamefont {F.}~\bibnamefont {Guinea}}, \bibinfo {author} {\bibfnamefont {N.~M.}\ \bibnamefont {Peres}}, \bibinfo {author} {\bibfnamefont {K.~S.}\ \bibnamefont {Novoselov}},\ and\ \bibinfo {author} {\bibfnamefont {A.~K.}\ \bibnamefont {Geim}},\ }\bibfield  {title} {\bibinfo {title} {The electronic properties of graphene},\ }\href@noop {} {\bibfield  {journal} {\bibinfo  {journal} {Reviews of modern physics}\ }\textbf {\bibinfo {volume} {81}},\ \bibinfo {pages} {109} (\bibinfo {year} {2009})}\BibitemShut {NoStop}%
\bibitem [{\citenamefont {Ma}\ \emph {et~al.}(2025)\citenamefont {Ma}, \citenamefont {Pizzochero},\ and\ \citenamefont {Chaudhary}}]{ma2025graphene}%
  \BibitemOpen
  \bibfield  {author} {\bibinfo {author} {\bibfnamefont {R.}~\bibnamefont {Ma}}, \bibinfo {author} {\bibfnamefont {M.}~\bibnamefont {Pizzochero}},\ and\ \bibinfo {author} {\bibfnamefont {G.}~\bibnamefont {Chaudhary}},\ }\bibfield  {title} {\bibinfo {title} {Graphene nanoribbons as a majorana platform},\ }\href@noop {} {\bibfield  {journal} {\bibinfo  {journal} {arXiv preprint arXiv:2506.14999}\ } (\bibinfo {year} {2025})}\BibitemShut {NoStop}%
\bibitem [{\citenamefont {Kaladzhyan}\ and\ \citenamefont {Bena}(2017)}]{kaladzhyan2017formation}%
  \BibitemOpen
  \bibfield  {author} {\bibinfo {author} {\bibfnamefont {V.}~\bibnamefont {Kaladzhyan}}\ and\ \bibinfo {author} {\bibfnamefont {C.}~\bibnamefont {Bena}},\ }\bibfield  {title} {\bibinfo {title} {Formation of majorana fermions in finite-size graphene strips},\ }\href@noop {} {\bibfield  {journal} {\bibinfo  {journal} {SciPost Physics}\ }\textbf {\bibinfo {volume} {3}},\ \bibinfo {pages} {002} (\bibinfo {year} {2017})}\BibitemShut {NoStop}%
\bibitem [{\citenamefont {Kaladzhyan}\ \emph {et~al.}(2017)\citenamefont {Kaladzhyan}, \citenamefont {Despres}, \citenamefont {Mandal},\ and\ \citenamefont {Bena}}]{kaladzhyan2017majorana}%
  \BibitemOpen
  \bibfield  {author} {\bibinfo {author} {\bibfnamefont {V.}~\bibnamefont {Kaladzhyan}}, \bibinfo {author} {\bibfnamefont {J.}~\bibnamefont {Despres}}, \bibinfo {author} {\bibfnamefont {I.}~\bibnamefont {Mandal}},\ and\ \bibinfo {author} {\bibfnamefont {C.}~\bibnamefont {Bena}},\ }\bibfield  {title} {\bibinfo {title} {Majorana fermions in finite-size strips with in-plane magnetic fields},\ }\href@noop {} {\bibfield  {journal} {\bibinfo  {journal} {The European Physical Journal B}\ }\textbf {\bibinfo {volume} {90}},\ \bibinfo {pages} {1} (\bibinfo {year} {2017})}\BibitemShut {NoStop}%
\bibitem [{\citenamefont {San-Jose}\ \emph {et~al.}(2015)\citenamefont {San-Jose}, \citenamefont {Lado}, \citenamefont {Aguado}, \citenamefont {Guinea},\ and\ \citenamefont {Fern\'andez-Rossier}}]{PhysRevX.5.041042}%
  \BibitemOpen
  \bibfield  {author} {\bibinfo {author} {\bibfnamefont {P.}~\bibnamefont {San-Jose}}, \bibinfo {author} {\bibfnamefont {J.~L.}\ \bibnamefont {Lado}}, \bibinfo {author} {\bibfnamefont {R.}~\bibnamefont {Aguado}}, \bibinfo {author} {\bibfnamefont {F.}~\bibnamefont {Guinea}},\ and\ \bibinfo {author} {\bibfnamefont {J.}~\bibnamefont {Fern\'andez-Rossier}},\ }\bibfield  {title} {\bibinfo {title} {Majorana zero modes in graphene},\ }\href@noop {} {\bibfield  {journal} {\bibinfo  {journal} {Phys. Rev. X}\ }\textbf {\bibinfo {volume} {5}},\ \bibinfo {pages} {041042} (\bibinfo {year} {2015})}\BibitemShut {NoStop}%
\bibitem [{\citenamefont {Laubscher}\ \emph {et~al.}(2020)\citenamefont {Laubscher}, \citenamefont {Loss},\ and\ \citenamefont {Klinovaja}}]{laubscher2020majorana}%
  \BibitemOpen
  \bibfield  {author} {\bibinfo {author} {\bibfnamefont {K.}~\bibnamefont {Laubscher}}, \bibinfo {author} {\bibfnamefont {D.}~\bibnamefont {Loss}},\ and\ \bibinfo {author} {\bibfnamefont {J.}~\bibnamefont {Klinovaja}},\ }\bibfield  {title} {\bibinfo {title} {Majorana and parafermion corner states from two coupled sheets of bilayer graphene},\ }\href@noop {} {\bibfield  {journal} {\bibinfo  {journal} {Physical Review Research}\ }\textbf {\bibinfo {volume} {2}},\ \bibinfo {pages} {013330} (\bibinfo {year} {2020})}\BibitemShut {NoStop}%
\bibitem [{\citenamefont {Wang}\ \emph {et~al.}(2018)\citenamefont {Wang}, \citenamefont {Castro},\ and\ \citenamefont {Lin}}]{wang2018strain}%
  \BibitemOpen
  \bibfield  {author} {\bibinfo {author} {\bibfnamefont {Z.-H.}\ \bibnamefont {Wang}}, \bibinfo {author} {\bibfnamefont {E.~V.}\ \bibnamefont {Castro}},\ and\ \bibinfo {author} {\bibfnamefont {H.-Q.}\ \bibnamefont {Lin}},\ }\bibfield  {title} {\bibinfo {title} {Strain manipulation of majorana fermions in graphene armchair nanoribbons},\ }\href@noop {} {\bibfield  {journal} {\bibinfo  {journal} {Physical Review B}\ }\textbf {\bibinfo {volume} {97}},\ \bibinfo {pages} {041414} (\bibinfo {year} {2018})}\BibitemShut {NoStop}%
\bibitem [{\citenamefont {Manesco}\ \emph {et~al.}(2019)\citenamefont {Manesco}, \citenamefont {Weber},\ and\ \citenamefont {Rodrigues~Jr}}]{manesco2019effective}%
  \BibitemOpen
  \bibfield  {author} {\bibinfo {author} {\bibfnamefont {A.~L.}\ \bibnamefont {Manesco}}, \bibinfo {author} {\bibfnamefont {G.}~\bibnamefont {Weber}},\ and\ \bibinfo {author} {\bibfnamefont {D.}~\bibnamefont {Rodrigues~Jr}},\ }\bibfield  {title} {\bibinfo {title} {Effective model for majorana modes in graphene},\ }\href@noop {} {\bibfield  {journal} {\bibinfo  {journal} {Physical Review B}\ }\textbf {\bibinfo {volume} {100}},\ \bibinfo {pages} {125411} (\bibinfo {year} {2019})}\BibitemShut {NoStop}%
\bibitem [{\citenamefont {Signorello}\ \emph {et~al.}(2017)\citenamefont {Signorello}, \citenamefont {Sant}, \citenamefont {Bologna}, \citenamefont {Schraff}, \citenamefont {Drechsler}, \citenamefont {Schmid}, \citenamefont {Wirths}, \citenamefont {Rossell}, \citenamefont {Schenk},\ and\ \citenamefont {Riel}}]{signorello2017manipulating}%
  \BibitemOpen
  \bibfield  {author} {\bibinfo {author} {\bibfnamefont {G.}~\bibnamefont {Signorello}}, \bibinfo {author} {\bibfnamefont {S.}~\bibnamefont {Sant}}, \bibinfo {author} {\bibfnamefont {N.}~\bibnamefont {Bologna}}, \bibinfo {author} {\bibfnamefont {M.}~\bibnamefont {Schraff}}, \bibinfo {author} {\bibfnamefont {U.}~\bibnamefont {Drechsler}}, \bibinfo {author} {\bibfnamefont {H.}~\bibnamefont {Schmid}}, \bibinfo {author} {\bibfnamefont {S.}~\bibnamefont {Wirths}}, \bibinfo {author} {\bibfnamefont {M.~D.}\ \bibnamefont {Rossell}}, \bibinfo {author} {\bibfnamefont {A.}~\bibnamefont {Schenk}},\ and\ \bibinfo {author} {\bibfnamefont {H.}~\bibnamefont {Riel}},\ }\bibfield  {title} {\bibinfo {title} {Manipulating surface states of iii--v nanowires with uniaxial stress},\ }\href@noop {} {\bibfield  {journal} {\bibinfo  {journal} {Nano letters}\ }\textbf {\bibinfo {volume} {17}},\ \bibinfo {pages} {2816} (\bibinfo {year} {2017})}\BibitemShut {NoStop}%
\bibitem [{\citenamefont {Zeng}\ \emph {et~al.}(2018)\citenamefont {Zeng}, \citenamefont {Gammer}, \citenamefont {Ozdol}, \citenamefont {Nordqvist}, \citenamefont {Nygard}, \citenamefont {Krogstrup}, \citenamefont {Minor}, \citenamefont {Jaager},\ and\ \citenamefont {Olsson}}]{zeng2018correlation}%
  \BibitemOpen
  \bibfield  {author} {\bibinfo {author} {\bibfnamefont {L.}~\bibnamefont {Zeng}}, \bibinfo {author} {\bibfnamefont {C.}~\bibnamefont {Gammer}}, \bibinfo {author} {\bibfnamefont {B.}~\bibnamefont {Ozdol}}, \bibinfo {author} {\bibfnamefont {T.}~\bibnamefont {Nordqvist}}, \bibinfo {author} {\bibfnamefont {J.}~\bibnamefont {Nygard}}, \bibinfo {author} {\bibfnamefont {P.}~\bibnamefont {Krogstrup}}, \bibinfo {author} {\bibfnamefont {A.~M.}\ \bibnamefont {Minor}}, \bibinfo {author} {\bibfnamefont {W.}~\bibnamefont {Jaager}},\ and\ \bibinfo {author} {\bibfnamefont {E.}~\bibnamefont {Olsson}},\ }\bibfield  {title} {\bibinfo {title} {Correlation between electrical transport and nanoscale strain in inas/in0. 6ga0. 4as core--shell nanowires},\ }\href@noop {} {\bibfield  {journal} {\bibinfo  {journal} {Nano Letters}\ }\textbf {\bibinfo {volume} {18}},\ \bibinfo {pages} {4949} (\bibinfo {year} {2018})}\BibitemShut {NoStop}%
\bibitem [{\citenamefont {Chen}\ \emph {et~al.}(2020)\citenamefont {Chen}, \citenamefont {Das}, \citenamefont {Aytan}, \citenamefont {Zhou}, \citenamefont {Horowitz}, \citenamefont {Satpati}, \citenamefont {Balandin}, \citenamefont {Lake},\ and\ \citenamefont {Wei}}]{chen2020strain}%
  \BibitemOpen
  \bibfield  {author} {\bibinfo {author} {\bibfnamefont {C.}~\bibnamefont {Chen}}, \bibinfo {author} {\bibfnamefont {P.}~\bibnamefont {Das}}, \bibinfo {author} {\bibfnamefont {E.}~\bibnamefont {Aytan}}, \bibinfo {author} {\bibfnamefont {W.}~\bibnamefont {Zhou}}, \bibinfo {author} {\bibfnamefont {J.}~\bibnamefont {Horowitz}}, \bibinfo {author} {\bibfnamefont {B.}~\bibnamefont {Satpati}}, \bibinfo {author} {\bibfnamefont {A.~A.}\ \bibnamefont {Balandin}}, \bibinfo {author} {\bibfnamefont {R.~K.}\ \bibnamefont {Lake}},\ and\ \bibinfo {author} {\bibfnamefont {P.}~\bibnamefont {Wei}},\ }\bibfield  {title} {\bibinfo {title} {Strain-controlled superconductivity in few-layer nbse2},\ }\href@noop {} {\bibfield  {journal} {\bibinfo  {journal} {ACS applied materials \& interfaces}\ }\textbf {\bibinfo {volume} {12}},\ \bibinfo {pages} {38744} (\bibinfo {year} {2020})}\BibitemShut {NoStop}%
\bibitem [{\citenamefont {Guinea}\ \emph {et~al.}(2010)\citenamefont {Guinea}, \citenamefont {Katsnelson},\ and\ \citenamefont {Geim}}]{guinea2010energy}%
  \BibitemOpen
  \bibfield  {author} {\bibinfo {author} {\bibfnamefont {F.}~\bibnamefont {Guinea}}, \bibinfo {author} {\bibfnamefont {M.~I.}\ \bibnamefont {Katsnelson}},\ and\ \bibinfo {author} {\bibfnamefont {A.}~\bibnamefont {Geim}},\ }\bibfield  {title} {\bibinfo {title} {Energy gaps and a zero-field quantum hall effect in graphene by strain engineering},\ }\href@noop {} {\bibfield  {journal} {\bibinfo  {journal} {Nature Physics}\ }\textbf {\bibinfo {volume} {6}},\ \bibinfo {pages} {30} (\bibinfo {year} {2010})}\BibitemShut {NoStop}%
\bibitem [{\citenamefont {Vozmediano}\ \emph {et~al.}(2010)\citenamefont {Vozmediano}, \citenamefont {Katsnelson},\ and\ \citenamefont {Guinea}}]{vozmediano2010gauge}%
  \BibitemOpen
  \bibfield  {author} {\bibinfo {author} {\bibfnamefont {M.~A.}\ \bibnamefont {Vozmediano}}, \bibinfo {author} {\bibfnamefont {M.}~\bibnamefont {Katsnelson}},\ and\ \bibinfo {author} {\bibfnamefont {F.}~\bibnamefont {Guinea}},\ }\bibfield  {title} {\bibinfo {title} {Gauge fields in graphene},\ }\href@noop {} {\bibfield  {journal} {\bibinfo  {journal} {Physics Reports}\ }\textbf {\bibinfo {volume} {496}},\ \bibinfo {pages} {109} (\bibinfo {year} {2010})}\BibitemShut {NoStop}%
\bibitem [{\citenamefont {Pereira}\ and\ \citenamefont {Castro~Neto}(2009)}]{pereira2009strain}%
  \BibitemOpen
  \bibfield  {author} {\bibinfo {author} {\bibfnamefont {V.~M.}\ \bibnamefont {Pereira}}\ and\ \bibinfo {author} {\bibfnamefont {A.}~\bibnamefont {Castro~Neto}},\ }\bibfield  {title} {\bibinfo {title} {Strain engineering of graphene’s electronic structure},\ }\href@noop {} {\bibfield  {journal} {\bibinfo  {journal} {Physical review letters}\ }\textbf {\bibinfo {volume} {103}},\ \bibinfo {pages} {046801} (\bibinfo {year} {2009})}\BibitemShut {NoStop}%
\bibitem [{\citenamefont {Levy}\ \emph {et~al.}(2010)\citenamefont {Levy}, \citenamefont {Burke}, \citenamefont {Meaker}, \citenamefont {Panlasigui}, \citenamefont {Zettl}, \citenamefont {Guinea}, \citenamefont {Neto},\ and\ \citenamefont {Crommie}}]{levy2010strain}%
  \BibitemOpen
  \bibfield  {author} {\bibinfo {author} {\bibfnamefont {N.}~\bibnamefont {Levy}}, \bibinfo {author} {\bibfnamefont {S.~A.}\ \bibnamefont {Burke}}, \bibinfo {author} {\bibfnamefont {K.}~\bibnamefont {Meaker}}, \bibinfo {author} {\bibfnamefont {M.}~\bibnamefont {Panlasigui}}, \bibinfo {author} {\bibfnamefont {A.}~\bibnamefont {Zettl}}, \bibinfo {author} {\bibfnamefont {F.}~\bibnamefont {Guinea}}, \bibinfo {author} {\bibfnamefont {A.~C.}\ \bibnamefont {Neto}},\ and\ \bibinfo {author} {\bibfnamefont {M.~F.}\ \bibnamefont {Crommie}},\ }\bibfield  {title} {\bibinfo {title} {Strain-induced pseudo--magnetic fields greater than 300 tesla in graphene nanobubbles},\ }\href@noop {} {\bibfield  {journal} {\bibinfo  {journal} {science}\ }\textbf {\bibinfo {volume} {329}},\ \bibinfo {pages} {544} (\bibinfo {year} {2010})}\BibitemShut {NoStop}%
\bibitem [{\citenamefont {Kang}\ \emph {et~al.}(2021)\citenamefont {Kang}, \citenamefont {Sun}, \citenamefont {Luo}, \citenamefont {Lu}, \citenamefont {Chen}, \citenamefont {Kim}, \citenamefont {Jung}, \citenamefont {Gao}, \citenamefont {Parluhutan}, \citenamefont {Ge} \emph {et~al.}}]{kang2021pseudo}%
  \BibitemOpen
  \bibfield  {author} {\bibinfo {author} {\bibfnamefont {D.-H.}\ \bibnamefont {Kang}}, \bibinfo {author} {\bibfnamefont {H.}~\bibnamefont {Sun}}, \bibinfo {author} {\bibfnamefont {M.}~\bibnamefont {Luo}}, \bibinfo {author} {\bibfnamefont {K.}~\bibnamefont {Lu}}, \bibinfo {author} {\bibfnamefont {M.}~\bibnamefont {Chen}}, \bibinfo {author} {\bibfnamefont {Y.}~\bibnamefont {Kim}}, \bibinfo {author} {\bibfnamefont {Y.}~\bibnamefont {Jung}}, \bibinfo {author} {\bibfnamefont {X.}~\bibnamefont {Gao}}, \bibinfo {author} {\bibfnamefont {S.~J.}\ \bibnamefont {Parluhutan}}, \bibinfo {author} {\bibfnamefont {J.}~\bibnamefont {Ge}}, \emph {et~al.},\ }\bibfield  {title} {\bibinfo {title} {Pseudo-magnetic field-induced slow carrier dynamics in periodically strained graphene},\ }\href@noop {} {\bibfield  {journal} {\bibinfo  {journal} {Nature Communications}\ }\textbf {\bibinfo {volume} {12}},\ \bibinfo {pages} {5087} (\bibinfo {year} {2021})}\BibitemShut {NoStop}%
\bibitem [{\citenamefont {Tewari}\ and\ \citenamefont {Sau}(2012)}]{tewari2012topological}%
  \BibitemOpen
  \bibfield  {author} {\bibinfo {author} {\bibfnamefont {S.}~\bibnamefont {Tewari}}\ and\ \bibinfo {author} {\bibfnamefont {J.~D.}\ \bibnamefont {Sau}},\ }\bibfield  {title} {\bibinfo {title} {Topological invariants for spin-orbit coupled superconductor nanowires},\ }\href@noop {} {\bibfield  {journal} {\bibinfo  {journal} {Physical review letters}\ }\textbf {\bibinfo {volume} {109}},\ \bibinfo {pages} {150408} (\bibinfo {year} {2012})}\BibitemShut {NoStop}%
\bibitem [{\citenamefont {Karoliya}\ \emph {et~al.}(2025)\citenamefont {Karoliya}, \citenamefont {Tewari},\ and\ \citenamefont {Sharma}}]{karoliya2025majorana}%
  \BibitemOpen
  \bibfield  {author} {\bibinfo {author} {\bibfnamefont {S.}~\bibnamefont {Karoliya}}, \bibinfo {author} {\bibfnamefont {S.}~\bibnamefont {Tewari}},\ and\ \bibinfo {author} {\bibfnamefont {G.}~\bibnamefont {Sharma}},\ }\bibfield  {title} {\bibinfo {title} {Majorana polarization in disordered quasi-one-dimensional hybrid nanowires},\ }\href@noop {} {\bibfield  {journal} {\bibinfo  {journal} {Physical Review B}\ }\textbf {\bibinfo {volume} {112}},\ \bibinfo {pages} {165410} (\bibinfo {year} {2025})}\BibitemShut {NoStop}%
\bibitem [{\citenamefont {Das~Sarma}\ \emph {et~al.}(2011)\citenamefont {Das~Sarma}, \citenamefont {Adam}, \citenamefont {Hwang},\ and\ \citenamefont {Rossi}}]{das2011electronic}%
  \BibitemOpen
  \bibfield  {author} {\bibinfo {author} {\bibfnamefont {S.}~\bibnamefont {Das~Sarma}}, \bibinfo {author} {\bibfnamefont {S.}~\bibnamefont {Adam}}, \bibinfo {author} {\bibfnamefont {E.}~\bibnamefont {Hwang}},\ and\ \bibinfo {author} {\bibfnamefont {E.}~\bibnamefont {Rossi}},\ }\bibfield  {title} {\bibinfo {title} {Electronic transport in two-dimensional graphene},\ }\href@noop {} {\bibfield  {journal} {\bibinfo  {journal} {Reviews of modern physics}\ }\textbf {\bibinfo {volume} {83}},\ \bibinfo {pages} {407} (\bibinfo {year} {2011})}\BibitemShut {NoStop}%
\bibitem [{\citenamefont {Sticlet}\ \emph {et~al.}(2012)\citenamefont {Sticlet}, \citenamefont {Bena},\ and\ \citenamefont {Simon}}]{sticlet2012spin}%
  \BibitemOpen
  \bibfield  {author} {\bibinfo {author} {\bibfnamefont {D.}~\bibnamefont {Sticlet}}, \bibinfo {author} {\bibfnamefont {C.}~\bibnamefont {Bena}},\ and\ \bibinfo {author} {\bibfnamefont {P.}~\bibnamefont {Simon}},\ }\bibfield  {title} {\bibinfo {title} {Spin and majorana polarization in topological superconducting wires},\ }\href@noop {} {\bibfield  {journal} {\bibinfo  {journal} {Physical Review Letters}\ }\textbf {\bibinfo {volume} {108}},\ \bibinfo {pages} {096802} (\bibinfo {year} {2012})}\BibitemShut {NoStop}%
\bibitem [{\citenamefont {Sedlmayr}\ and\ \citenamefont {Bena}(2015)}]{sedlmayr2015visualizing}%
  \BibitemOpen
  \bibfield  {author} {\bibinfo {author} {\bibfnamefont {N.}~\bibnamefont {Sedlmayr}}\ and\ \bibinfo {author} {\bibfnamefont {C.}~\bibnamefont {Bena}},\ }\bibfield  {title} {\bibinfo {title} {Visualizing majorana bound states in one and two dimensions using the generalized majorana polarization},\ }\href@noop {} {\bibfield  {journal} {\bibinfo  {journal} {Physical Review B}\ }\textbf {\bibinfo {volume} {92}},\ \bibinfo {pages} {115115} (\bibinfo {year} {2015})}\BibitemShut {NoStop}%
\bibitem [{\citenamefont {Sedlmayr}\ \emph {et~al.}(2016)\citenamefont {Sedlmayr}, \citenamefont {Aguiar-Hualde},\ and\ \citenamefont {Bena}}]{sedlmayr2016majorana}%
  \BibitemOpen
  \bibfield  {author} {\bibinfo {author} {\bibfnamefont {N.}~\bibnamefont {Sedlmayr}}, \bibinfo {author} {\bibfnamefont {J.}~\bibnamefont {Aguiar-Hualde}},\ and\ \bibinfo {author} {\bibfnamefont {C.}~\bibnamefont {Bena}},\ }\bibfield  {title} {\bibinfo {title} {Majorana bound states in open quasi-one-dimensional and two-dimensional systems with transverse rashba coupling},\ }\href@noop {} {\bibfield  {journal} {\bibinfo  {journal} {Physical Review B}\ }\textbf {\bibinfo {volume} {93}},\ \bibinfo {pages} {155425} (\bibinfo {year} {2016})}\BibitemShut {NoStop}%
\bibitem [{\citenamefont {Bena}(2017)}]{bena2017testing}%
  \BibitemOpen
  \bibfield  {author} {\bibinfo {author} {\bibfnamefont {C.}~\bibnamefont {Bena}},\ }\bibfield  {title} {\bibinfo {title} {Testing the formation of majorana states using majorana polarization},\ }\href@noop {} {\bibfield  {journal} {\bibinfo  {journal} {Comptes Rendus. Physique}\ }\textbf {\bibinfo {volume} {18}},\ \bibinfo {pages} {349} (\bibinfo {year} {2017})}\BibitemShut {NoStop}%
\bibitem [{\citenamefont {Awoga}\ and\ \citenamefont {Cayao}(2024)}]{awoga2024identifying}%
  \BibitemOpen
  \bibfield  {author} {\bibinfo {author} {\bibfnamefont {O.~A.}\ \bibnamefont {Awoga}}\ and\ \bibinfo {author} {\bibfnamefont {J.}~\bibnamefont {Cayao}},\ }\bibfield  {title} {\bibinfo {title} {Identifying trivial and majorana zero-energy modes using the majorana polarization},\ }\href@noop {} {\bibfield  {journal} {\bibinfo  {journal} {Physical Review B}\ }\textbf {\bibinfo {volume} {110}},\ \bibinfo {pages} {165404} (\bibinfo {year} {2024})}\BibitemShut {NoStop}%
\bibitem [{\citenamefont {Jackiw}\ and\ \citenamefont {Rebbi}(1976)}]{jackiw1976solitons}%
  \BibitemOpen
  \bibfield  {author} {\bibinfo {author} {\bibfnamefont {R.}~\bibnamefont {Jackiw}}\ and\ \bibinfo {author} {\bibfnamefont {C.}~\bibnamefont {Rebbi}},\ }\bibfield  {title} {\bibinfo {title} {Solitons with fermion number $1/2$},\ }\href@noop {} {\bibfield  {journal} {\bibinfo  {journal} {Physical Review D}\ }\textbf {\bibinfo {volume} {13}},\ \bibinfo {pages} {3398} (\bibinfo {year} {1976})}\BibitemShut {NoStop}%
\bibitem [{\citenamefont {de~Juan}\ \emph {et~al.}(2012)\citenamefont {de~Juan}, \citenamefont {Sturla},\ and\ \citenamefont {Vozmediano}}]{de2012space}%
  \BibitemOpen
  \bibfield  {author} {\bibinfo {author} {\bibfnamefont {F.}~\bibnamefont {de~Juan}}, \bibinfo {author} {\bibfnamefont {M.}~\bibnamefont {Sturla}},\ and\ \bibinfo {author} {\bibfnamefont {M.~A.}\ \bibnamefont {Vozmediano}},\ }\bibfield  {title} {\bibinfo {title} {Space dependent fermi velocity in strained graphene},\ }\href@noop {} {\bibfield  {journal} {\bibinfo  {journal} {Physical review letters}\ }\textbf {\bibinfo {volume} {108}},\ \bibinfo {pages} {227205} (\bibinfo {year} {2012})}\BibitemShut {NoStop}%
\end{thebibliography}%
\clearpage

\appendix
\begin{widetext}
 \normalsize

\section{Symmetric and Asymmetric Strain} \label{A1}
\paragraph*{(i) Unstrained system ($\epsilon_0=0$).}
Figure~\ref{fig:A1} presents the reference case in the absence of strain already discussed in the main text. In Fig.~\ref{fig:A1}(a-i), the lowest-energy state is strongly localized at the two opposite ends of the nanowire, consistent with spatially separated Majorana-bound states in the topological regime. The corresponding spectral evolution in Fig.~\ref{fig:A1}(b-i) shows a clear bulk-gap closing and reopening near the critical Zeeman field $h/\Delta \approx 4$, signaling a topological phase transition. Simultaneously, the nonlocal polarization amplitude $|\mathcal{P}^{l}|$ evolves sharply from trivial values toward nearly unity in magnitude, while the energy splitting $\delta E$ approaches zero inside the reopened bulk-gap $\xi$. This strain-free case therefore, serves as the benchmark topological Majorana phase against which the strained configurations are compared.

\begin{figure}
    \centering
    \includegraphics[width=\linewidth]{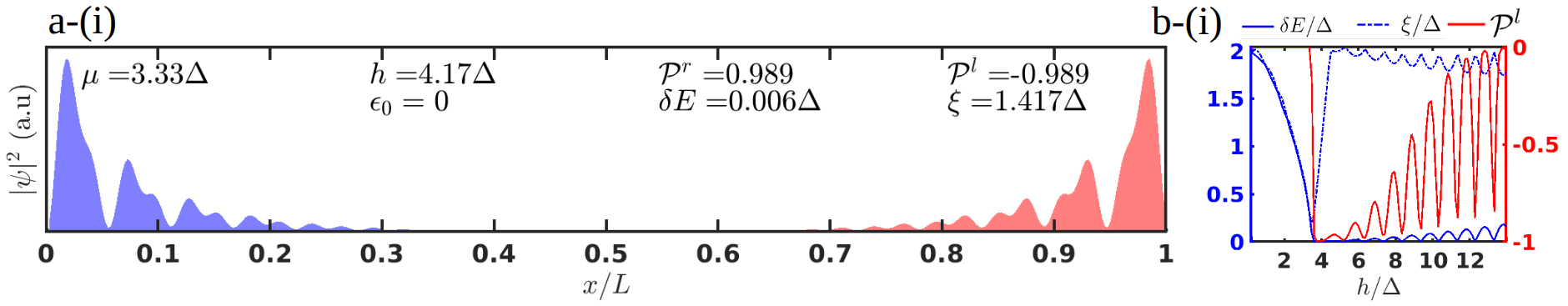}
    \caption{Low-energy bound states in a clean (no disorder) nanowire without strain ($\epsilon_{0}=0$). Panel (a-i) shows the spatial probability density of the lowest-energy mode, exhibiting end-localized Majorana-bound states. Panel (b-i) shows the corresponding evolution of the energy splitting $\delta E$, bulk-gap $\xi$, and nonlocal polarization $\mathcal{P}^{l}$ as functions of Zeeman field $h/\Delta$, indicating the topological phase transition through gap closing and reopening.}
    \label{fig:A1}
\end{figure}

\paragraph*{(ii) One-sided strain.}
Figure~\ref{fig:A2} illustrates the effect of an asymmetric strain profile applied from one end of the nanowire, mimicking a device fixed at the left boundary and stretched from the right boundary. Panels~\ref{fig:A2}(a-i)--(a-vi) show that the left-end mode remains largely pinned near the unstrained boundary, whereas the right-end mode is progressively displaced toward the interior as the strain amplitude $\epsilon_0$ increases. This demonstrates that nonuniform strain acts as a spatial control knob capable of selectively moving only the Majorana component located near the strained side. The accompanying spectra in panels~\ref{fig:A2}(b-i)--(b-vi) reveal that the dotted blue bulk-gap curve no longer exhibits a sharp closing-reopening structure but just acquires a finite minimum. The gap evolves smoothly while low-energy states remain near zero energy over an extended parameter window. Correspondingly, the polarization remains finite but no longer coincides with a conventional bulk topological transition. These features indicate a crossover from a topological Majorana pair to partially separated Andreev bound states, where one Majorana component is shifted into the bulk while the opposite component remains edge localized. 

\begin{figure}
    \centering
    \includegraphics[width=\linewidth]{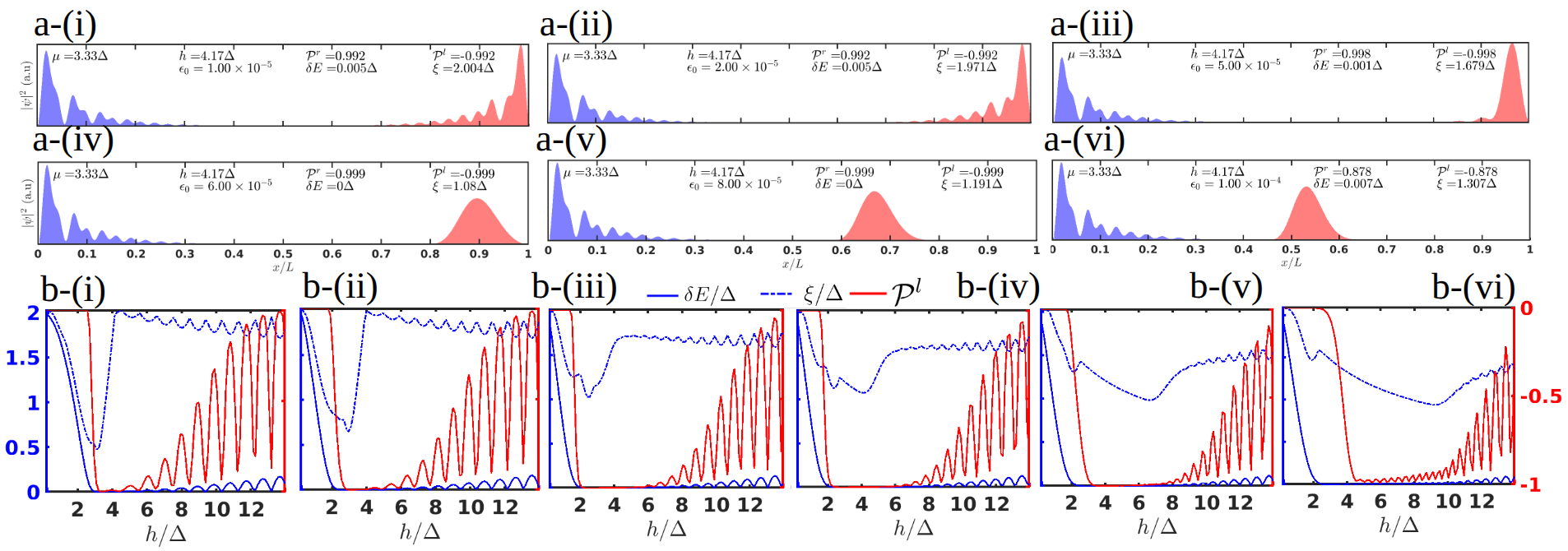}
    \caption{Effect of one-sided strain applied from a single end of the nanowire. Panels (a-i)--(a-vi) show the spatial profiles of the lowest-energy mode for increasing strain strength $\epsilon_{0}$, demonstrating the selective inward motion of the mode near the strained end. Panels (b-i)--(b-vi) display the corresponding evolution of $\delta E$, $\xi$, and $\mathcal{P}^{l}$ versus $h/\Delta$.}
    \label{fig:A2}
\end{figure}
\begin{figure}
    \centering
    \includegraphics[width=\linewidth]{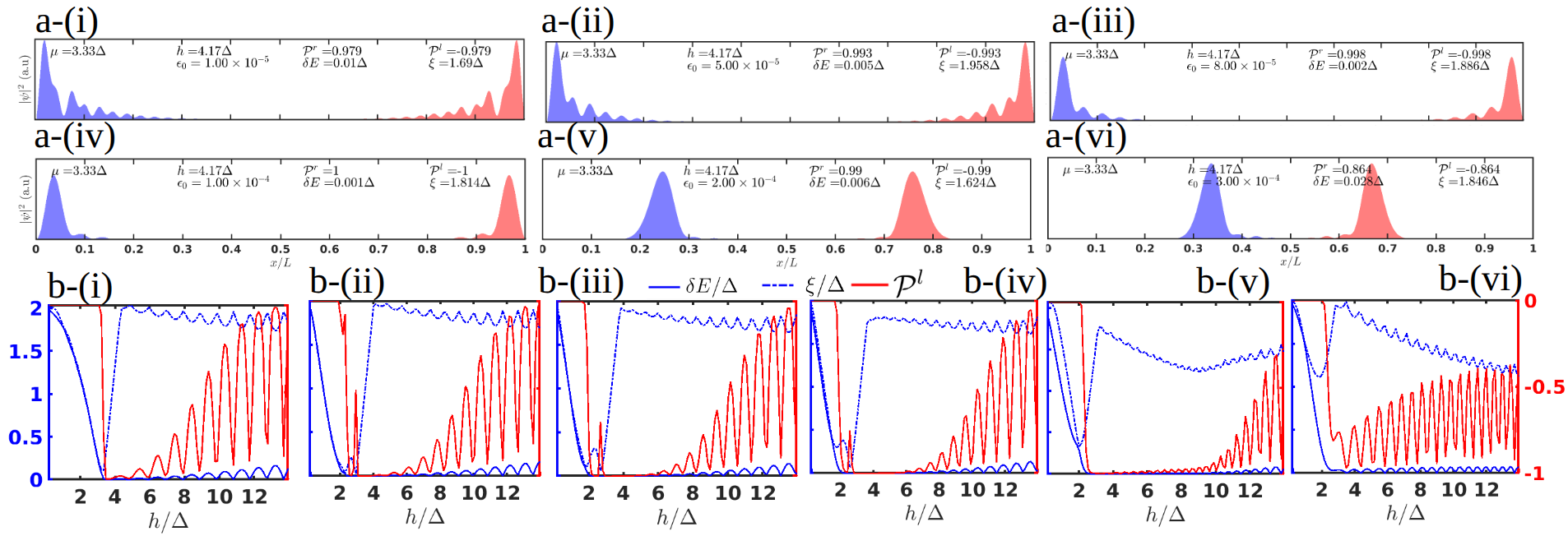}
    \caption{Effect of symmetric strain applied from both ends of the nanowire. Panels (a-i)--(a-vi) show the spatial profiles of the lowest-energy mode for increasing strain strength $\epsilon_{0}$, where both end-localized modes move toward the interior while maintaining approximate left-right symmetry. Panels (b-i)--(b-vi) show the corresponding evolution of $\delta E$, $\xi$, and $\mathcal{P}^{l,r}$ as functions of $h/\Delta$.}
    \label{fig:A3}
\end{figure}

\paragraph*{(iii) Symmetric two-sided strain.}
Figure~\ref{fig:A3} shows the response to a symmetric strain profile applied from both ends, corresponding to a nanowire pulled simultaneously from the two boundaries. In contrast to the one-sided case, panels~\ref{fig:A3}(a-i)--(a-vi) demonstrate that both end-localized modes move symmetrically inwards as the strain strength increases. An important distinction from the one-sided strain configuration is that the spectral transition is considerably more gradual. As seen in panels~\ref{fig:A3}(b-i)--(b-iii), the gap-closing region near the topological transition point remains only weakly perturbed at low to moderate strain, with merely a slight distortion around the critical field. Thus, symmetric strain initially leaves the underlying topological gap structure largely intact. Only for stronger strain amplitudes [panels~\ref{fig:A3}(b-iv)--(b-vi)] does a more pronounced reopening of the low-energy gap emerge, accompanied by a clear reshaping of the spectrum. This behavior indicates that two-sided strain modifies the bound-state spectrum more gently than one-sided strain, where the asymmetry drives a faster reconstruction of the transition region. Finite polarization persists over a broad field range throughout this evolution, showing that the low-energy modes retain substantial Majorana character even as their spatial profiles are displaced inwards. However, the continued reduction of separation between the two components enhances their overlap, leading to stronger hybridization and increasingly visible oscillatory energy splittings at larger fields. For sufficiently large strain, the system crosses over into a regime best described as a pair of Andreev bound states localized away from the physical ends. 

\section{Quasi-one-dimensional strip under symmetric strain}\label{A1_2}
We now examine the effect of a spatially symmetric strain profile on a quasi-one-dimensional proximitized strip consisting of five coupled one-dimensional chains. The system is described by the transverse hopping and Rashba coupling terms of Eq.~\ref{Ham_q1D_long} in addition to the longitudinal nanowire Hamiltonian described in Eq.~\ref{Eq:Ham1D}. Because the transverse Rashba term breaks chiral symmetry, the chiral definition of Majorana polarization is no longer applicable. Therefore, throughout this analysis, we employ the generalized particle-hole based Majorana polarization defined in Eq.~\ref{Eq:P_region2_PRB}, identical to the diagnostic used for the graphene strip geometries and in our earlier quasi-1D study~\ref{graphene_section}. 
Figure~\ref{fig:quasi1d} (c) shows the representative disorder realization for parameters $V_0=2~\mathrm{meV}$, $\lambda=10~\mathrm{nm}$, and impurity density $n_d=0.05/\mathrm{site}$. 
A symmetric strain is applied along the longitudinal ($x$) direction, corresponding physically to pulling the left and right ends of the strip away from the center, so that the induced hopping and spin-orbit renormalization is strongest near the two boundaries and weaker toward the middle.
The upper-left set of panels, Fig.~\ref{fig:quasi1d}(a-i)--(a-v), illustrates a crossover from partially separated Andreev bound states to topological Majorana modes as the strain amplitude increases. In the absence of strain [panel (a-i)], the low-energy state is predominantly localized near one end of the system, with negligible weight at the opposite boundary, characteristic of a trivial psABS. As the strain strength is increased, spectral splitting decreases and the wave function reorganizes progressively toward both ends of the strip. For intermediate strain [panels (a-iii) and (a-iv)], the left and right Majorana polarization magnitudes $|\mathcal{P}^{l}|$ and $|\mathcal{P}^{r}|$ become comparable, with the finite nonlocal correlator $\mathcal{P}_{l}\mathcal{P}_{r}^{*}$. At larger strain, the mode becomes clearly end-localized on opposite sides of the strip, signaling a strain-induced conversion of an initially trivial psABS into a Majorana pair.
In contrast, Fig.~\ref{fig:quasi1d}(b-i)--(b-v) presents the opposite evolution, namely a Majorana-to-psABS crossover. In the low-strain regime [panels (b-i)--(b-iii)], the low-energy mode is strongly localized at the two ends with nearly symmetric polarization weights and a large correlation $\mathcal{P}_{l}\mathcal{P}_{r}^{*}$, consistent with robust nonlocal Majorana modes. Upon further increasing the strain, the two end modes are pushed inward from the boundaries. In the strongest-strain regime [panel (b-v)], the weight is concentrated at two interior locations rather than at the physical ends.
These results demonstrate that the qualitative role of strain in the quasi-1D strip closely parallels that found in strictly one-dimensional nanowires: strain can either separate overlapping trivial states into genuine Majorana end modes or, conversely, drive initially topological Majorana modes inward and convert them into psABSs through enhanced overlap. Thus, despite the absence of chiral symmetry and the multichannel nature of the quasi-1D geometry, the principal strain-induced phenomenology remains unchanged.
For completeness, we also examined symmetric strain applied along the transverse direction, corresponding to pulling the upper and lower edges of the five-chain strip away from each other. In contrast to longitudinal strain, this deformation produces negligible changes to the low-energy wave functions and the polarization structure. Since the strip contains only five chains across its width, the transverse degree of freedom is too limited to generate substantial spatial redistribution of the bound states. Consequently, strain applied along the long direction remains the dominant and physically relevant tuning parameter for quasi-one-dimensional strips.
\begin{figure}
    \centering
    \includegraphics[width=\linewidth]{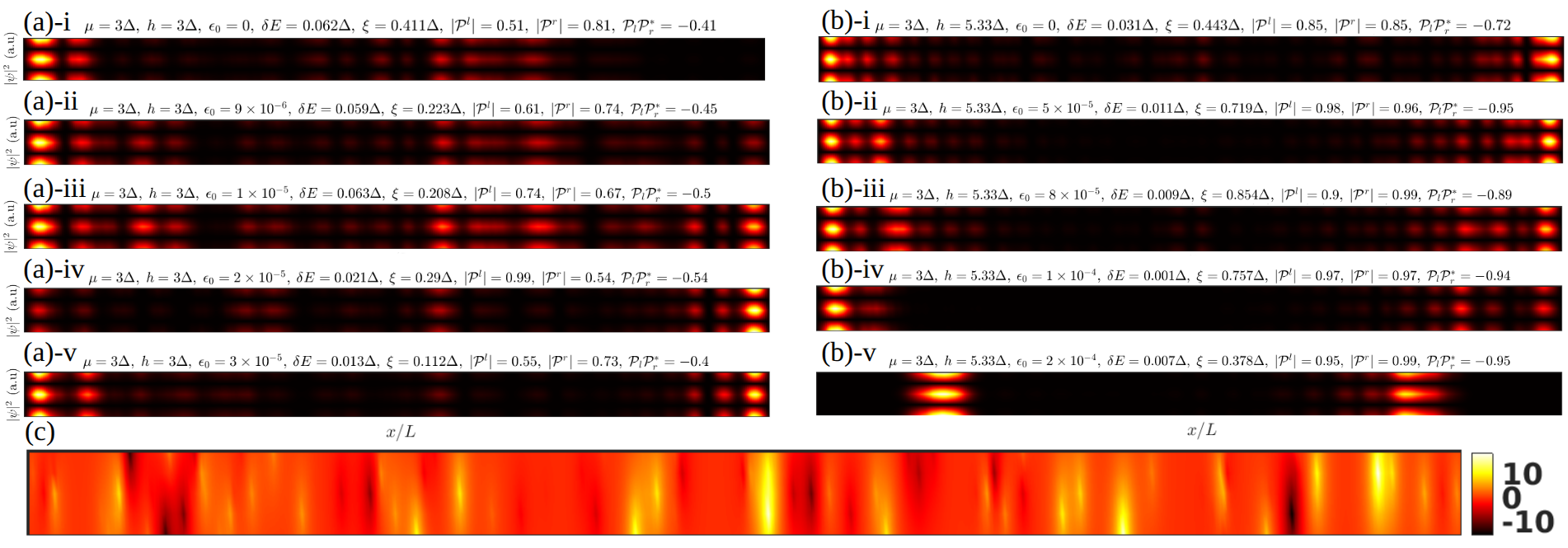}
    \caption{Low-energy wave-function probability density $|\psi|^2$ in a disordered quasi-one-dimensional strip comprising five coupled chains under a symmetric longitudinal strain profile. Panels (a-i)–(a-v) show a crossover from partially separated Andreev bound states to Majorana modes as the strain strength increases, while panels (b-i)–(b-v) display the opposite evolution from Majorana modes to psABS. The quantities listed above each panel denote the strain amplitude $\epsilon_0$, the energy splitting between the two lowest modes $\delta E$, bulk-gap $\xi$, left/right Majorana polarizations $|\mathcal{P}^{l,r}|$, and the nonlocal correlator $\mathcal{P}_{l}\mathcal{P}_{r}^{*}$. Panel (c) shows the corresponding disorder potential profile for $V_0=2~\mathrm{meV}$, $\lambda=10~\mathrm{nm}$, and impurity density $n_d=0.05/\mathrm{site}$.}
    \label{fig:quasi1d}
\end{figure}

\section{Zigzag graphene}~\label{App:zzg}
In this section, we investigate the combined effects of lattice disorder and strain (both uniaxial and symmetric) on the electronic and topological properties of vertically oriented zigzag graphene nanoribbons featuring short armchair edges.

\paragraph*{(i) Clean zigzag with uniaxial strain}
For clean zigzag ribbons, uniaxial strain applied along the $+y$-direction fails to open a robust topological gap, yielding a dense accumulation of low-energy states across the entire strain regime considered (e.g., for $B_y = 0.7t$ and $B_y = 0.8t$). Figure~\ref{clean_zigzag} illustrates the variation of the band gap and Majorana polarization across the strip. Unlike in armchair nanoribbons, where strain induces a systematic, directional displacement of well-localized modes away from the strained region, the modes in the zigzag geometry do not exhibit controlled spatial relocation. Instead, the LDOS confirms that these modes remain heavily delocalized across the bulk of the system [Fig ~\ref{clean_zigzag}a(iii)-a(vi)] and [Fig ~\ref{clean_zigzag}b(iii)-b(vi)]. Consequently, despite the presence of large Majorana polarization values and the absence of any disorder, the small band gap and delocalisation of modes throughout the strip indicate that these near-zero-energy states lack topological protection and do not constitute well-defined Majorana zero modes.

\begin{figure}
    \centering
    \includegraphics[width=1\columnwidth]{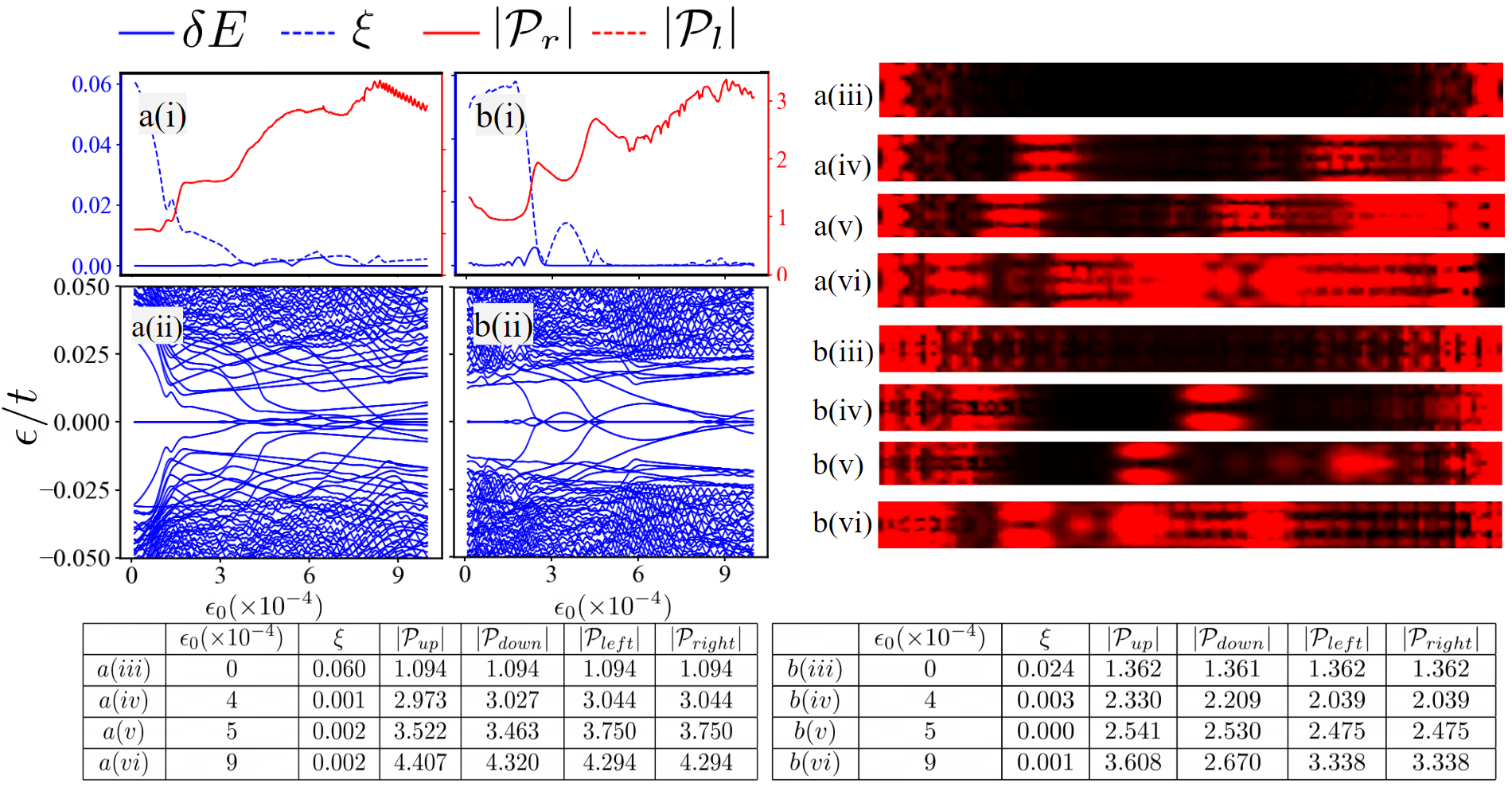}
    \caption{Majorana polarization and spectral evolution in a finite-size \textbf{clean zigzag} graphene strip under non-uniform longitudinal uniaxial strain (along the $y$-direction) for Zeeman fields (a) $B_y = 0.4t$ and (b) $B_y = 0.6t$, at a fixed chemical potential $\mu = 2.0t$. Panels (a-i) and (b-i) show the mode splitting $\delta E$ (solid blue), bulk gap $\xi$ (dashed blue), and absolute Majorana polarization at the right ($|P_r|$, solid red) and left ($|P_l|$, dashed red) boundaries as a function of uniaxial strain $\epsilon_0$. Panels (a-ii) and (b-ii) display the low-energy spectrum corresponding to the $100$ eigenvalues closest to zero energy. Panels (a-iii)--(a-vi) and (b-iii)--(b-vi) illustrate the spatial profiles of the local density of states (LDOS) near zero energy for the representative strain values listed below.}
    \label{clean_zigzag}
\end{figure}
\paragraph*{(ii) Normal disorder with uniaxial strain}
Here, we introduce random on-site potential disorder by distributing $N_d = 20$ impurity sites drawn from a Gaussian distribution, as depicted in Fig.~\ref{z_dis4}(c). The introduction of disorder strongly hybridizes the edge and bulk states, driving the low-energy modes deeper into the interior of the nanoribbon. When uniaxial strain is subsequently applied, this bulk delocalization is significantly amplified [Fig.~\ref{z_dis4}(a(ii)--b(ii))]. The band gap collapses almost entirely, generating a high density of states near zero energy. Although the polarization values remain nominally large, the LDOS spatial profiles [Fig.~\ref{z_dis4}(a(iii)--a(vi)) and Fig.~\ref{z_dis4}(b(iii)--b(vi))] clearly demonstrate that the low-energy states are entirely extended. Ultimately, these results demonstrate a lack of a controlled tuning mechanism in zigzag strips; uniaxial strain in the presence of disorder is not capable of isolating well-defined, localized edge states.

\paragraph*{(iii) Normal disorder with symmetric strain}
We further explore the application of symmetric longitudinal strain, where both ends of the disordered zigzag ribbon are stretched symmetrically away from the center [see Fig.~\ref{fig:hopping_strain}-(c)]. Similar to the uniaxial case, the system exhibits a vanishingly small band gap across the considered parameter space [Fig.~\ref{z_dis4_bothy}(a(ii)--b(ii))]. The symmetric strain actively promotes the inward migration of states from both edges, maximizing bulk delocalization [Fig.~\ref{z_dis4_bothy}(a(iii)--a(vi)) and Fig.~\ref{z_dis4_bothy}(b(iii)--b(vi))]. The persistence of a near-zero gap, coupled with broadly distributed LDOS profiles and large polarization values, demonstrates that the low-energy states remain trivially extended. Therefore, for zigzag strips with short armchair edges, symmetric strain is an ineffective parameter for engineering or stabilizing topologically protected localized modes.

\begin{figure}
    \centering
    \includegraphics[width=1\columnwidth]{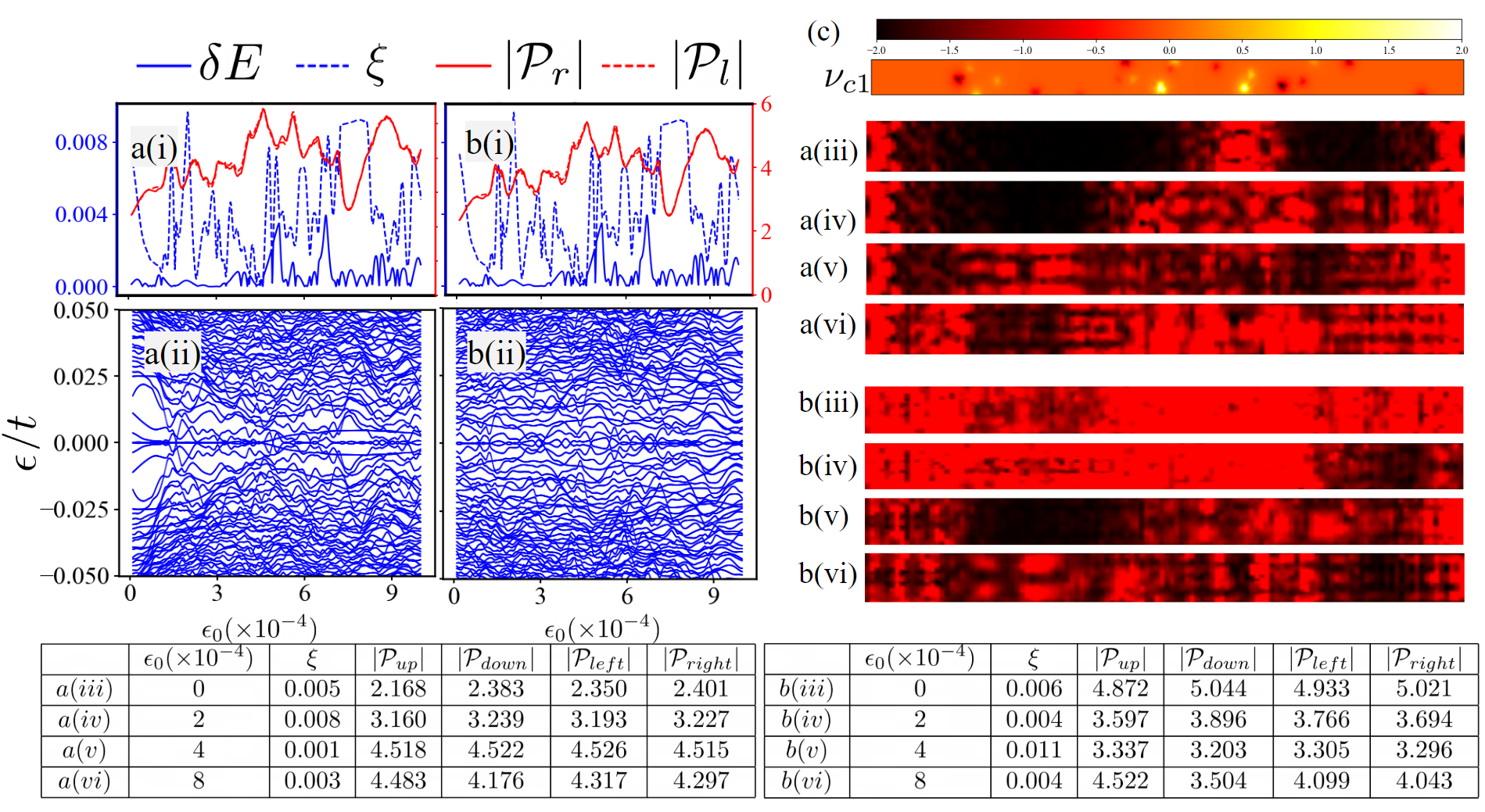}
     \caption{Majorana polarization and spectral evolution in a finite-size \textbf{disordered zigzag} graphene strip for Zeeman fields (a) $B_y = 0.4t$ and (b) $B_y = 0.6t$, at fixed chemical potential $\mu = 2.0t$. The disorder configuration $\nu_{c1}$, depicted in panel (c), is generated using Eq.~\ref{eq:disorder} with parameters $V_0 = 1.0t$, $\lambda = 1.0t$, and on-site amplitudes $A_i \in [-2.0,\,2.0]t$ applied across $N_d = 20$ lattice sites. Panels (a-i) and (b-i) detail the mode splitting $\delta E$ (solid blue), bulk gap $\xi$ (dashed blue), and absolute Majorana polarization ($|P_r|$ solid red; $|P_l|$ dashed red) versus uniaxial strain $\epsilon_0$. Panels (a-ii) and (b-ii) plot the $100$ eigenvalues closest to zero energy. The spatial LDOS profiles near zero energy are shown in panels (a-iii)--(a-vi) and (b-iii)--(b-vi) for selected strain parameters.}
    \label{z_dis4}
\end{figure}
\section{Armchair with symmetric strain}\label{App:armchairsymm}
\paragraph*{ Normal disorder with symmetric strain in armchair strip}
We contrast the behavior of a disordered armchair strip featuring shorter zigzag edges under uniaxial strain, discussed in section ~\ref{B-ii}, with symmetric strain (Fig.~\ref{dis_arm_bothy}). In the unstrained, disordered regime, the system maintains some localization to edges along, even though disorder induces partial delocalization into the bulk [Fig.~\ref{dis_arm_bothy}(a(iii), b(iii))]. 
\begin{figure}
    \centering
    \includegraphics[width=1\columnwidth]{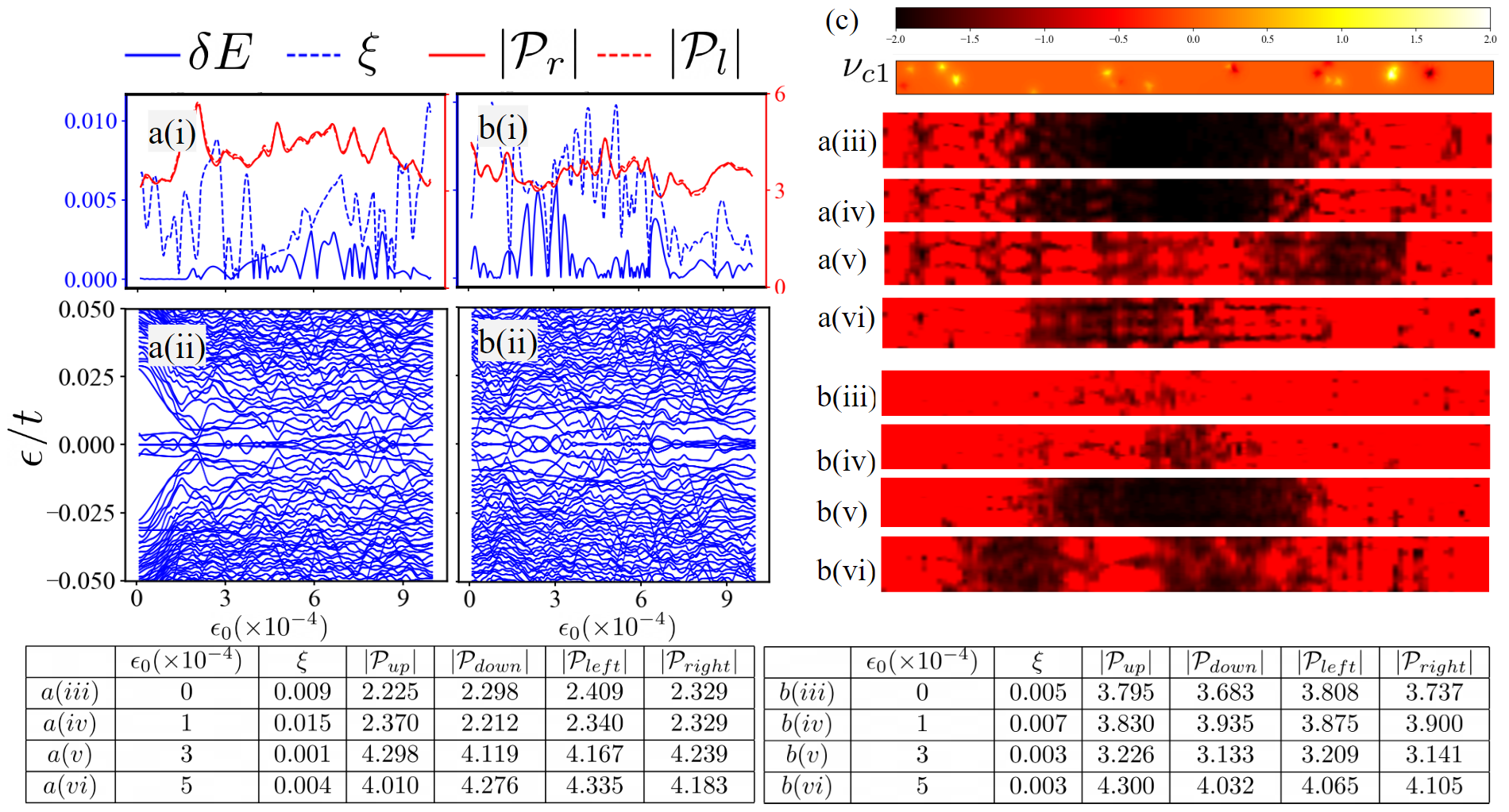}
     \caption{Effects of symmetric longitudinal strain on a finite-size \textbf{disordered zigzag} graphene strip for Zeeman fields (a) $B_y = 0.4t$ and (b) $B_y = 0.6t$, at $\mu = 2.0t$. The disorder configuration $\nu_{c1}$ (panel c) follows Eq.~\ref{eq:disorder} with $V_0 = 1.0t$, $\lambda = 1.0t$, $A_i \in [-2.0,\,2.0]t$, and $N_d = 20$. Panels (a-i) and (b-i) display mode splitting $\delta E$ (solid blue), bulk gap $\xi$ (dashed blue), and absolute boundary polarizations ($|P_r|$, $|P_l|$ in red) as a function of symmetric strain $\epsilon_0$. Low-energy spectra ($100$ eigenvalues nearest zero) are shown in (a-ii) and (b-ii), while panels (a-iii)--(a-vi) and (b-iii)--(b-vi) map the corresponding LDOS profiles near zero energy.}
    \label{z_dis4_bothy}
\end{figure}

\begin{figure}
    \centering
    \includegraphics[width=1\columnwidth]{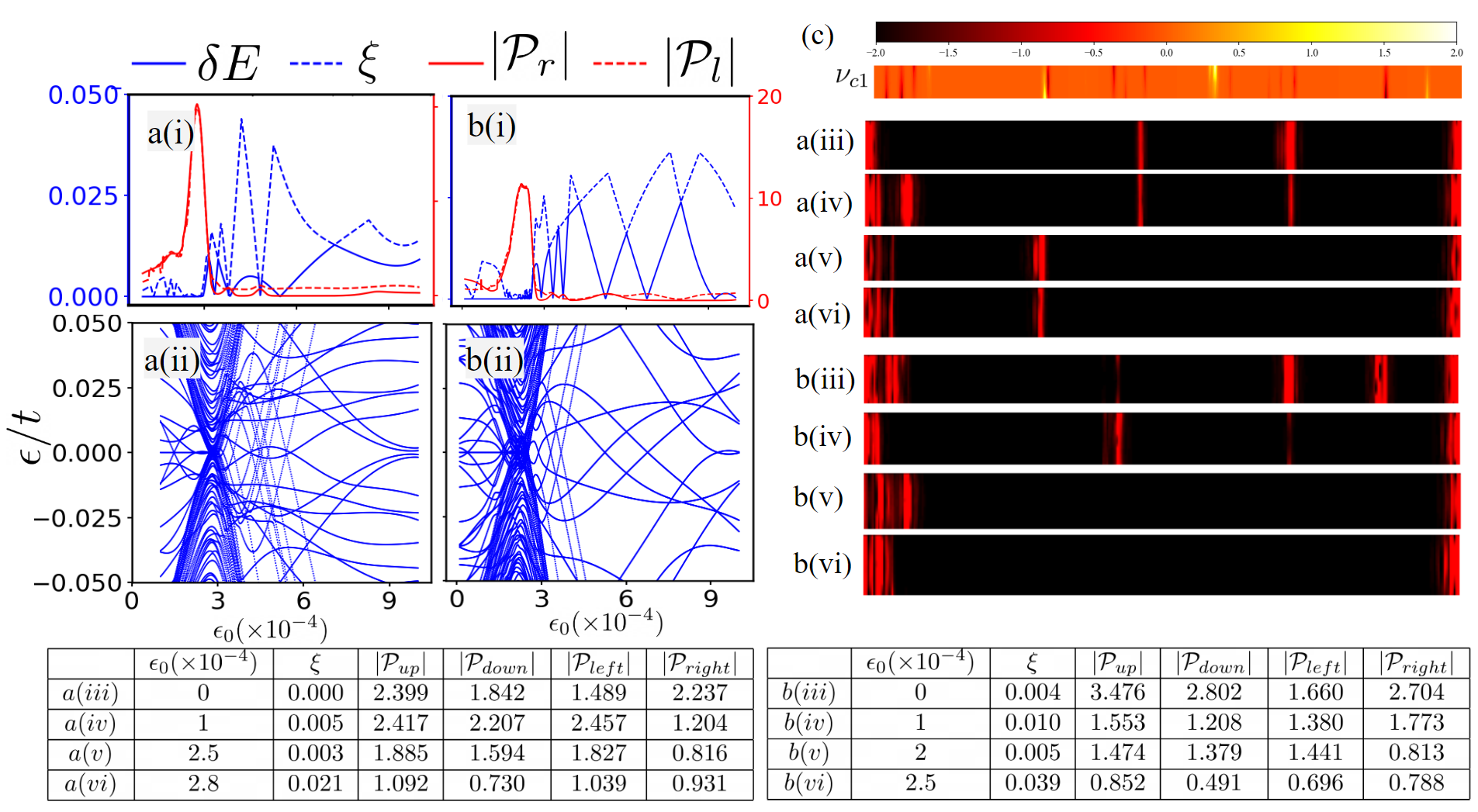}
     \caption{Effects of symmetric longitudinal strain on a finite-size \textbf{disordered armchair} graphene strip for Zeeman fields (a) $B_x = 0.7t$ and (b) $B_x = 0.8t$, at $\mu = 2.0t$. The disorder configuration $\nu_{c1}$ (panel c) follows Eq.~\ref{eq:disorder} utilizing $V_0 = 1.0t$, $\lambda = 1.0t$, $A_i \in [-2.0,\,2.0]t$, and $N_d = 20$. Panels (a-i) and (b-i) track the mode splitting $\delta E$ (solid blue), bulk gap $\xi$ (dashed blue), and boundary Majorana polarizations ($|P_r|$, $|P_l|$ in red) against symmetric strain $\epsilon_0$. The $100$ eigenvalues closest to zero energy are plotted in (a-ii) and (b-ii). Spatial LDOS distributions near zero energy for specific strain parameters are provided in panels (a-iii)--(a-vi) and (b-iii)--(b-vi).}
    \label{dis_arm_bothy}
\end{figure}

However, the application of symmetric strain destabilizes this regime and fails to induce a robust topological phase. As depicted in [Fig.~\ref{dis_arm_bothy}a(i)) and (b(i)], the splitting $\delta E$ is not clearly sustained; instead, it undergoes repeated closures. Concurrently, the energy spectra in Fig.~\ref{dis_arm_bothy}(a(ii)) and (b(ii)) illustrate a dense accumulation of energy modes near $E=0$. Beyond a certain strain threshold, these modes do not settle into well-defined, isolated zero-energy states but rather remain at finite energies. Although this strain-induced spectral weight near zero energy is accompanied by significant polarization, the fundamental lack of a hard, finite bulk gap critically undermines topological protection. Furthermore, achieving controlled spatial localization of these modes proves unsuccessful. The local density of states mappings in Fig.~\ref{dis_arm_bothy}(a(iii)--a(vi)) demonstrate that symmetric strain does not systematically drive the modes to the system boundaries. 

While there are specific parameter windows where the modes appear to shift toward the edges as seemingly suggested by [Fig.~\ref{dis_arm_bothy}(b(iii)--b(vi)], this apparent edge localization is fragile and misleading. Cross-referencing these spatial profiles with the corresponding energy spectrum in [Fig.~\ref{dis_arm_bothy}(b(ii)] reveals that in these exact regions, the bulk gap is collapsed. The modes accumulating near zero energy in this regime form a continuum rather than isolated edge states. Consequently, the rigorous criteria for true, topologically protected Majorana states—namely, strict zero energy, a finite bulk gap, robust edge localization, and finite polarization—are not simultaneously satisfied. This indicates that symmetric strain does not act as a reliable or controlled tuning parameter for stabilizing topological edge states in this configuration.

\end{widetext}
\end{document}